\documentclass[aps,nofootinbib,noshowpacs,showkeys,preprintnumbers,amsmath,amssymb]{revtex4}
\usepackage{graphicx,color}
\usepackage{graphics}
\usepackage{rotating}
\usepackage{amssymb}

\usepackage{color}

\usepackage{epsfig}
\usepackage{dcolumn}
\usepackage{bm}
\usepackage{multirow}

\def\be{\begin{equation}}
\def\ee{\end{equation}}
\def\ba{\begin{eqnarray}}
\def\ea{\end{eqnarray}}
\def\bea{\begin{eqnarray}}
\def\eea{\end{eqnarray}}
\def\bes{\begin{subequations}}
\def\ees{\end{subequations}}
\def\bear{\begin{array}}
\def\eear{\end{array}}

\newcommand{\A}{{\mathcal{A}}}
\newcommand{\tA}{{\widetilde {\mathcal{A}}}}
\newcommand{\ta}{{\widetilde a}}
\newcommand{\ths}{{\widetilde {\mathfrak h}}}
\newcommand{\tH}{{\widetilde {\mathfrak H}}}
\newcommand{\tB}{{\widetilde B}}
\newcommand{\tal}{{\widetilde \alpha}}

\newcommand{\ts}{{\widetilde s}}
\newcommand{\tS}{{\widetilde S}}
\newcommand{\tZ}{{\widetilde Z}}
\newcommand{\tK}{{\widetilde K}}
\newcommand{\td}{{\widetilde d}}
\newcommand{\cC}{{\cal C}}
\newcommand{\cO}{{\cal O}}
\newcommand{\cD}{{\cal D}}
\newcommand{\cDo}{{{\cal D}^{(1)}}}
\newcommand{\cF}{{\cal F}}
\newcommand{\cFo}{{{\cal F}^{(1)}}}
\newcommand{\cK}{{\cal K}}
\newcommand{\tcK}{{\widetilde {\cal K}}}
\newcommand{\tcKo}{{{\widetilde {\cal K}}^{(1)}}}
\newcommand{\tk}{{\widetilde k}}
\newcommand{\tf}{{\widetilde f}}
  
\newcommand{\MSbar}{\overline{\rm MS}}

\begin{document}

\title{Renormalon-based resummation for spacelike and timelike QCD quantities whose perturbation expansion has general form}
\author{C\'esar Ayala$^a$}
\email{c.ayala86@gmail.com}
\author{Gorazd Cveti\v{c}$^b$}
\email{gorazd.cvetic@gmail.com}
\author{Reinhart K\"ogerler$^c$}
\email{koeg@physik.uni-bielefeld.de}
\affiliation{$^a$Departamento de Ingenier\'ia y Tecnolog\'ias, Sede La Tirana, Universidad de Tarapac\'a, Av.~La Tirana 4802, Iquique, Chile}
\affiliation{$^b$Department of Physics, Universidad T{\'e}cnica Federico Santa Mar{\'\i}a, Avenida España 1680, Valpara{\'\i}so, Chile} 
\affiliation{$^c$Department of Physics, Universit\"at Bielefeld, 33501 Bielefeld, Germany}

\date{\today}

\begin{abstract}
We present a generalisation of our previous approach of a renormalon-motivated resummation of the QCD observables. Previously it was applied to the spacelike observables whose perturbation expansion was ${\cal D}(Q^2) = a(Q^2) + {\cal O}(a^2)$, where $a(Q^2) \equiv \alpha_s(Q^2)/\pi$ is the running QCD coupling. Now we generalise the resummation to spacelike quantities ${\cal D}(Q^2) = a(Q^2)^{\nu_0} + {\cal O}(a^{\nu_0+1})$ and timelike quantities ${\cal F}(\sigma) = a(\sigma)^{\nu_0} + {\cal O}(a^{\nu_0+1})$, where $\nu_0$ is in general a noninteger number ($0<\nu_0 \leq 1$). We evaluate with this approach a timelike quantity, namely the scheme-invariant factor of the Wilson coefficient of the chromomagnetic operator in the heavy-quark effective Lagrangian, and related quantities.
\end{abstract}
\maketitle

\section{Introduction}
\label{sec:intr}

The study of QCD properties at low-momentum transfers ($\lesssim 1$ GeV) is challenging due to the nonperturbative nature of QCD. There are some efforts to try to unify the high- and low-momentum regimes, and most of them lead to noncontinuous and/or nonholomorphic transition between these two regimes. Recently, in \cite{BrodECH,ACECH}, certain effective charges with holomorphic connection between the two regimes are supposed to contain information of the theory, but they are unfortunately tied to only one specific observable.
  
In this work, we present a resummation formalism that evaluates the (leading-twist part of the) QCD observables with a single integral, keeping some of the ideas of the effective charge, but with clear differences. An early version of this resummation was constructed by Neubert \cite{Neubert} some time ago for the large-$\beta_0$ approximation (cf.~also \cite{NC}). Subsequently, the resummation formalism was extended to any loop-level \cite{renmod}, leading to an integration involving the (exact) running coupling and an observable-dependent characteristic (weight) function, where the latter can be (approximately) determined by the knowledge of the renormalon structure of the considered observable. These resummations were constructed for the case when the perturbation expansions of the considered observables were series of integer powers of the coupling. In the present work, we extend these resummations to the case of observables whose perturbation expansions have, in general, noninteger powers of the couplings. In principle, such resummations are for the spacelike QCD observables, but here we also extend this formalism to the general case of timelike QCD observables, and present a phenomenological application of this formalism. \textcolor{black}{Expansions in noninteger powers appear in general for observables whose one-loop associated anomalous dimensions are nonzero. }  

In Ref.~\cite{renmod}, a renormalon-motivated resummation procedure has been developed for spacelike QCD observables whose perturbation expansion is ${\cal D}(Q^2) = a(Q^2) + {\cal O}(a^2)$, where $a(Q^2) \equiv \alpha_s(Q^2)/\pi$ and $Q^2 \equiv - q^2$ ($=- (q^0)^2 + {\vec q}^2$) is in the spacelike (i.e., non-timelike) regime in the complex plane, i.e., $Q^2 \in \mathbb{C} \backslash (-\infty,0)$. Furthermore, a simple extension to timelike variables ${\cal F}(\sigma)$ ($\sigma > 0$) was made there, for the case when ${\cal F}(\sigma)$ is a contour integral of the aforementioned spacelike ${\cal D}(Q^2)$. The expansion of ${\cal D}(Q^2)$ was considered to be known exactly. In \cite{renmod} the resummation was applied to the Adler function ${\cal D}(Q^2)=d(Q^2)_{\rm Adl}$ and to the semihadronic $\tau$-lepton decay ratio ${\cal F}(\sigma) = r_{\tau}(\sigma)$ ($\sigma = m_{\tau}^2$). The resummation method was applied with perturbative (pQCD) coupling $a(Q^2)$, i.e., a coupling that has Landau singularities in the spacelike region $0 \leq Q^2 \leq {\Lambda}_c^2$. Thereafter, the same procedure was also performed there with holomorphic QCD couplings $\A(Q^2)$ (i.e., when $a(Q^2)$ is replaced by $\A(Q^2)$), i.e., couplings that have no Landau singularities and are thus holomorphic for $Q^2 \in  \mathbb{C} \backslash (-\infty,0)$ and hence reflect correctly the holomorphic structure of the spacelike QCD observables ${\cal D}(Q^2)$ in the complex $Q^2$-plane.

Later, we applied this method specifically to the (spacelike) QCD quantity Bjorken polarised sume rule (BSR), $d(Q^2)_{\rm BSR}$, with the coupling being either pQCD \cite{BSRpQCD} or holomorphic \cite{BSRAQCD}. With the holomorphic coupling, we can describe successfully the experimental data for $d(Q^2)_{\rm BSR}$ in a wider interval of positive $Q^2$ that includes also $Q^2 < 1 \ {\rm GeV}^2$.

In this work, we extend the described method to the resummation of the spacelike observables whose perturbation expansion is ${\cal D}= a(Q^2)^{\nu_0} + {\cal O}(a^{\nu_0+1})$ and timelike quantities with perturbation expansion ${\cal F}(\sigma) = a(\sigma)^{\nu_0} + {\cal O}(a^{\nu_0+1})$, where $\nu_0$ has in general a noninteger value  ($0<\nu_0 \leq 1$).
\textcolor{black}{Thus, this formalism allows us to evaluate in a consistent way timelike quantities such as decay rates, heavy quark masses and other quantities related with the heavy meson masses, etc., i.e., quantities especially interesting due to their active experimental development.}
As an illustration, we will apply this approach to a specific timelike observable $\cF(\sigma)$ (with $\nu_0=1/3$), and will perform the renormalon-motivated resummation by using both the pQCD coupling $a$ and a holomorphic coupling $\A$.

\textcolor{black}{In Sec.~\ref{sec:space} we present the resummation formalism for spacelike observables $\cD(Q^2)$ with noninteger $\nu_0$ power index. There we also explain how to obtain, from the knowledge of the renormalon structure of the (usual) Borel transform of such an observable, the characteristic function needed for the resummation of $\cD(Q^2)$. In Sec.~\ref{sec:time} we extend this formalism to timelike observables $\cF(\sigma)$. For such observables, the procedure to obtain the characteristic function needed in the resummation, becomes more complicated than in the spacelike case, as it involves relations between several Borel and modified Borel transforms of various related quantities. In Sec.~\ref{sec:AQCD} we present a specifically chosen QCD framework (3$\delta$AQCD) whose running QCD coupling is free from the Landau singularities and agrees at high momenta with pQCD coupling; such regularised coupling is needed to perform the mentioned resummations (formally) without ambiguities. In Sec.~\ref{sec:impl} we then apply the described formalism and coupling to the evaluation (resummation) of a specific timelike quantity, namely the renormalisation-scheme invariant factor of the Wilson coefficient of the chromomagnetic operator that appears in the heavy-quark effective theory (HQET) for hadronic bound states with one heavy quark ($c$ or $b$). In Sec.~\ref{sec:concl} we summarise the conclusions of this work. Proofs of some Theorems appearing in the main text were relegated to Appendices \ref{app:T2} and \ref{app:T4}, scheme transformations are summarised in Appendix \ref{app:RSch}, and an extended discussion of some mathematical and numerical aspects of the formalism are in Appendix \ref{app:matnum}.}

\section{Resummation for the case of the spacelike observable ${\cal D}(Q^2)$}
\label{sec:space}

We will describe the formalism by first using the pQCD coupling $a(Q^2)$, and thereafter comment on the simple changes that have to be made when the holomorphic coupling $\A(Q^2)$ is used.

The perturbation expansion of the considered spacelike obervable ${\cal D}(Q^2)$ is
\bes
\label{Dptlpt}
\bea
{\cal D}(Q^2) &=&  \sum_{n=0}^{\infty} d_n(\nu_0;\kappa) a(\kappa Q^2)^{\nu_0+n}
\label{Dpt}
\\
& = & \sum_{n=0}^{\infty} \td_n(\nu_0;\kappa) \ta_{\nu_0+n}(\kappa Q^2).
\label{Dlpt} \eea \ees
Here, $\nu_0$ is the index of the first expansion term ($0 < \nu_0 \leq 1$), and $\kappa \equiv \mu^2/Q^2$ denotes the renormalisation scale parameter which is positive and chosen to be $\kappa \sim 1$. The first expansion is in powers of $a(\kappa Q^2)$, with the expansion coefficients $d_n(\nu_0;\kappa)$ (note: $d_0=1$ by convention). The second expansion is in couplings  $\ta_{\nu_0+n}$ which are the noninteger generalisations of the logarithmic derivatives of $a$, with the expansion coefficients $\td_n(\nu_0;\kappa)$ (note: $\td_0=1$).\footnote{\label{ftnu01} For the integer $\nu_0=1$ we have: $\ta_{n+1}(Q^2) \equiv (-1)^n (n! \beta_0^n)^{-1} (d/d \ln Q^2)^n a(Q^2)$.} The noninteger powers and the noninteger logarithmic derivatives are related \cite{GCAK}
\bes
\label{taaata}
\bea
\ta_{\nu} &=& \sum_{m=0}^{\infty} k_m(\nu) a^{\nu + m} \qquad (k_0(\nu)=1)
\label{tanuanu}
\\
a^{\nu} & =& \sum_{m=0}^{\infty} {\tk}_m(\nu) \ta_{\nu+m} \qquad (\tk_0(\nu)=1).
\label{anutanu}
\eea \ees
Explicit expressions for the coefficients $k_m(\nu)$ and ${\tk}_m(\nu)$, valid for any $\nu>0$ ($\nu$ could be integer) and for $m \leq 4$, are given in \cite{GCAK}\footnote{In the first line of Eq.~(A.11) of App.~A in \cite{GCAK} there is an (obvious) typo, a sign is missing, the correct equation is ${\tilde k}_1(\nu)=-k_1(\nu)$.} and are independent of the momentum $Q^2$ or $\kappa Q^2$ in $a \equiv a(\kappa Q^2)$. On the other hand, when the coupling is instead holomorphic ($a \mapsto \A$), the general explicit expression for the holomorphic analog $\tA_{\nu}(Q^2)$ of $\ta_{\nu}(Q^2)$, for any $\nu > -1$ and for any holomorphic QCD (AQCD) framework, was derived in Ref.~\cite{GCAK} (see later).\footnote{\label{ftAQCDvspQCD}We point out already here that in AQCD frameworks the couplings $\A$ ($=\tA_1$) and $\tA_{\nu}$ are the basic couplings, and from them we derive the power analogs $\A_{\nu}$ (of $a^{\nu}$). This is just the opposite to the situation in pQCD, where the basic couplings are $a$ and the powers $a^{\nu}$, and from them the couplings $\ta_{\nu}$ are derived, cf.~Eq.~(\ref{tanuanu}).} 

When we use the relation (\ref{anutanu}) in the expansion (\ref{Dpt}), and take into account the notation (\ref{Dlpt}), we obtain the following relations between the coefficients $\td_n$ and $d_k$:
\be
\td_n(\nu_0,\kappa) = \sum_{s=0}^{n} \tk_{n-s}(\nu_0+s) d_s(\nu_0;\kappa).
\label{tdd} \ee
Analogously, when we use the relation (\ref{tanuanu}) in the expansion (\ref{Dlpt}), and take into account the notation (\ref{Dpt}), we obtain the relations between the coefficients $d_n$ and $\td_k$
\be
d_n(\nu_0,\kappa) = \sum_{s=0}^{n} k_{n-s}(\nu_0+s) \td_s(\nu_0;\kappa).
\label{dtd} \ee
The (generalised) logarithmic derivatives obey the differential (recursive) relation \cite{GCAK}
\be
\frac{d}{d \ln \kappa} \ta_{\nu}(\kappa Q^2) = (-\beta_0) \nu \ta_{\nu+1}(\kappa Q^2),
\label{dertanu} \ee
where $\beta_0=(11 - 2 n_f/3)/4$ is the one-loop beta coefficient of the renormalisation group equation (RGE) of the running $a(\kappa Q^2)$
\bes
\label{RGE}
\bea
\frac{d {a}(\mu^2)}{d \ln \mu^2} &=& - \beta_0 a(\mu^2)^2 - \beta_1 a(\mu^2)^3 - \beta_2 a(\mu^2)^4  - \beta_3 a(\mu^2)^5  - \ldots  
\label{RGEa} \\
& = &  - \beta_0 a(\mu^2)^2 \left[ 1 + c_1 a(\mu^2) + c_2 a(\mu^2)^2 + c_3 a(\mu^2)^3 + \ldots \right].
\label{RGEb} \eea \ees
In the mass-independent schemes (such as $\MSbar$ or MiniMOM), the first two coefficients ($\beta_0$, $\beta_1$) are universal, i.e., scheme-independent.\footnote{We have: $\beta_0=(11 - 2 n_f/3)/4$ and $\beta_1=(102-38 n_f/3)/16$, where $n_f$ is the number of active massless quark flavours. At low $|Q^2| \lesssim 1 \ {\rm GeV}^2$ we have $n_f=3$. The coefficients $\beta_j$ (or $c_j \equiv \beta_j/\beta_0$) with $j=2,3,\ldots$ then define the renormalisation scheme.}
Since ${\cal D}(Q^2)$ is an observable, it is independent of the renormalisation scale parameter $\kappa$ ($\equiv \mu^2/Q^2$), i.e., $(d/d \ln \kappa) {\cal D}(Q^2) = 0$. Applying this condition to the expansion (\ref{Dlpt}) for ${\cal D}(Q^2)$ and using the recursive relation (\ref{dertanu}) leads to
\bea
\lefteqn{
  \frac{d}{d \ln \kappa} {\cal D}(Q^2) = 0
}
\nonumber\\
& = & \ta_{\nu_0}(\kappa Q^2) \left[ \frac{d}{d \ln \kappa} \td_0(\nu_0,\kappa) \right] + \sum_{n=1}^{\infty} \ta_{\nu_0+n}(\kappa Q^2) \left[ (- \beta_0) (\nu_0+n-1) \td_{n-1}(\nu_0;\kappa) + \frac{d}{d \ln \kappa} \td_n(\nu_0;\kappa) \right],
\label{dD} \eea
which implies that the coefficient at each $ \ta_{\nu_0+n}$ in this expression must be zero
\be
\frac{d}{d \ln \kappa} \td_n(\nu_0,\kappa) = \beta_0 (n+ \nu_0-1) \td_{n-1}(\nu_0;\kappa)
\qquad (n=1,2,\ldots),
\label{dtdn}
   \ee
   and for $n=0$ we get $(d/d \ln \kappa) \td_0(\nu_0,\kappa)=0$, which allows us to choose $\td_0(\nu_0;\kappa)=1$, i.e., the canonical choice. In Eq.~(\ref{dtdn}) we see that the $\kappa$-derivative of the coefficients $\td_n(\nu_0;\kappa)$ yields a factor on the right-hand side (RHS) where the index $\nu_0$ explicitly appears.

The next step is to improve the evaluation of the observable ${\cal D}(Q^2)$, from using the simple (truncated) expansions Eqs.~(\ref{Dptlpt}) to using a renormalon-motivated resummation procedure. This is done by generalising the resummation method which has been developed for the integer case of $\nu_0$ ($\nu_0=1$) in Ref.~\cite{renmod} to the case of noninteger $\nu_0$. This suggests that we now consider a modified quantity, with modified coefficients $\td_n(\nu_0;\kappa) \mapsto \td_n(1;\kappa)$ where the resulting factor on the RHS of Eq.~(\ref{dtdn}) becomes $(n+\nu_0-1) \mapsto n$. This is achieved with the following rescaling of the original coefficients $\td_n(\nu_0;\kappa)$ that appear in the expansion of (\ref{Dlpt}):
\be
\td_n(1;\kappa) \equiv \frac{\Gamma(\nu_0) \Gamma(1+n)}{\Gamma(\nu_0+n)} \td_n(\nu_0;\kappa).
\label{tdn1} \ee
When we use this relation in the recursive relations (\ref{dtdn}), we obtain indeed
\be
\frac{d}{d \ln \kappa} \td_n(1,\kappa) = \beta_0 n \td_{n-1}(1;\kappa)
\qquad (n=1,2,\ldots), 
\label{dtdn1}
\ee
and $\td_0(1;\kappa)=1$.

If we now define a new auxiliary quantity $\cDo(Q^2)$ whose perturbation expansion in logarithmic derivatives involves these rescaled coefficients
\be
\cDo(Q^2) \equiv \sum_{n=0}^{\infty} \td_n(1;\kappa) \ta_{n+1}(\kappa Q^2),
\label{D1def} \ee
we can see immediately that this quantity is a quasiobservable, i.e., it is independent of the renormalisation scale parameter $\kappa$: $(d/d \ln \kappa) \cDo(Q^2) = 0$. We note that the expansion (\ref{D1def}) starts with $\ta_1(\kappa Q^2)$ which is identical with $a(\kappa Q^2)$. The logarithmic derivatives $\ta_{n+1}(\kappa Q^2)$ appearing in the expansion (\ref{D1def}) have now integer indices $n+1$, and are literally logarithmic derivatives of $a(\kappa Q^2)$ (cf.~footnote \ref{ftnu01})
\be
\ta_{n+1}(Q^2) \equiv \frac{(-1)^n}{n! \beta_0^n} \left(\frac{d}{d \ln Q^2} \right)^n a(Q^2).
\label{tan} \ee
Following Ref.~\cite{renmod}, we now define the modified Borel transform $\tB$ of this quantity $\cDo(Q^2)$
\be
\tB[\cDo](u;\kappa) \equiv \sum_{n=0}^{\infty} \frac{\td_n(1;\kappa)}{n! \beta_0^n} u^n,
\label{tBD1def} \ee
for which we know, from our treatment of the $\nu_0=1$ case \cite{renmod} (cf.~also \cite{BSRpQCD}), that it has the following simple (one-loop type) $\kappa$-dependence:\footnote{This property is exact and is the direct consequence of the definition (\ref{tBD1def}) and the recursion relations (\ref{dtdn1}), where the latter are the consequence of the $\kappa$-independence of $\cDo(Q^2)$ [$\Leftrightarrow$ $\kappa$-independence of the original quantity $\cD(Q^2)$ Eqs.~(\ref{Dpt})-(\ref{Dlpt})].}
\be
\tB[\cDo](u;\kappa) = \kappa^u \tB[\cDo](u),
\label{tBD1kap} \ee
and that we can resum the quantity $\cDo(Q^2)$ with a characteristic function $F_{\cDo}(t)$
\be
\cDo(Q^2)_{\rm res.} = \int_0^{\infty} \frac{dt}{t}  F_{\cDo}(t) a(t Q^2),
\label{D1res} \ee
where the characteristic function is the inverse Mellin transformation of the modified Borel $\tB[\cDo]$  
\be
F_{\cDo}(t) = \frac{1}{2 \pi i} \int_{u_0 - i \infty}^{u_0 + \infty} du \tB[\cDo](u) t^u.
\label{FD1} \ee
Here, $u_0$ is zero, or any real number closer to zero than the first renormalon singularity of $\tB[\cDo](u)$.

We can check that the resummation (\ref{D1res}) is correct, if we Taylor-expand $a(t Q^2)$ there around the point $\kappa Q^2$ [in fact: around the point $\ln (\kappa Q^2)$]
\be
a(t Q^2) = \sum_{n=0}^{\infty} (- \beta_0)^n \ln^n \left( \frac{t}{\kappa} \right) \ta_{n+1}(\kappa Q^2),
\label{atintakap} \ee
insert this expansion in Eq.~(\ref{D1res}), exchange the order of the summation and integration, and require that we obtain the expansion (\ref{D1def}) of $\cDo$, which then gives at each $\ta_{n+1}(\kappa Q^2)$ the condition
\be
\td_n(1;\kappa) = (- \beta_0)^n \int_0^{\infty} \frac{dt}{t} F_{\cDo}(t) \ln^n \left( \frac{t}{\kappa} \right) \qquad (n=0,1,2,\ldots).
\label{tdn1SR} \ee
We then multiply each of these sum rules by $u^n/(n! \beta_0^n)$ and sum over $n$; we must obtain in this way the modified Borel $\tB[\cDo](u;\kappa)$ Eq.~(\ref{tBD1def}); this then implies\footnote{\label{ftexp}We note: $\sum_{n=0}^{\infty} (-1)^n w^n/n! = e^{-w}$, where we have in our case $w=u \ln(t/\kappa)$.}
\be
\tB[\cDo](u;\kappa) = \kappa^u \int_0^{\infty} \frac{dt}{t} F_{\cDo}(t) t^{-u},
\label{tBD1ukap} \ee
where we have expressed the obtained exponential as $\exp(- u \ln(t/\kappa)) = \kappa^u t^{-u}$. The $\kappa$-dependence in this expression is consistent with the property (\ref{tBD1kap}) that reflects the $\kappa$-independence of $\cDo(Q^2)$ and $\cD(Q^2)$. When $\kappa=1$, we thus obtain
\be
\tB[\cDo](u) = \int_0^{\infty} \frac{dt}{t} F_{\cDo}(t) t^{-u},
\label{tBD1u} \ee
which means that $\tB[\cDo](u)$ is the Mellin transform of the characteristic function $F_{\cDo}(t)$ that appeared in the resummation of $\cDo(Q^2)$, Eq.~(\ref{D1res}). Thus the (sought) characteristic function for the resummation of $\cDo(Q^2)$ is the inverse Mellin of the modified Borel $\tB[\cDo](u)$, i.e., Eq.~(\ref{D1res}) is proven.\footnote{A similar formalism of resummation, applicable to observables with expansion in powers of the perturbative running coupling in the one-loop approximation, $a_{\rm (1\ell)}(Q^2)$, was developed first by Neubert \cite{Neubert} and later applied also in \cite{NC}. The formalism of \cite{renmod}, however, works when the running coupling $a(Q^2)$ is taken at an arbitrary loop-level. We note that, at the one-loop level, the powers and the logarithmic derivatives (for integer $n$) coincide, $a_{\rm (1\ell)}^n = \ta_n^{\rm (1\ell)}$.}

\textcolor{black}{In Appendix \ref{app:matnum} we present an extended discussion on the consistency aspects of this formalism, concentrating on several mathematical and numerical considerations. The Appendix takes into account that the couplings used in practice in this formalism are not pQCD couplings (which have Landau singularities), but are holomorphic couplings (they are free from Landau singularities) discussed in Secs.~\ref{sec:AQCD}-\ref{sec:impl}, as such couplings allow us to perform the resummations (formally) without ambiguity.}  

Until now, we presented and explained relations for $\cDo(Q^2)$, i.e., for the case $\nu_0 \mapsto 1$, which is a recapitulation of the results already obtained in \cite{renmod} and \cite{BSRpQCD,BSRAQCD}.
Now we proceed to the resummation of the original observable ${\cal D}(Q^2)$ which has $\nu_0 \not= 1$ (i.e., $\nu_0$ in general noninteger), i.e., whose expansion is given in Eqs.~(\ref{Dptlpt}). It turns out that this step consists in simply replacing, in the resummation Eq.~(\ref{D1res}) for the auxiliary quasiobservable $\cDo(Q^2)$, the factor $a(t Q^2)$ by $\ta_{\nu_0}(t Q^2)$:

\vspace{0.5cm}

\noindent
 {\bf Theorem 1}:

 \noindent
The resummation of $\cD(Q^2)$ observable as characterised by its expansions Eqs.~(\ref{Dpt})-(\ref{Dlpt}) yields
\be
\cD(Q^2)_{\rm res.} = \int_0^{\infty} \frac{dt}{t}  F_{\cDo}(t) \ta_{\nu_0}(t Q^2),
\label{Dres} \ee
where $F_{\cDo}(t)$ is the inverse Mellin, Eq.~(\ref{FD1}), of the auxiliary quasiobservable $\cDo(Q^2)$ which is defined by the perturbation expansion (\ref{D1def}) with the rescaled coefficients $d_n(1;\kappa)$ defined via relations (\ref{tdn1}). The generalised logarithmic derivative $\ta_{\nu_0}(t Q^2)$ in the result Eq.~(\ref{Dres}) is a linear combination of powers $a(t Q^2)^{\nu_0+n}$ appearing in Eq.~(\ref{tanuanu}) where the general coefficients $k_m(\nu_0)$ were obtained in \cite{GCAK}. 

\vspace{0.3cm}

\noindent
{\bf Proof:}

\noindent
In order to prove the resummation formula (\ref{Dres}), we perform the Taylor-expansion of $\ta_{\nu_0}(t Q^2)$ around $\kappa Q^2$ [analogous to the expansion Eq.~(\ref{atintakap})], by using the recursive relations (\ref{dertanu})
\be
\ta_{\nu_0}(t Q^2) = \sum_{n=0}^{\infty} (- \beta_0)^n \ln^n \left( \frac{t}{\kappa} \right) \frac{\Gamma(\nu_0+n)}{\Gamma(\nu_0) \Gamma(n+1)} \ta_{\nu_0+n}(\kappa Q^2),
\label{tatintakap} \ee
We use this expansion on the RHS of Eq.~(\ref{Dres}), and exchange the order of the summation and integration. The term at $\ta_{\nu_0+n}(\kappa Q^2)$ must then give the coefficient $\td_n(\nu_0;\kappa)$ of the expansion (\ref{Dlpt})
\bes
\label{tdnSR}
\bea
\td_n(\nu_0;\kappa) &=&  \frac{\Gamma(\nu_0+n)}{\Gamma(\nu_0) \Gamma(n+1)} (- \beta_0)^n \int_0^{\infty} \frac{dt}{t} F_{\cDo}(t) \ln^n \left( \frac{t}{\kappa} \right) \Rightarrow
\label{tdnSRa}
\\
\td_n(1;\kappa) \left[ \equiv \frac{\Gamma(\nu_0) \Gamma(n+1)}{\Gamma(\nu_0+n)} \td_n(\nu_0;\kappa) \right] & = & (- \beta_0)^n \int_0^{\infty} \frac{dt}{t} F_{\cDo}(t) \ln^n \left( \frac{t}{\kappa} \right) \qquad (n=0,1,2,\ldots)
\label{tdnSRb} \eea \ees
The form (\ref{tdnSRb}) is indeed true, by the construction of the characteristic function $F_{\cDo}(t)$, cf.~the sum rules (\ref{tdn1SR}). This then proves that the result (\ref{Dres}) is the correct resummation because it implies the original expansion (\ref{Dlpt}) of the considered observable $\cD(Q^2)$.

\vspace{0.5cm}

In practice, in order to obtain the characteristic function $F_{\cDo}(t)$ needed in the resummation Eq.~(\ref{Dres}) of the spacelike QCD observable $\cD(Q^2)$ [whose expansions are in Eqs.~(\ref{Dptlpt})], we will need to know the modified Borel transform $\tB[\cDo](u)$ of the auxiliary quantity $\cDo(Q^2)$. Therefore, we face the following problem: if we know the renormalon structure of this observable $\cD(Q^2)$,\footnote{For a review of renormalon physics, see \cite{Renormalons}.} i.e., the behaviour of the (usual) Borel transform $B^{(\nu_0)}[\cD](u)$ of $\cD$, what is the corresponding behaviour of the modified Borel $\tB^{(\nu_0)}[\cD](u)$ of $\cD$, and of the modified Borel $\tB[\cDo](u)$? Namely, we need the behaviour of the latter quantity in order to get the characteristic function $F_{\cDo}(t)$ that is the inverse Mellin of $\tB[\cDo](u)$ Eq.~(\ref{FD1}), i.e., the characteristic function appearing also in the resummation (\ref{Dres}) of the obervable $\cD(Q^2)$. We will adopt for the observable $\cD(Q^2)$, Eqs.~(\ref{Dptlpt}), the following definitions of the Borel and modified Borel transforms:
\bes
\label{BDtBDdef}
\bea
B^{(\nu_0)}[\cD](u;\kappa) & \equiv & \sum_{n=0}^{\infty} \frac{d_n(\nu_0;\kappa)}{n! \beta_0^n} u^n,
\label{BDdef} \\
\tB^{(\nu_0)}[\cD](u;\kappa) & \equiv & \sum_{n=0}^{\infty} \frac{\td_n(\nu_0;\kappa)}{n! \beta_0^n} u^n,
\label{tBDdef} \eea \ees
and the definition of $\tB[\cDo](u;\kappa)$ is given in Eq.~(\ref{tBD1def}).\footnote{The modified Borel transforms of the quantities with $\nu_0=1$ (such as $\cDo$) will be denoted, for simplicity, without the superscript '$(1)$' at $\tB$.} 

The above problem is addressed by the following Theorems 2 and 3. We first assume that the renormalon structures are dominated by an infrared (IR) renormalon at $u=p$ ($p>0$), and then comment on the modification for the case of an ultraviolet (UV) renormalon at $u=-p$. \textcolor{black}{
  The location of the (renormalon) singularities of the (usual) Borel transform of a QCD observable can be obtained by the large-$\beta_0$ approach. The strength of these singularities (the power of the poles) is modified, at the level beyond the large-$\beta_0$, by the RGE for the strong coupling [$\beta$-function, cf.~Eq.~(\ref{RGEa})] and the RGE for the local operator matrix element (anomalous dimension $\gamma$). The RGEs add to the large-$\beta_0$ part of the power a term proportional to the two-loop $\beta$-coefficient $\beta_1/\beta_0^2$, and the term $-\gamma_0^{(1)}/\beta_0$ containing the leading-order anomalous dimension coefficient $\gamma_0^{(1)}$, respectively (cf., for example, \cite{JPGour}). The general term in a (usual) Borel transform $B(u)$ that represents an IR renormalon at $u=p$ ($p>0$) is schematically $\sim 1/(p-u)^{s_0}$ where $s_0>0$ is the (noninteger) power of the IR renormalon. A general form of an UV renormalon in $B(u)$ at $u=-p$ ($-p < 0$) is $\sim 1/(p+u)^{s_0}$.}

\vspace{0.5cm}

\noindent
{\bf Theorem 2:}

\noindent
If the (IR) renormalon structure of the Borel transform, Eq.~(\ref{BDdef}), of the observable $\cD(Q^2)$ [Eq.~(\ref{Dptlpt})] has the following form:
\be
B^{(\nu_0)}[\cD](u;\kappa) = \frac{\cK(\kappa)}{(p-u)^{s_0}} \left[ 1 + \cO((p-u)) \right],
\label{BDren} \ee
where $p>0$ (in practice: $p=1/2, 1, 3/2, \ldots$), then the modified Borel transform (\ref{tBDdef}) of this observable is
\be
\tB^{(\nu_0)}[\cD](u;\kappa) = \frac{\tcK(\kappa)}{(p-u)^{\ts_0}} \left[ 1 + \cO((p-u)) \right],
\label{tBDren} \ee
where the two indices are related by
\be
s_0 = \ts_0+ p \frac{\beta_1}{\beta_0^2},
\label{s0vsts0} \ee
where $\beta_0$ and $\beta_1$ are the two leading coefficients in the $\beta$-function (\ref{RGE}).

\vspace{0.3cm}

The proof of this theorem is relatively long, and we refer to Appendix \ref{app:T2} for the proof. The residues $\cK(\kappa)$ and $\tcK(\kappa)$ are also related, but we will not need their relation in our application.\footnote{An approximate relation between these two residues can be inferred by comparing Eq.~(\ref{dnT2b}) with Eqs.~(\ref{tdnc1asfin}) and (\ref{tdnfin}).}

If the renormalon is at $u=-p$ (i.e., UV renormalon), then we replace in Eqs.~(\ref{BDren})-(\ref{tBDren}): $(p-u) \mapsto (p+u)$; and in the relation (\ref{s0vsts0}): $p \mapsto -p$. The proof can be repeated in this case (with: $p \mapsto -p$)

The validity of this Theorem was checked numerically for quantities with $\nu_0=1$ and several integer $\ts_0$ and integer $p$ in Ref.~\cite{renmod}; and for noninteger $\ts_0=0.778$ and $\ts_0=0.375$ (with $p=3$ and $\nu_0=1$) in Ref.~\cite{JPGour}.\footnote{At the time, the general validity of Theorem 2 (i.e., for noninteger $\nu_0$, $\ts_0$ and $p$) was not known or used in those references.}

\vspace{0.5cm}

\noindent
{\bf Theorem 3:}
    
\noindent
If the modified Borel transform $\tB^{(\nu_0)}[\cD](u;\kappa)$ Eq.~(\ref{tBDdef}) of the observable $\cD(Q^2)$ Eq.~(\ref{Dptlpt}) has the renormalon form as given in Eq.~(\ref{tBDren}), then the modified Borel transform $\tB[\cDo](u;\kappa)$ Eq.~(\ref{tBD1def}) of the auxiliary quasiobservable $\cDo(Q^2)$ Eq.~(\ref{D1def}) [cf.~Eq.~(\ref{tdn1})] is
\be
\tB[\cDo](u;\kappa) = \frac{\tcKo(\kappa)}{(p-u)^{\ts_0-\nu_0+1}} \left[ 1 + \cO((p-u)) \right],
\label{tBD1ren} \ee

\vspace{0.3cm}

\noindent
{\bf Proof:}

\noindent
We will use the general expansion ($s$ can be noninteger)
\be
(p -u)^{-s} =p^{-s} \sum_{n=0}^{\infty} \frac{\Gamma(s+n)}{\Gamma(s) n!} \left( \frac{u}{p} \right)^n,
\label{binexp} \ee
and the asymptotic formula for the $\Gamma$ function (for large $n$)
\be
\frac{\Gamma(s+n)}{n!} = n^{s-1} \left[ 1 + \cO(1/n) \right].
\label{Gammaasym} \ee
Using this, the asymptotic expansion of Eq.~(\ref{tBDren}) is
\be
\tB^{(\nu_0)}[\cD](u;\kappa) =\tcK(\kappa) \frac{p^{-\ts_0}}{\Gamma(\ts_0)}
\sum_{n} \frac{n^{\ts_0-1}}{p^n} \left( 1 + \cO(1/n) \right) u^n,
\label{tBDrenexp} \ee
which implies the following asymptotic behavior of the corresponding coefficients $\td_n(\nu_0;\kappa)$:
\bea
\td_n(\nu_0;\kappa) & = & \tcK(\kappa)  \frac{p^{-\ts_0}}{\Gamma(\ts_0)} n! \; n^{\ts_0-1} \left( \frac{\beta_0}{p} \right)^n \left( 1 + \cO(1/n) \right).
\label{tdnasym}
\eea
The asymptotic behavior of the corresponding modified coefficients $\td_n(1;\kappa)$ [as defined by Eq.~(\ref{tdn1})] is then
\bes
\label{td1nasym}
\bea
\td_n(1;\kappa) &=&  \frac{\Gamma(\nu_0) n!}{\Gamma(\nu_0+n)} \tcK(\kappa)  \frac{p^{-\ts_0}}{\Gamma(\ts_0)} n! \; n^{\ts_0-1} \left( \frac{\beta_0}{p} \right)^n \left( 1 + \cO(1/n) \right)
\label{td1nasyma} 
\\
& = & \tcK(\kappa)  \frac{\Gamma(\nu_0) p^{-\ts_0}}{\Gamma(\ts_0)} n! \; n^{(\ts_0-\nu_0+1)-1} \left( \frac{\beta_0}{p} \right)^n \left( 1 + \cO(1/n) \right),
\label{td1nasymb}
\eea \ees
where we used in the last identity (\ref{td1nasymb}) the asymptotic formula (\ref{Gammaasym}). When we compare the expression (\ref{td1nasymb}) with (\ref{tdnasym}), we see that the coefficients $\td_n(1;\kappa)$ give in the Borel transform the same structure as $\td_n(\nu_0;\kappa)$, except that $\ts_0 \mapsto \ts_0 - \nu_0+1$ in the exponent of $n$. This then implies, that the renormalon structure in $\tB[\cDo](u;\kappa)$ is the same as in $\tB^{(\nu_0)}[\cD](u;\kappa)$, when we replace in the index $\ts_0 \mapsto \ts_0 - \nu_0+1$. This then proves Theorem 3.\footnote{\label{ft1n} The relative corrections $\cO(1/n)$ in Eq.~(\ref{td1nasymb}) correspond to the relative corrections $\cO(p-u)$ in the expression (\ref{tBD1ren}), because $\Gamma((s-1)+n)= (\Gamma(s+n)/n)$ $\times (1 + \cO(1/n))$.}

If the renormalon is at $u=-p$ (UV renormalon), then we replace in the theorem Eq.~(\ref{tBD1ren}): $(p-u) \mapsto (p+u)$, and the proof can be repeated (with: $p \mapsto -p$).

\vspace{0.5cm}

In order to perform in practice the resummation Eq.~(\ref{Dres}) for general spacelike QCD observables $\cD(Q^2)$, Eqs.~(\ref{Dptlpt}), we need to obtain the associated characteristic function $F_{\cDo}(t)$, Eq.~(\ref{FD1}), that is the inverse Mellin transform of the modified Borel $\tB[\cDo](u)$.
The following theorem enables us to evaluate the characteristic function $F_{\cDo}(t)$ if we know the renormalon structure of the Borel $B^{(\nu_0)}[\cD](u)$ Eq.~(\ref{BDren}) [and thus $\tB[\cDo](u)$ by Theorems 2 and 3]:

\vspace{0.5cm}

\noindent
{\bf Theorem 4:}
    
\noindent
If the modified Borel transform $\tB[\cDo](u;\kappa)$ (with $\kappa=1$) has the (IR) renormalon contribution
\be
\tB[\cDo](u)_{(p,\ts)} = \frac{\pi}{(p-u)^{\ts}},
\label{tBD1rencan} \ee
where $p>0$ and $0 < \ts \leq 1$, then the corresponding inverse Mellin transform Eq.~(\ref{FD1}), i.e., the corresponding characteristic function $F_{\cDo}(t)_{(p,\ts)}$, is
\be
F_{\cDo}(t)_{(p,\ts)} = \Theta(1 - t) \pi \frac{t^p}{\Gamma(\ts) (- \ln t)^{1 - \ts}},
\label{FD1pts} \ee
where $\Theta(1-t)$ is the Heaviside function (i.e., it is unity for $0 \leq t \leq 1$, and is zero for $t>1$).

\vspace{0.3cm}

\noindent
We refer for a formal proof of this theorem to Appendix \ref{app:T4}.
This theorem, together with Theorems 2 and 3, implies that the characteristic function $F_{\cDo}(t)$ appearing in the resummation (\ref{Dres}) of the spacelike observable $\cD(Q^2)$ Eqs.~(\ref{Dptlpt}) with the IR renormalon structure (\ref{BDren}) is
\be
F_{\cDo}(t) = \Theta(1 - t) \tcKo \frac{t^p}{\Gamma(\ts) (- \ln t)^{1 - \ts}}
  \qquad {\rm with:}\;\; \ts =s_0 - p \frac{\beta_1}{\beta_0^2} - \nu_0 + 1,
\label{FD1IRp} \ee
where $\tcKo$ is a constant.\footnote{\label{ftkapdep} We took here $\kappa=1$. We note that then: $\tcKo(\kappa) = \kappa^p \tcKo$, because $\kappa^u = \kappa^p [1 + \cO((p-u)) ]$.}
We point out that in our approach we take into account only the leading (i.e., the most singular) renormalon contributions, i.e., we neglect the relative corrections $\cO( (p-u)^1 )$ in the (modified) Borel transforms. According to Eq.~(\ref{FD1IRp}), these relative corrections then contribute to the characteristic function $F_{\cDo}(t)$ the relative corrections $\cO(1/|\ln t|)$ (we note that $1/|\ln t| < 1$ for $0 \leq t < 1/e$ ($\approx 0.37$).

When, however, we have in the modified Borel transform $\tB[\cDo](u)$ the effects beyond the leading term (\ref{tBD1rencan}) contained in the rescaling factor $\exp(\tK_e u)$ [reflecting possible redefinition of the momentum scale, according to Eq.~(\ref{tBD1kap})]
\be
\tB[\cDo](u)_{(p,\ts,\tK_e)} = \frac{\pi \exp(\tK_e u)}{(p-u)^{\ts}},
\label{tBD1rencan2} \ee
then it is straightforward to check that the inverse Mellin (\ref{FD1}) implies
for the characteristic function in such a case
\be
F_{\cDo}(t)_{(p,\ts,\tK_e)} = F_{\cDo}(t e^{\tK_e})_{(p,\ts)},
\label{FD1ptstK} \ee
where $F_{\cDo}(t)_{(p,\ts)}$ is given in Eq.~(\ref{FD1pts}).
This then implies, after a simple redefinition of the integration variable $t$, that the resummation expressions (\ref{D1res}), (\ref{Dres}), in the case of the modified Borel (\ref{tBD1rencan2}), have the form
\bes
\label{restK}
\bea
\cDo(Q^2)_{\rm res.} &=& \int_0^{\infty} \frac{dt}{t}  F_{\cDo}(t)_{(p,\ts)} a(t e^{-\tK_e} Q^2),
\label{D1restK} \\
\cD(Q^2)_{\rm res.} &=& \int_0^{\infty} \frac{dt}{t}  F_{\cDo}(t)_{(p,\ts)} \ta_{\nu_0}(t e^{-\tK_e} Q^2).
\label{DrestK}
\eea \ees

\section{Resummation for timelike observable $\cF(\sigma)$}
\label{sec:time}

We will now consider the timelike observable ${\cF}(\sigma)$ ($\sigma>0$) that is associated with the spacelike observable $\cD(Q^2)$ of Eqs.~(\ref{Dptlpt}), and the auxiliary timelike quasiobservable ${\cFo}(\sigma)$ that is associated with the auxiliary spacelike quasiobservable $\cDo(Q^2)$ of Eq.~(\ref{D1def}) [in conjunction with Eq.~(\ref{tdn1})]
\bes
\label{FF1def}
\bea
\cF(\sigma) &=& \frac{1}{2 \pi} \int_{-\pi}^{+\pi} d \phi \; \cD(\sigma e^{i \phi}), 
\label{Fdef} \\
\cFo(\sigma) &=& \frac{1}{2 \pi} \int_{-\pi}^{+\pi} d \phi \; \cDo(\sigma e^{i \phi}) .  
\label{F1def}
\eea \ees
We recall that the inverse relations have the somewhat more familiar form
\bes
\label{DD1vsFF1}
\bea
\cD(Q^2) & = & Q^2 \int_0^{\infty} \frac{d \sigma \cF(\sigma)}{(\sigma + Q^2)^2},
\label{DvsF} \\
\cDo(Q^2) & = & Q^2 \int_0^{\infty} \frac{d \sigma \cFo(\sigma)}{(\sigma + Q^2)^2}
\label{D1vsF1} \eea \ees
\textcolor{black}{Relations (\ref{DD1vsFF1}) are similar to once-subtracted dispersion relations. We wrote them here for completeness because they are known to be the exact inverse relations of the relations (\ref{FF1def}).}

The corresponding expansions in generalised logarithmic derivatives for these timelike quantities, analogous to the expansions (\ref{Dlpt}) and (\ref{D1def}), are denoted analogously as
\bes
\label{FF1lpt}
\bea
\cF(\sigma) &=& \sum_{n=0}^{\infty} \tf_n(\nu_0;\kappa) \ta_{\nu_0+n}(\kappa \sigma),
\label{Flpt} \\
\cFo(\sigma) &=& \sum_{n=0}^{\infty} \tf_n(1;\kappa) \ta_{1+n}(\kappa \sigma).
\label{F1lpt} \eea \ees  

\begingroup \color{black}
When we apply the resummation of $\cD(Q^2)$, as given in Theorem 1, Eq.~(\ref{Dres}), to the integral transformation (\ref{Fdef}), and exchange the order of integration over $t$ and $\phi$, we obtain
\be
\cF(\sigma)_{\rm res.} = \int_0^{\infty}  \frac{dt}{t}  F_{\cDo}(t) \ths_{\nu_0}(t \sigma),
\label{Fres} \ee
where $\ths_{\nu}(\sigma)$ is the timelike analog of the spacelike (generalised logarithmic derivative) $\ta_{\nu}(Q^2)$, defined and investigated in \cite{GCAK}
\be
\ths_{\nu}(\kappa \sigma) \equiv \frac{1}{2 \pi} \int_{-\pi}^{\pi} d \phi \; \ta_{\nu}(\kappa \sigma e^{i \phi}).
\label{thnudef} \ee
We will comment on the justification for such an exchange of the order of integration within the holomorphic QCD (AQCD) later on in Sec.~\ref{sec:impl}.
\endgroup

First we will prove the following theorem which relates the modified Borel transform $\tB$ of the quantity $\cFo(\sigma)$ with $\tB$ of $\cDo(Q^2)$:

\vspace{0.5cm}

\noindent
{\bf Theorem 5:}
We have
\be
\tB[\cDo](u;\kappa) \frac{\sin(\pi u)}{\pi u} = \tB[\cFo](u;\kappa),
\label{tBD1vstBF1} \ee
where $\tB[\cDo](u;\kappa)$ was defined through its expansion in Eq.~(\ref{tBD1def}), and $\tB[\cFo](u\kappa)$ is defined via the corresponding expansion
\be
\tB[\cFo](u;\kappa) = \sum_{n=0}^{\infty} \frac{\tf_n(1;\kappa)}{n! \beta_0^n} u^n.
\label{tBF1def} \ee

\vspace{0.3cm}

\noindent
{\bf Proof:}

\noindent
In this proof we follow closely the steps applied in Appendix A of our previous paper \cite{EPJC81} [for $\delta_{x^n}^{(d)}$ there, with $n=0$].

The idea is to operationally replace the logarithmic-derivative couplings $\ta_{1+n}(\kappa Q^2)$ and $\ta_{1+n}(\kappa \sigma)$ in the expansions (\ref{D1def}) and (\ref{F1lpt}) by the simple powers $a_{\rm (1\ell)}(\kappa Q^2)^{1+n}$ and $a_{\rm (1\ell)}(\kappa \sigma)^{1+n}$, where $a_{\rm (1\ell)}(Q^2)$ is the one-loop running coupling. For the purpose of the present proof, this is legitimate, because the momentum dependence of those logarithmic-derivative couplings is exactly the same as that of one-loop coupling powers. This can be seen directly from the relation (\ref{dertanu}) (when $\nu=1+n$).\footnote{We point out, however, that in our approach the logarithmic derivatives are considered at any loop level.}$^{,}$\footnote{As a consequence, the scale-dependence relation (\ref{tatintakap}) remains valid also when we replace there everywhere $\ta_{\nu_0+n} \mapsto a_{\rm (1\ell)}^{\nu_0+n}$, and $\nu_0$ can be either integer or noninteger.}
We can also interpret this replacement by the fact that in the one-loop approximation ($\beta_j=0$ for $j \geq 1$) we have clearly $\ta_{1+n} = a^{n+1}$.

Following this approach, we introduce expansions in powers of the one-loop coupling analogous to the exact expansions (\ref{D1def}) and (\ref{F1lpt})
\bes
\label{D1F11l} 
\bea
\cDo_{\rm (1\ell,pow.)}(Q^2) &=& \sum_{n=0}^{\infty} \td_n(1;\kappa) a_{\rm (1\ell)}(\kappa Q^2)^{1+n},
\label{D11l} \\
\cFo_{\rm (1\ell,pow.)}(\sigma) &=& \sum_{n=0}^{\infty} \tf_n(1;\kappa) a_{\rm (1\ell)}(\kappa \sigma)^{1+n}.
\label{F11l} \eea \ees
We point out that these power expansions contain exactly the same coefficients $\td_n(1;\kappa)$ and $\tf_n(1;\kappa)$ of the (exact) expansions (\ref{D1def}) and (\ref{F1lpt}) of the quantities $\cDo(Q^2)$ and $\cFo(\sigma)$. The quantities (\ref{D11l}) and (\ref{F11l}) are exactly independent of the renormalisation scale parameter $\kappa$ [as are $\cDo(Q^2)$ and $\cFo(\sigma)$]. Now, the {\it usual\/} formal Borel transforms of $B[\cDo_{\rm (1\ell,pow.)}]$ and $B[\cFo_{\rm (1\ell,pow.)}]$ are
\bes
\label{BD1F11l}
\bea
B[\cDo_{\rm (1\ell,pow.)}](u;\kappa) &=& \sum_{n=0}^{\infty} \frac{\td_n(1;\kappa)}{n! \beta_0^n} u^n  \quad \left[ = \tB[\cDo](u;\kappa) \right],
\label{BD11l} \\
B[\cFo_{\rm (1\ell,pow.)}](u;\kappa) &=& \sum_{n=0}^{\infty} \frac{\tf_n(1;\kappa)}{n! \beta_0^n} u^n  \quad \left[ = \tB[\cFo](u;\kappa) \right],
\label{BF11l} 
\eea \ees
which, as indicated, are identical to the {\it modified\/} Borel transforms $\tB[\cDo](u;\kappa)$ Eq.~(\ref{tBD1def}) and $\tB[\cFo](u;\kappa)$ Eq.~(\ref{tBF1def}), respectively. The inverse Borel transformation of the usual Borel $B[\cDo_{\rm (1\ell,pow.)}](u;\kappa)$ is then
\be
\cDo_{\rm (1\ell,pow.)}(Q^2) = \frac{1}{\beta_0} \int_0^{\infty} du \; \exp \left( - \frac{u}{\beta_0 a_{\rm (1\ell)}(\kappa Q^2)} \right) \tB[\cDo](u;\kappa).
\label{D11linvB} \ee
This then implies that the corresponding timelike quantity $\cFo_{\rm (1\ell,pow.)}(\sigma)$ is [cf.~Eqs.~(\ref{F1def}) and (\ref{F11l})]
\bes
\label{F11linvB}
\bea
\cFo_{\rm (1\ell,pow.)}(\sigma) &=&  \frac{1}{2 \pi} \int_{-\pi}^{+\pi} d \phi \; \cDo_{\rm (1\ell,pow.)}(\sigma e^{i \phi})
\label{F11la} \\
& = & \frac{1}{2 \pi \beta_0} \int_{-\pi}^{+\pi} d \phi \; \int_0^{\infty} du \; \exp \left( - \frac{u}{\beta_0 a_{\rm (1\ell)}(\kappa \sigma e^{i \phi})} \right) \tB[\cDo](u;\kappa)
\label{F11lb} \\
& = & \frac{1}{2 \pi \beta_0} \int_0^{\infty} \; du  \exp \left( - \frac{u}{\beta_0 a_{\rm (1\ell)}(\kappa \sigma)} \right) \tB[\cDo](u;\kappa) \;  \int_{-\pi}^{+\pi} d \phi \exp \left( - \frac{i \beta_0 u}{\beta_0} \phi \right),
\label{F11lc} \eea \ees
where in the last identity we exchanged the order of integrations and used the one-loop RGE-running relation
\be
\frac{1}{a_{\rm (1\ell)}(\kappa \sigma e^{i \phi})} = \frac{1}{a_{\rm (1\ell)}(\kappa \sigma)} + i \beta_0 \phi.
\label{a1l} \ee
The integral over $\phi$ in Eq.~(\ref{F11lc}) is trivial, equal to $2 \sin(\pi u)/u$, leading to
\be
\cFo_{\rm (1\ell,pow.)}(\sigma) = \frac{1}{\beta_0} \int_0^{\infty} du \; \exp \left( - \frac{u}{\beta_0 a_{\rm (1\ell)}(\kappa \sigma)} \right)  \left( \tB[\cDo](u;\kappa) \frac{\sin(\pi u)}{\pi u} \right),
\label{F11lf} \ee
and thus
\be
B[\cFo_{\rm (1\ell,pow.)}](u;\kappa) = \tB[\cDo](u;\kappa) \frac{\sin(\pi u)}{\pi u}.
\label{BF1vsBD1} \ee
Here we have on the left-hand side the quantity Eq.~(\ref{BF11l}), i.e., this is identical to the modified Borel $\tB[\cFo](u;\kappa)$.

This concludes the proof of Theorem 5.

In this context, we mention that an analogous relation to Eq.~(\ref{tBD1vstBF1}) was obtained in \cite{BY1992} between the (usual) Borel transforms of the $R_{ee}(\sigma)$ ratio and of the Adler function in the large-$\beta_0$ approximation, and in \cite{BenekeDis} between the Borel transforms of the $\tau$-lepton semihadronic decay ratio $R_{\tau}$ and of the Adler function in the large-$\beta_0$ approximation.

\vspace{0.5cm}

We recall that Theorem 5 above relates the modified Borel of the timelike and spacelike auxiliary quantities $\cFo(\sigma)$ Eq.~(\ref{F1def}) and $\cDo(Q^2)$ Eq.~(\ref{D1def}).

We will now prove the following analogous Theorem 6, which relates the modified Borel of the full timelike and spacelike quantities $\cF(\sigma)$ Eq.~(\ref{Fdef}) and $\cD(Q^2)$ Eq.~(\ref{Dptlpt}):

\vspace{0.5cm}

\noindent
{\bf Theorem 6:}
Let $\cD(Q^2)$ be the spacelike observable whose expansions are written in Eqs.~(\ref{Dptlpt}), and $\cF(\sigma)$ be the corresponding timelike quantity defined in Eq.~(\ref{Fdef}), and let us assume that the modified Borel of $\cD(Q^2)$ has the form\footnote{
\textcolor{black}{The form (\ref{tBD}) for the modified Borel transform $\tB^{(\nu_0)}[\cD](u)$ was chosen because it corresponds, according to Theorem 2, to the associated general form of the IR renormalon term $\sim 1/(p-u)^{s_0}$ (at $u=p >0$ and with power $s_0$) in the usual Borel transform $B^{(\nu_0)}[\cD](u)$, cf.~Eqs.~(\ref{BDren})-(\ref{s0vsts0}).}} 
\bea
\tB^{(\nu_0)}[\cD](u;\kappa) \left[ =\sum_{n=0}^{\infty} \frac{ \td_n(\nu_0;\kappa)}{n! \beta_0^n} u^n \right] & = & \frac{\tcK(\kappa)}{(p - u)^{\ts_0}}.
\label{tBD}
\eea
Then the modified Borel of $\cF(\sigma)$ has the following form:
\bea
\tB^{(\nu_0)}[\cF](u;\kappa) \left[ =\sum_{n=0}^{\infty} \frac{ \tf_n(\nu_0;\kappa)}{n! \beta_0^n} u^n \right] & = & \frac{\sin(\pi p)}{\pi p} \frac{\tcK(\kappa)}{(p - u)^{\ts_0}} \left[ 1 + \cO \left((p-u) \right) \right].
\label{tBF}
\eea
In this context, we recall that the coefficients $\td_n(\nu_0;\kappa)$ and $\tf_n(\nu_0;\kappa)$ that define the expansions of these modified Borel transforms appear in the expansions (\ref{Dlpt}) and (\ref{Flpt}) in generalised logarithmic derivatives $\ta_{\nu_0+n}$.

\vspace{0.3cm}

\noindent
{\bf Proof:}

\noindent
As shown in \cite{GCAK}, the timelike coupling Eq.~(\ref{thnudef}) obeys the following differential recursion relation [completely analogous to that of $\ta_{\nu}$ Eq.~(\ref{dertanu})]:
\be
\frac{d}{d \ln \kappa} \ths_{\nu}(\kappa \sigma) = (-\beta_0) \nu \ths_{\nu+1}(\kappa \sigma).
\label{derthnu} \ee
This allows us to Taylor-expand $\ths_{\nu_0}(t \sigma)$ appearing in the integral in Eq.~(\ref{Fres}) around $\kappa \sigma$ [analogous to Eq.~(\ref{tatintakap})]
\be
\ths_{\nu_0}(t Q^2) = \sum_{n=0}^{\infty} (- \beta_0)^n \ln^n \left( \frac{t}{\kappa} \right) \frac{\Gamma(\nu_0+n)}{\Gamma(\nu_0) \Gamma(n+1)} \ths_{\nu_0+n}(\kappa Q^2).
\label{thtinthkap} \ee
When we insert this expansion in the integral (\ref{Fres}), exchange the order of integration and summation, and use the sum rules Eq.~(\ref{tdnSRa}) to express the obtained integrals through the coefficinets $\td_n(\nu_0;\kappa)$, we obtain the following expansion for $\cF(\sigma)$:
\be
\cF(\sigma) = \sum_{n=0}^{\infty} \td_n(\nu_0;\kappa) \ths_{\nu_0+n}(\kappa \sigma),
\label{Finth} \ee
which is completely analogous to the expansion (\ref{Dlpt}) for $\cD(Q^2)$, with the only difference being that the spacelike couplings $\ta_{\nu_0+n}(\kappa Q^2)$ is now replaced by the timelike couplings $\ths_{\nu_0+n}(\kappa \sigma)$.

We now want to express in the expansion (\ref{Finth}) the couplings $\ths_{\nu_0+n}(\kappa \sigma)$ in terms of the couplings $\ta_{\nu_0+m}(\kappa \sigma)$, in order to relate the coefficients $\td_n(\nu_0;\kappa)$ appearing in Eq.~(\ref{Finth}) with the coefficients $\tf_n(\nu_0;\kappa)$ appearing in Eq.~(\ref{Flpt}), since the latter appear in the definition (expansion) of the sought modified Borel $\tB^{(\nu_0)}[\cF](u;\kappa)$ Eq.~(\ref{tBF}). This can be obtained, when we use in the definition (\ref{thnudef}) of $\ths_{\nu}(\kappa \sigma)$, for the coupling $\ta_{\nu}(\kappa \sigma e^{i \phi})$ the Taylor-expansion around $\kappa \sigma$ [cf.~the expansion Eq.~(\ref{tatintakap})]
\be
\ta_{\nu}(\kappa \sigma e^{i \phi}) = \sum_{m=0}^{\infty} (- \beta_0)^m (i \phi)^m \frac{\Gamma(\nu+m)}{\Gamma(\nu) \Gamma(m+1)} \ta_{\nu+m}(\kappa \sigma).
\label{taphiinta} \ee
We insert this expansion in the integral Eq.~(\ref{thnudef}), exchange the order of integration and summation, and obtain the relation we sought
\be
\ths_{\nu}(\kappa \sigma) \equiv \sum_{m=0}^{\infty}  (- \beta_0)^m  \frac{\Gamma(\nu+m)}{\Gamma(\nu) \Gamma(m+1)} I_{m}  \ta_{\nu+m}(\kappa \sigma),
\label{thinta1} \ee
where the integrals $I_{m}$ are
\be
I_{m} \equiv \frac{1}{2 \pi} \int_{-\pi}^{+\pi} d \phi \; (i \phi)^m .
\label{Im} \ee
These integrals are nonzero only for even $m= 2 r$
\be
I_{2 r} = (-1)^r \frac{\pi^{2 r}}{(2 r+1)} \qquad (I_{2 r +1}=0).
\label{I2r} \ee
The relation (\ref{thinta1}) and the expressions (\ref{I2r}) imply
\be
\ths_{\nu}(\kappa \sigma) \equiv \sum_{r=0}^{\infty}  (-\beta_0^2 \pi^2)^r \frac{1}{(2 r+1)} \frac{\Gamma(\nu+2 r)}{\Gamma(\nu) \Gamma(2 r+1)} \ta_{\nu+2 r}(\kappa \sigma).
\label{thinta2} \ee
We now use this formula in the expansion (\ref{Finth}) of $\cF(\sigma)$, resulting in the following expansion of $\cF(\sigma)$ in terms of $\ta_{\nu_0+n}(\kappa \sigma)$:
\bes
\label{Fexp}
\bea
\cF(\sigma) & = & \sum_{n \geq 0} \sum_{r \geq 0} \td_n(\nu_0;\kappa) (- \beta_0^2 \pi^2)^r \frac{\Gamma(\nu_0+n+2 r)}{\Gamma(\nu_0+n) (2 r+1)!} \ta_{\nu_0+n+2 r}(\kappa \sigma) 
\label{Fexpa} \\
& =& \sum_{N \geq 0} \ta_{\nu_0+N}(\kappa \sigma) \left[ \sum_{r=0}^{[N/2]} (- \beta_0^2 \pi^2)^r \frac{\Gamma(\nu_0+N)}{\Gamma(\nu_0+N-2 r) (2 r+1)!} \td_{N-2 r}(\nu_0;\kappa) \right].
\label{Fexpb} \eea \ees
In the second identity (\ref{Fexpb}) we used the notation $n=N-2 r$ ($=0,1,\ldots$), and $[N/2]$ is the integer part of $N/2$.
The result (\ref{Fexpb}), in conjunction with the expansion (\ref{Flpt}), then finally gives us the sought relation between the $\tf_n$ and $\td_n$ coefficients
\be
\tf_n(\nu_0;\kappa) = \sum_{r=0}^{[n/2]}  (- \beta_0^2 \pi^2)^r \frac{\Gamma(\nu_0+n)}{\Gamma(\nu_0+n-2 r) (2 r+1)!} \td_{n-2 r}(\nu_0;\kappa).
\label{tfnvstdn} \ee
Since $\tB^{(\nu_0)}[\cD](u;\kappa)$ has the the form (\ref{tBD}), this implies that $\td_n(\nu_0;\kappa)$ has the asymptotic form as given in Eq.~(\ref{tdnasym}). We use this form in the relations (\ref{tfnvstdn}) in order to obtain the asymptotic form for the coefficients $\tf_n(\nu_0;\kappa)$
\bes
\label{tfnasym}
\bea
\tf_n(\nu_0;\kappa) &=& \frac{\tcK(\kappa) p^{-\ts_0}}{\Gamma(\ts_0)}
\sum_{r=0}^{[n/2]} (-\beta_0^2 \pi^2)^r \left( \frac{\beta_0}{p} \right)^{n-2 r} \frac{(n - 2 r)^{\ts_0-1}}{(2 r +1)!} (n- 2 r)! (\nu_0+n-1) \cdots (\nu_0+n-2 r) \left[ 1 + \cO \left( \frac{1}{(n-2 r)} \right) \right]
\nonumber
\label{tfnasym0} \\
& = & \frac{\tcK(\kappa) p^{-\ts_0}}{\Gamma(\ts_0)}
\left( \frac{\beta_0}{p} \right)^n n! n^{\ts_0-1}
{\Bigg \{}  1  + \ldots +
\frac{(-\beta_0^2 \pi^2)^r}{(2 r+1)!} \left( \frac{p}{\beta_0} \right)^{2 r}
\left( 1 - \frac{(1-\nu_0)}{n} \right) \cdots \left( 1 - \frac{(1-\nu_0)}{(n- 2 r +1)} \right)
\nonumber\\
&& \times \left[ 1 + \cO \left(\frac{1}{(n-2 r)} \right) \right]   + \ldots
{\Bigg \}}
\label{tfnasyma} \\
& = & \frac{\tcK(\kappa) p^{-\ts_0}}{\Gamma(\ts_0)}
\left( \frac{\beta_0}{p} \right)^n n! n^{\ts_0-1}
\left\{ 1 + \ldots + \frac{(-\beta_0^2 \pi^2)^r}{(2 r+1)!} \left( \frac{p}{\beta_0} \right)^{2 r} \left[ 1 + \cO \left(\frac{1}{(n-2 r)} \right)  + \cO \left(\frac{r}{(n-2 r)} \right) \right] + \ldots \right\}
\label{tfnasymb} \\
& \approx  &
\frac{\tcK(\kappa) p^{-\ts_0}}{\Gamma(\ts_0)}
\left( \frac{\beta_0}{p} \right)^n n! n^{\ts_0-1}   
\left\{ 1 + \ldots + \frac{(-p \pi)^{2 r}}{(2 r+1)!} + \ldots \right\} \left[ 1 + \cO \left(\frac{1}{n} \right) \right]
\label{tfnasymc} \\
& = &
\frac{\tcK(\kappa) p^{-\ts_0}}{\Gamma(\ts_0)}
\left( \frac{\beta_0}{p} \right)^n n! n^{\ts_0-1}  \frac{\sin(\pi p)}{\pi p} 
\left[ 1 + \cO \left(\frac{1}{n} \right) \right].
\label{tfnasymd} \eea \ees
In Eq.~(\ref{tfnasymc}) we took into account that $n \gg 1$, and that in the sum over $r$ only terms $r \ll n$ contribute, because of the strong suppression of the terms by the factor $1/(2 r+1)!$. In fact, as we see in Eqs.~(\ref{tfnasymc})-(\ref{tfnasymd}), the sum over $r$ converges fast to $\sin(\pi p)/(\pi p)$. When we compare this result with the asymptotic behaviour of $\td_n(\nu_0;\kappa)$ Eq.~(\ref{tdnasym}), we see that the result (\ref{tfnasymd}) implies the relation
\be
\tf_n(\nu_0;\kappa) = \frac{\sin(\pi p)}{\pi p} \td_n(\nu_0;\kappa) \left(1 + \cO (1/n) \right),
\label{tfnvstdnasym} \ee
and this then immediately implies Eq.~(\ref{tBF}). This concludes the proof of Theorem 6.

\vspace{0.5cm}

One of the consequences of Theorems 5 and 6 is the following lemma:

\vspace{0.5cm}

\noindent
{\bf Lemma:}
The timelike quantities $\cF$ and $\cFo$ are defined in Eqs.~(\ref{FF1def}), with the expansions in the generalised logarithmic derivatives as denoted in Eqs.~(\ref{FF1lpt}). If the modified Borel of the coresponding spacelike quantity $\cD$ has the IR renormalon form Eq.~(\ref{tBD}), i.e., $\tB^{(\nu_0)}[\cD](u;\kappa) \sim (p-u)^{-\ts_0}$, then we have the following  relation between the coefficients of these expansion:
\be
\tf_n(1;\kappa) = \frac{\Gamma(\nu_0) \Gamma(1+n)}{\Gamma(\nu_0+n)} \tf_n(\nu_0;\kappa) \left( 1 + \cO(1/n) \right),
\label{tfn1} \ee
which is asymptotically analogous to the relations (\ref{tdn1}) of the $d_n$ coefficients of the corresponding spacelike quantities $\cD$ and $\cDo$.

\vspace{0.3cm}

\noindent
{\bf Proof:}

\noindent
The form Eq.~(\ref{tBD}) for $\tB^{(\nu_0)}[\cD](u;\kappa)$ implies, according to Theorem 3, Eq.~(\ref{tBD1ren}), a specific form for $\tB[\cDo](u;\kappa)$, namely
\bes
\label{tBDD1}
\bea
\tB^{(\nu_0)}[\cD](u;\kappa) &=& \frac{\tcK(\kappa)}{(p-u)^{\ts_0}} \left[ 1 + \cO((p-u)) \right], \; \Rightarrow
\label{tBDb} \\
\tB[\cDo](u;\kappa) &=& \frac{\tcKo(\kappa)}{(p-u)^{\ts_0-\nu_0+1}} \left[ 1 + \cO((p-u)) \right].
\label{tBD1} \eea \ees
In conjunction with Theorem 5, Eq.~(\ref{tBD1vstBF1}), this implies
\bes
\label{tBF1p}
\bea
\tB[\cFo](u;\kappa) &=& \frac{\sin(\pi u)}{\pi u} \frac{\tcKo(\kappa)}{(p-u)^{\ts_0-\nu_0+1}} \left[ 1 + \cO((p-u)) \right]
\label{tBF1pa} \\
& = & 
\frac{\sin(\pi p)}{\pi p} \frac{\tcKo(\kappa)}{(p-u)^{\ts_0-\nu_0+1}} \left[ 1 + \cO((p-u)) \right],
\label{tBF1pb} \eea \ees
where in the last identity we took into account that
\be
\frac{\sin(\pi u)}{\pi u} =
\frac{\sin(\pi p)}{\pi p} \left( 1 + \cO(p-u) \right).
\label{sinexp} \ee
The relation (\ref{tBF1pb}), in conjunction with the notations (\ref{D1def}) and (\ref{F1lpt}) for the expansions of $\cDo$ and $\cFo$, respectively, then implies
\be
\tf_n(1;\kappa) = \frac{\sin(\pi p)}{\pi p} \td_n(1;\kappa) \left(1 + \cO(1/n) \right).
\label{tfn1vstdn1asym} \ee
This relation, together with the analogous relation (\ref{tfnvstdnasym}) for
$\tf_n(\nu_0;\kappa)$ and $\td_n(\nu_0;\kappa)$ [that reflects Theorem 6, Eqs.~(\ref{tBD})-(\ref{tBF})], and the original relation (\ref{tdn1}) between  $\td_n(1;\kappa)$ and $\td_n(\nu_0;\kappa)$, then immediately implies the relation (\ref{tfn1}), i.e., the claim of the Lemma. This concludes the proof of the Lemma.

\begingroup \color{black}
{\bf Lemma\/} Eq.~(\ref{tfn1}) has the following consequence: {\bf Theorem 3\/} [cf.~Eqs.~(\ref{tBDren}) and (\ref{tBD1ren})], that was applied to $\tB^{(\nu_0)}[\cD](u)$ and $\tB[\cDo](u)$, can be applied also to $\tB^{(\nu_0)}[\cF](u)$ and $\tB[\cFo](u)$:
\bes
\label{tBFtBF1}
\bea
\tB^{(\nu_0)}[\cF](u;\kappa) &=& \frac{\tcK(\kappa)}{(p-u)^{\ts_0}} \left[ 1 + \cO((p-u)) \right] \quad \Rightarrow
\label{tBFc} \\
\tB[\cFo](u;\kappa) &=& \frac{\tcKo(\kappa)}{(p-u)^{\ts_0-\nu_0+1}} \left[ 1 + \cO((p-u)) \right].
\label{tBF1} \eea \ees
This means that, if we know $\tB^{(\nu_0)}[\cF](u)$, then $\tB[\cFo](u)$ can be obtained. When using subsequently {\bf Theorem 5\/}, Eq.~(\ref{tBD1vstBF1}), the modified Borel transform $\tB[\cDo](u)$ is obtained.

Further, we note that {\bf Theorem 2\/} [Eqs.~(\ref{BDren})-(\ref{s0vsts0})] is valid also for timelike observables, because in the proof of that Theorem we did not use the spacelike (or timelike) character of the associated observable. Therefore
\bes
\label{BFtBF}
\bea
B^{(\nu_0)}[\cF](u;\kappa) &=& \frac{\cK(\kappa)}{(p-u)^{s_0}} \left[ 1 + \cO((p-u)) \right] \quad \Rightarrow
\label{BFb} \\
\tB^{(\nu_0)}[\cF](u;\kappa) &=& \frac{\tcK(\kappa)}{(p-u)^{\ts_0}} \left[ 1 + \cO((p-u)) \right],
\label{tBFb} \eea \ees
where the two indices are related
\be
s_0 = \ts_0+ p \frac{\beta_1}{\beta_0^2}.
\label{s0vsts0b} \ee

Therefore, we conclude that from the knowledge of the structure of the (usual) Borel transform $B^{(\nu_0)}[\cF](u)$ of a timelike observable $\cF(\sigma)$, we obtain the structure of the modified Borel transform $\tB[\cDo](u)$ of the auxiliary spacelike quantity $\cDo(Q^2)$, via the following steps:
\be
B^{(\nu_0)}[\cF](u) \mapsto \tB^{(\nu_0)}[\cF](u) \mapsto \tB[\cFo](u) \mapsto \tB[\cDo](u),
\label{Bsteps} \ee
where the first step is via Eqs.~(\ref{BFtBF}), the second step is via Eqs.~(\ref{tBFtBF1}), and the third step is via {\bf Theorem 5\/} [Eq.~(\ref{tBD1vstBF1})].

We recall that we usually (or sometimes) know the renormalon structure of the (usual) Borel transform of the considered timelike quantity $\cF(\sigma)$, but in order to resum it we need the characteristic function $F_{\cDo}(t)$ [cf.~Eq.~(\ref{Fres})], and to obtain the latter we need to know $\tB[\cDo](u)$ because $F_{\cDo}(t)$ is the inverse Mellin transform of  $\tB[\cDo](u)$ [cf.~Eq.~(\ref{FD1})]. And the structure of $\tB[\cDo](u)$ is obtained from the structure of $B^{(\nu_0)}[\cF](u)$ via the steps (\ref{Bsteps}).
\endgroup

\section{The use of holomorphic QCD (AQCD) couplings in the formalism}
\label{sec:AQCD}

The holomorphic (AQCD) coupling $\A(Q^2)$ [i.e., when $a(Q^2)$ is replaced by $\A(Q^2)$] is such that $\A(Q^2)$ is a holomorphic (analytic) function of $Q^2$ in the entire complex $Q^2$-plane with the exception of the negative semiaxis, i.e., for all $Q^2 \in  \mathbb{C} \backslash (-\infty,-M^2_{\rm thr})$ where $M_{\rm thr} \sim m_{\pi} \sim 0.1 \ {\rm GeV}$ is a threshold mass. Such a behaviour qualitatively reflects the holomorphic properties of the (QCD) spacelike observables $\cD(Q^2)$, where the latter properties are a direct consequence of the locality, unitarity and causality of Quantum Field Theories \cite{BookShirkov}. This is in contrast to the properties of the pQCD coupling $a(Q^2)$ which has (Landau) cuts within the Euclidean regime of $Q^2$, usually on the positive axis: $0 \leq Q^2 \leq \Lambda^2_{\rm Lan}$; $\Lambda_{\rm Lan} \sim 0.1 \ {\rm GeV}$. A basic property of $\A(Q^2)$ is that it should effectively coincide with the perturbative $a(Q^2)$ for large Euclidean $Q^2$ ($Q^2 > 1 {\rm GeV}^2$).

Usually the AQCD coupling framework is defined via the specification of the form of the spectral (discontinuity) function of the coupling $\A$ along its cut: $\rho_{\A}(\sigma) \equiv {\rm Im} \; \A(-\sigma - i \varepsilon)$ for all $\sigma \geq M^2_{\rm thr}$. At large squared timelike momenta $\sigma$, this function should effectively coincide with its pQCD counterpart, $\rho_{\A}(\sigma) = \rho_{a}(\sigma)$. Here, $a(Q^2)$ is called the underlying pQCD coupling; $a(Q^2)$ and $\A(Q^2)$ are in the same renormalisation scheme.

This approach suggests that the construction of an AQCD framework reduces basically to the modelling of the spectral function $\rho_{\A}(\sigma)$ at low $\sigma$: $0 < \sigma < 1 \ {\rm GeV}^2$.\footnote{The most prominent early model of holomorphic coupling has the spectral function $\rho_{\A}(\sigma)$ equal to the pQCD  $\rho_{a}(\sigma)$ for all $\sigma>0$, including low values $\sigma \lesssim 1 \ {\rm GeV}^2$. This model can be called the Minimal Analytic framework (MA; named also (F)APT) \cite{Shirkov,SMS,ShirRev,FAPT,SMnu,BakRev}, and thus has $\rho_{\A}^{\rm (MA)}(\sigma) = \Theta(\sigma) \rho^{\rm (pt)}(\sigma)$. The difference $\A(Q^2) - a(Q^2)$ in the MA at large $|Q^2|$ is $\sim (\Lambda^2_{\rm Lan}/Q^2)^1$, i.e., a relatively slow fall-off when $|Q^2|$ increases. A generalisation of (F)APT was proposed in \cite{AM1,AM2} in the context of the pion form factor.}
In this regime, it is reasonable to represent the parametrisation of  $\rho_{\A}(\sigma)$  in terms of a sum of a finite number of Dirac deltas, e.g. two deltas \cite{2dQCD1,2dQCD2} or three deltas \cite{3dAQCD}.

In the present work we choose an ansatz with three Dirac $\delta$-functions (the resulting version of QCD is denoted as 3$\delta$AQCD), i.e., we make the following ansatz for the corresponding spectral function:
\be
\rho_{\A}(\sigma) = \pi \sum_{j=1}^3 {\cF}_j \delta(\sigma - M_j^2) + \Theta(\sigma - M_0^2) \rho_{a}(\sigma),
\label{rhoA} \ee
($\Theta$ being the Heaviside step function) and we assume the following hierarchy of the squared masses: $0 < M_1^2 < M_2^2 < M_3^2 < M_0^2$. Here, $M_1^2=M^2_{\rm thr}$ is the IR-threshold scale (i.e., $\sigma_{\rm min}=M_1^2$), and $M_0^2$ is the pQCD-onset scale ($M_0 \sim 1$ GeV). The holomorphic coupling is then obtained, with the use of the Cauchy theorem, as a dispersion integral involving $\rho_{\A}(\sigma)$
\bes
\label{A}
\bea
\A(Q^2) & = & \frac{1}{\pi} \int_{-M^2_{\rm thr}- \eta}^{\infty} \frac{d \sigma \; \rho_{\A}(\sigma)}{(\sigma + Q^2)} \qquad (\eta \to +0)
\label{Aa} \\
& = &  \sum_{j=1}^3 \frac{{\cF}_j}{(Q^2 + M_j^2)} + \frac{1}{\pi} \int_{M^2_0}^{\infty} \frac{d \sigma  \; \rho_{a}(\sigma)}{(\sigma + Q^2)}.
\label{Ab} \eea \ees
The pQCD coupling $a(Q^2)$ (in the LMM scheme) has $n_f=3$ or $n_f=4$, if we are interested in low-energy QCD phenomenology. This $a(Q^2)$, and thus also its spectral function $\rho_a(\sigma) \equiv {\rm Im} \; a(-\sigma - i \varepsilon)$, is completely fixed by specifying the value of $\alpha_s^{(\MSbar)}(M_Z^2)$.

For specifying numerically the other seven parameters ($\cF_j$, $M_j^2$ for $j=1,2,3$; and $M_0^2$), we use the following inputs: For large $Q^2$ ($|Q^2| > 1 \ {\rm GeV}^2$) we require $\A(Q^2)$ to approach the underlying perturbative coupling $a(Q^2)$ quickly, specifically $\A(Q^2)-a(Q^2) \sim (\Lambda_{\rm Lan}^2/Q^2)^5$, and this gives us four conditions;  whereas at low $Q^2$ the required behaviour is as suggested by large-volume lattice calculations \cite{3dLVLattice,Lattb,Latt2b,Latt2c}: $\A(Q^2)$ at positive $Q^2$ has a local maximum at $Q^2 \approx 0.135 \ {\rm GeV}^2$ and for $Q^2 \to 0$ it behaves as $\A(Q^2) \sim Q^2$ ($\to 0$). One of the seven parameters is then still free, which we choose to be the threshold scale of the spectral function, namely $\sigma_{\rm thr.}=M_1^2$, which is expected to be of the order of the square of the lowest hadronic mass: $M_1^2 \approx m_{\pi}^2$ ($\sim 0.15^2 \ {\rm GeV}^2$). Note that the renormalisation scheme we use in this construction is MiniMOM \cite{MiniMOM} because the large-volume lattice calculations (\cite{3dLVLattice,Lattb}) are performed in this scheme. In addition, we rescale the momenta to the usual $\MSbar$-type scheme scales (scaled with $\Lambda^2_{\MSbar}$), and this is called Lambert MiniMOM (LMM) scheme. We refer for details of construction of $\A(Q^2)$ in such framework to \cite{3dAQCD,amu3dAQCD}.\footnote{
\textcolor{black}{In principle, it is possible to include even more than three Dirac $\delta$-functions in the ansatz for $\rho_{\A}(\sigma)$. The number of equations fixing all the parameters then increases, e.g., by requiring very high level of coincidence of $\A(Q^2)$ with the underlying pQCD coupling $a(Q^2)$ in the high-momentum perturbative regime $|Q^2| > 1 \ {\rm GeV}^2$, such as $\A(Q^2) - a(Q^2) \sim  (\Lambda_{\rm Lan}^2/Q^2)^{N_{\rm max}}$ with $N_{\rm max} > 5$. However, all these equations are nonlinear, and it becomes increasing hard to solve them numerically as their number increases. Further, in our case of (only) three Dirac delta functions, we can already fulfill the (two) mentioned conditions in the IR regime, and in the UV regime we have $N_{\rm max}=5$ which is sufficient for the coupling $\A(Q^2)$ to practically coincide with the underlying pQCD coupling $a(Q^2)$ for $Q^2 > 3 \ {\rm GeV}^2$, cf.~Fig.~\ref{FigAs}.}}

In AQCD, the generalised logarithmic derivatives in pQCD, $\ta_{\nu}(Q^2)$ [cf.~Eq.~(\ref{tanuanu})], get replaced by their AQCD analogs $\tA_{\nu}(Q^2)$ whose expression turns out to be, conveniently, determined entirely by a dispersive integral involving the spectral function $\rho_{\A}(\sigma)$ \cite{GCAK}\footnote{The use of the logarithmic derivatives $\tA_{\nu}$ for integer $\nu=n$ for any AQCD framework was developed in \cite{CV}, and was extended to noninteger $\nu$ in \cite{GCAK} (for any AQCD). For the Minimal Analytic (MA) QCD, the extended logarithmic derivatives $\tA_{\nu}^{\rm (MA)}(Q^2)$ were constructed, in a MA-specific way, as explicit functions at one-loop order in \cite{FAPT} and at any loop order in \cite{Kotikov}. We point out that the general approach \cite{GCAK}, Eq.~(\ref{tAnu}), can also be applied in the MA case and it gives the same numerical result as the MA-specific approach.}
\bea
\tA_{\nu}(Q^2) & = & \frac{1}{\pi} \frac{(-1)}{\beta_0^{\nu-1} \Gamma(\nu)}
  \int_{0}^{\infty} \frac{d \sigma}{\sigma} \rho_{\A}(\sigma)
  {\rm Li}_{-\nu+1} \left(-\frac{\sigma}{Q^2} \right) \quad (\nu > 0).
  \label{tAnu} \eea
This formula can even be modified, by a subtraction approach \cite{GCAK}, so that is becomes valid for even lower values of the index $\nu$ ($\nu > -1$).

Furthermore, the timelike analog of this coupling, namely $\tH_{\nu}(\sigma)$ which is the AQCD analog of $\ths_{\nu}$ of Eq.~(\ref{thnudef}), can also be obtained in terms of an integral involving the spectral function $\rho_{\A}$ \cite{GCAK}
\bes
\label{tH}
\bea
\tH_{\nu}(\kappa \sigma) &\equiv& \frac{1}{2 \pi} \int_{-\pi}^{\pi} d \phi \; \tA_{\nu}(\kappa \sigma e^{i \phi}).
\label{tHa} \\
& = & - \frac{\sin (\pi \nu)}{\pi^2 (\nu-1) \beta_0^{\nu-1}} \int_{0}^{\infty} \frac{d w}{w^{\nu-1}} \rho_{\A}(\sigma e^w) \quad (0 < \nu < 2),
\label{tHb} \eea \ees
where $\sigma > 0$.
The expressions of $\tH_{\nu}$ for higher indices $\nu \geq 2$ are also given in \cite{GCAK}. In the case of 3$\delta$AQCD, Eq.~(\ref{rhoA}), the integration Eq.~(\ref{tHb}) obtains the form
\be
\tH_{\nu}(\kappa \sigma) =\frac{\sin (\pi \nu) \beta_0^{1-\nu}}{\pi^2 (1-\nu)}
\left\{ \pi \sum_{j=1}^3 \frac{\cF_j}{M_j^2}
\Theta (M_j^2 -\sigma)
\ln^{1-\nu} \left( \frac{M_j^2}{\sigma} \right)
+ \int_{\Theta(M_0^2-\sigma) \ln(M_0^2/\sigma)}^{\infty} dw \; w^{1-\nu}  \rho_a(\sigma e^w) \right\}.
\label{tHb3d} \ee
\textcolor{black}{The differential recursion relation (\ref{derthnu}) remains valid in AQCD \cite{GCAK}, i.e., with the substitutions $\ths \mapsto \tH$.}

All the results of the previous Sections with pQCD couplings can now be simply rewritten in AQCD, by the replacements:\footnote{We point out, though, that these results are evaluated in practice in (any) AQCD more simply than in pQCD, because in AQCD the basic elements are $\A$ and the generalised logarithmic derivatives $\tA_{\nu}$ and $\tH_{\nu}$ (determined entirely by $\rho_{\A}(\sigma)$), while this is not true for pQCD (where the powers of $a$ are the basic elements).}  $a \mapsto \A$, $\ta_{\nu} \mapsto \tA_{\nu}$, and $\ths_{\nu} \mapsto \tH_{\nu}$. The central results for renormalon-motivated  resummations, Eqs.~(\ref{restK}), are now simply rewritten for (any) AQCD in the following form:
\bes
\label{restKA}
\bea
\cDo(Q^2)_{\rm res.} &=& \int_0^{\infty} \frac{dt}{t}  F_{\cDo}(t)_{(p,\ts)} \A(t e^{-\tK_e} Q^2),
\label{D1restKA} \\
\cD(Q^2)_{\rm res.} &=& \int_0^{\infty} \frac{dt}{t}  F_{\cDo}(t)_{(p,\ts)} \tA_{\nu_0}(t e^{-\tK_e} Q^2),
\label{DrestKA} \\
\cF(\sigma)_{\rm res.} &=& \int_0^{\infty}  \frac{dt}{t}  F_{\cDo}(t)_{(p,\ts)} \tH_{\nu_0}(t e^{-\tK_e} \sigma).
\label{FrestKA}
\eea \ees
Furthermore, due to the Landau cuts of the pQCD coupling $a(Q^2)$ at low positive $Q^2$  ($0 \leq Q^2 \leq \Lambda^2_{\rm Lan}$) we have to regularise the pQCD resummations Eqs.~(\ref{restK}) in some way to avoid those cuts in the integration of $t$ at low $t$ values. One possibility is a PV-type regularisation
\bes
\label{restKr}
\bea
\cDo(Q^2)_{\rm res.} &=& {\rm Re} \; \int_0^{\infty} \frac{dt}{t}  F_{\cDo}(t)_{(p,\ts)} a(t e^{-\tK_e} Q^2 + i \varepsilon),
\label{D1restKr} \\
\cD(Q^2)_{\rm res.} &=& {\rm Re} \;\int_0^{\infty} \frac{dt}{t}  F_{\cDo}(t)_{(p,\ts)} \ta_{\nu_0}(t e^{-\tK_e} Q^2 + i \varepsilon),
\label{DrestKr} \\
\cF(\sigma)_{\rm res.} &=& {\rm Re} \;\int_0^{\infty}  \frac{dt}{t}  F_{\cDo}(t)_{(p,\ts)} \ths_{\nu_0}(t e^{-\tK_e} \sigma + i \varepsilon),
\label{FrestKr}
\eea \ees
where $\varepsilon \to +0$.

In pQCD, one may argue that for the resummation of a timelike quantity $\cF(\sigma)$ we may use exactly the same approach as for a spacelike quantity $\cD(Q^2)$: namely, if the perturbation expansions of these quantities, one in powers of $a(\sigma)$ and the other in powers of $a(Q^2)$, have the same structure, cf.~Eqs.~(\ref{Dptlpt}) and (\ref{Flpt}),\footnote{Eq.~(\ref{Flpt}) is a series in generalised logarithmic derivatives $\ta_{\nu_0+n}(\kappa \sigma)$, but it can be rewritten in powers, namely as $\sum f_n(\nu_0;\kappa) a(\kappa \sigma)^{\nu_0+n}$, in exactly the same way that $\cD(Q^2)$ was written in powers of $a(\kappa Q^2)^{\nu_0+n}$ in Eq.~(\ref{Dpt}).} then we can construct the auxiliary quantity $\cF^{(1)}(\sigma)$ and the corresponding characteristic function $F_{\cF^{(1)}}(t)$ that would then appear in the integral of $t$ involving also the factor $\ta_{\nu_0}(t e^{-\tK_e} \sigma)$ [$\mapsto \tA_{\nu_0}(t e^{-\tK_e} \sigma)$]. However, the work \cite{GCAK} makes it clear that the couplings $a$ and $\ta_{\nu_0}$ have their AQCD analogs $\A$, $\tA_{\nu_0}$ that are holomorphic functions of $Q^2$ in the complex-$Q^2$ plane (with the exception of the negative semiaxis), and that they reflect in this way the holomorphic properties of the spacelike QCD observables. The timelike QCD observables $\cF(\sigma)$ have no such holomorphic properties, they are defined only for $\sigma >0$, and they or their derivatives are in general not even continuous functions of $\sigma$. Therefore, the use of $\A$ and $\tA_{\nu_0}$ (in pQCD: of $a$ and $\ta_{\nu_0}$) couplings for $\cF(\sigma)$ is not warranted. 
  
As mentioned, the underlying pQCD coupling of the 3$\delta$AQCD is in the Lambert MiniMOM (LMM) renormalisation scheme, where the first two scheme parameters $c_j$($\equiv \beta_j/\beta_0$) ($j=2,3$) are known \cite{MiniMOM}. For practical reasons, we use the pQCD coupling whose $\beta$-function has a specific form of the $\beta(a)$, namely P[4/4]$(a)$ Pad\'e,\footnote{When this $\beta$-function is expanded in powers of $a$, the first four terms are reproduced [cf.~Eq.~(\ref{RGEb})], with the correct MiniMOM values of $c_2$ and $c_3$.}, because then the pQCD running coupling can be expressed as an explicit function involving Lambert function $W_{\pm 1}(z)$, and where $z$ is a dimensionless quantity scaled by a specific scale which we call Lambert scale $\Lambda_{\rm L}$: $z = - (\Lambda_{\rm L}^2/Q^2)^{\beta_0/c_1}$ $\times 1/(e c_1)$ (where $e=2.71828$). The Lambert scale (either at $n_f=3$ or $n_f=4$) is uniquely determined by the value of the $\MSbar$ pQCD coupling at the canonical high scale $M_Z^2$ (at $n_f=5$), $\alpha_s^{\MSbar}(M_Z^2)$. We refer for details to \cite{3dAQCD}.

When we change the renormalisation scheme (i.e., we change the scheme parameters $c_j$, $j \geq 2$), e.g., from $\MSbar$ to LMM, the expansion coefficients $f_n$ (and $d_n$) transform according to the rules as given in Appendix \ref{app:RSch}. We point out that the LMM scheme involves, in our convention, no rescaling of the momenta (i.e., it remains in that sense a $\MSbar$-type scheme), only the scheme parameters $c_2$ and $c_3$ change. 

In Tables \ref{tab3dnf3} and \ref{tab3dnf4} we present the parameters of the 3$\delta$AQCD coupling (\ref{A}) for $n_f=3$ and $n_f=4$, respectively, for various values of the coupling $\alpha_s^{\MSbar}(M_Z^2)=0.1180 \pm 0.0009$ (i.e., the world average values \cite{PDG2024}), and for various threshold scales $M_1 \sim m_{\pi}$, namely $M_1=(0.150^{+0.100}_{-0.050})$ GeV.
We note that in the $n_f=4$ case, the underlying coupling is the (LMM-scheme) pQCD coupling $a(Q^2;n_f=4)$.
\begin{table}
  \caption{Values of the parameters of the 3$\delta$AQCD coupling ($n_f=3$, LMM scheme), for various values of the input parameters: the IR-threshold scale $M_1=(0.150^{+0.100}_{-0.050})$ GeV and $\alpha_s^{\MSbar}(M_Z^2)=0.1180 \pm 0.0009$. The dimensionless parameters are $s_j=M_j^2/\Lambda_L^2$ and $f_j={\cal F}_j/\Lambda_L^2$. $\Lambda_L$ is the Lambert scale $(n_f=3)$ of the underlying pQCD coupling, determined by the value of  $\alpha_s^{\MSbar}(M_Z^2)$ ($n_f=5$). Our central case is $M_1=0.150$ GeV and $\alpha_s^{\MSbar}(M_Z^2)=0.1180$.}
\label{tab3dnf3}
\begin{ruledtabular}
\centering
\begin{tabular}{r|r|llllllll}
$M_1$ [GeV] & $\alpha_s^{\MSbar}(M_Z^2)$ & $\Lambda_{{\rm L}}$ [GeV] & $f_1$ & $f_2$ & $f_3$  & $s_1$ & $s_2$ & $s_3$ &  $s_0$ 
\\
\hline
0.100 & 0.1180 &  0.112500 & -0.168833 & 12.21307 & 7.50079 & 0.790117 & 87.4086 & 858.412 &  1158.76 \\
0.150 & 0.1180 & 0.112500 & -0.583466 & 10.64786 & 6.055440 & 1.77776 & 42.6800 & 605.184 & 824.850 \\
0.250 & 0.1180 &  0.112500 & -4.35108 & 13.24663 & 5.22514 &  4.93823 & 16.58907 & 469.278  & 645.569 \\
\hline
0.150 & 0.1171 & 0.108036 & -0.609272 & 10.82440 & 6.16694 & 1.92773 & 45.5255 & 623.641 &  849.268 \\
0.150 & 0.1189 & 0.117067 & -0.559701 & 10.48340 & 5.95174 & 1.64178 & 40.0474 & 588.119 &  802.275 \\
\hline
\end{tabular}
\end{ruledtabular}
\end{table}
\begin{table}
  \caption{The same as in Table \ref{tab3dnf3}, but now with the underlying pQCD coupling (in the LMM scheme) with $n_f=4$.}
 \label{tab3dnf4}
\begin{ruledtabular}
\centering
\begin{tabular}{r|r|llllllll}
 $M_1$ [GeV] & $\alpha_s^{\MSbar}(M_Z^2)$ & $\Lambda_{{\rm L}}$ [GeV] & $f_1$ & $f_2$ & $f_3$  & $s_1$ & $s_2$ & $s_3$ &  $s_0$ 
\\
\hline
0.100 & 0.1180 &  0.0960725 &  -0.225523 & 14.44965 & 8.98645 & 1.08343 & 112.1859 & 981.749  & 1319.90 \\
0.150 & 0.1180 &  0.0960725 & -0.757927 & 12.28852 & 7.00460 & 2.43772 & 54.4228 & 659.170 & 894.669 \\
0.250 & 0.1180 &  0.0960725 & -5.615187 & 15.55630 & 5.86556 &  6.77145 & 21.0441 & 488.475  & 669.434 \\
\hline
0.150 & 0.1171 & 0.0917892 & -0.799402 & 12.56215 & 7.17931 & 2.67054 & 58.4764 & 685.697 &  929.797 \\
0.150 & 0.1189 & 0.100482 & -0.719985 & 12.03468 & 6.84274 & 2.22845 & 50.6968 & 634.799 &  862.397 \\
\hline
\end{tabular}
\end{ruledtabular}
\end{table}

\begingroup \color{black}
In Fig.~\ref{FigAs} we present the 3$\delta$AQCD running coupling $\A(Q^2)$, Eq.~(\ref{Ab}), for $n_f=3$ and $\alpha_s^{\MSbar}(M_Z^2)=0.1180$, for positive $Q^2$ and various threshold scales $M_1$, cf.~Table \ref{tab3dnf3}. These couplings were first constructed and used in \cite{3dAQCD}. Similar couplings were also obtained numerically by large-volume lattice calculations \cite{3dLVLattice,Lattb,Latt2b,Latt2c}.\footnote{
\textcolor{black}{Our construction was guided by the lattice results of \cite{3dLVLattice,Lattb,Latt2b,Latt2c}.}}
Within the framework of the Curci-Ferrari version of QCD, the authors of \cite{Pelaez} obtained the ghost-gluon coupling that also has a similar form.
\begin{figure}[htb] 
\centering\includegraphics[width=80mm]{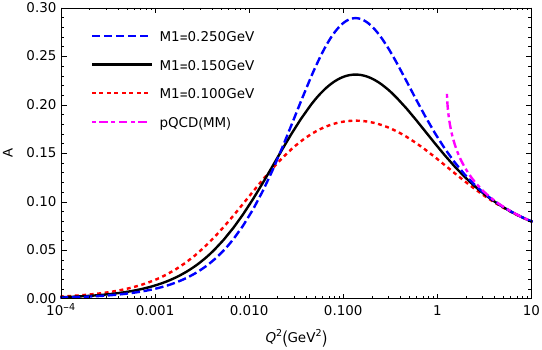}
 \vspace{4pt}
 \caption{\footnotesize \textcolor{black}{$\A(Q^2)$ of 3$\delta$AQCD for positive $Q^2$ and various values of the threshold scale parameter $M_1$. For comparison, the underlying pQCD coupling $a(Q^2)$ (in the same LMM scheme) is included.}}
 \label{FigAs}
\end{figure}
\endgroup

\begin{figure}[htb] 
\begin{minipage}[b]{.49\linewidth}
\includegraphics[width=80mm,height=50mm]{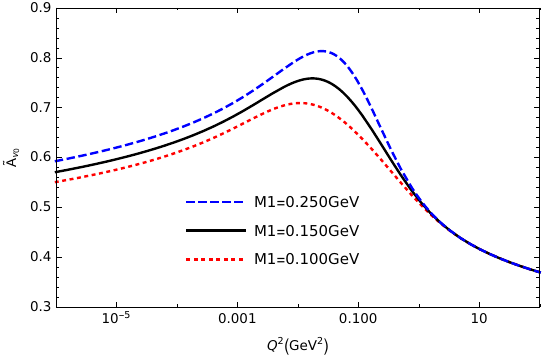}
\end{minipage}
\begin{minipage}[b]{.49\linewidth}
  \includegraphics[width=80mm,height=50mm]{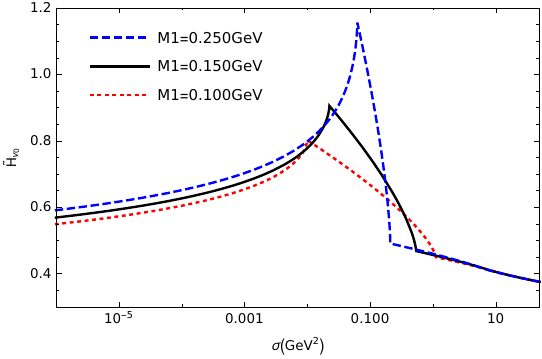}
\end{minipage} \vspace{4pt}
\caption{\footnotesize The spacelike running coupling $\tA_{\nu_0}(Q^2)$ for positive $Q^2$ (left-hand figure), and the timelike running coupling $\tH_{\nu_0}(\sigma)$ (right-hand figure), in 3$\delta$AQCD, with $n_f=3$, $\alpha_s^{\MSbar}(M_Z^2)=0.1180$, and three different values of the IR-threshold scale parameter $M_1$.} 
\label{FigtAtHnu0}
\end{figure}

In Figs.~\ref{FigtAtHnu0}(a), (b) we present the spacelike and timelike running coupling, $\tA_{\nu_0}(Q^2)$ and $\tH_{\nu_0}(\sigma)$, respectively, in 3$\delta$AQCD with $n_f=3$ (and $\nu_0=1/3$), for $\alpha_s^{\MSbar}(M_Z^2)=0.1180$ and for three different IR-threshold scales $M_1=(0.150^{+0.100}_{-0.050})$ GeV.
We recall that the spacelike couplings, such as $\tA_{\nu_0}(Q^2)$, are holomorphic functions of $Q^2$ in the $Q^2$-complex plane $Q^2 \in \mathbb{C} \backslash (-\infty,-M^2_{\rm thr})$, where $M_{\rm thr}=M_1$ is the IR-threshold scale of the spectral function $\rho_{\A}(\sigma)$, Eq.~(\ref{rhoA}). On the other hand, the timelike couplings, such as $\tH_{\nu_0}(\sigma)$, are defined only for positive $\sigma \geq 0$, and are in general neither holomorphic nor are their derivatives continuous, especially when the low-$\sigma$ regime of $\rho_{\A}(\sigma)$ is parametrised by Dirac delta functions. In fact, the derivatives of $\tH_{\nu_0}(\sigma)$ are discontinuous at $\sigma=M_1^2, M_2^2, M_3^2$ (although the discontinuity at the highest squared mass $M_3^2 \sim 10 \ {\rm GeV}^2$ is weak and cannot be seen by simple inspection).

\section{An example: Implementation of resummation, timelike case}
\label{sec:impl}

Here we will show how to implement the renormalon-motivated resummation in a case of a timelike observable. The considered observable will be $\cF(\sigma) = {\hat C}(m)$ of Ref.~\cite{GandN}, which is the renormalisation-scheme invariant factor of the Wilson coefficient of the chromomagnetic operator in the heavy-quark effective theory (HQET) for hadronic bound states containing one heavy quark ($c$ or $b$). Strictly speaking, $\cF(\sigma) = \pi^{-\nu_0} {\hat C}(m)$, where $\sigma=m$ is the (pole) mass of the heavy quark, and $\nu_0=\gamma_0/(8 \beta_0)$ is the (noninteger) power index in the expansion of $\cF(\sigma)$ Eq.~(\ref{Flpt}). Here, $\gamma_0=2 C_A=6$ is the one-loop coefficient of the anomalous dimension of the chromomagnetic operator, and $\beta_0=(11 - 2 n_f/3)/4$ is the aforementioned one-loop coefficient of the $\beta$-function [cf.~Eqs.~(\ref{RGE})], where $n_f$ is the number of active light quark flavours. For example, in the case of $n_f=3$ ($m=m_c \approx 1.67 \ {\rm GeV}$) we have $\nu_0=1/3$.

The expansion in (generalised) logarithmic derivatives of $\cF(\sigma)$ is given in Eq.~(\ref{Flpt}). Here we write this expansion, together with the usual expansion in powers of $a$:
\bea
\cF(\sigma) &=&  \sum_{n=0}^{\infty} f_n(\nu_0;\kappa) a(\kappa \sigma)^{\nu_0+n}
\; \left[ =\sum_{n=0}^{\infty} \tf_n(\nu_0;\kappa) \ta_{\nu_0+n}(\kappa \sigma) \right].
\label{Fpt}
\eea
We recall that these two expansions are completely analogous with those of the corresponding spacelike observable $\cD(Q^2)$, Eqs.~(\ref{Dptlpt}).

It turns out that the coefficients $f_n(\nu_0;\kappa)$ are known\footnote{
\label{GandNnot} In \cite{GandN} the relevant coefficients are denoted as $c_n$, and they are related to our $f_n(\nu_0;\kappa)$ ($\kappa=1$) as: $f_n(\nu_0;1)=c_n/4^n$  ($n=0,1,\ldots$). Their quantity ${\hat C}(m_q)$ is related to $\cF(\sigma)$ as: ${\hat C}(m_q)=$ $\pi^{\nu_0} \cF(m_q^2)$.}
for $n=0,1$ \cite{GandN}:
\bes
\label{f0f1}
\bea
f_0(\nu_0;\kappa) & = & 1,
\label{f0} \\
f_1(\nu_0;1) &=& \frac{1}{4} \left(\frac{91}{6}- \frac{189}{8 \beta_0} + \frac{321}{16 \beta_0^2} \right) \; \left[
  = \frac{233}{108} (n_f=3); \; \frac{7921}{3750} (n_f=4) \right]
\label{f1} \eea \ees
The leading coefficient $f_0$ is $\kappa$-independent.
The higher order coefficients $f_2(\nu_0;1)$ and $f_3(\nu_0;1)$ can be obtained using the results of \cite{Getal}, and turn out to be
\bes
\label{f2f3}
\bea
f_2(\nu_0;1) & = &  16.89993447125 \; (n_f=3); \;  14.6762041125 \;  (n_f=4);
\label{f2} \\
f_3(\nu_0;1) & = &  193.419605571875 \;  (n_f=3); \;  150.11891031953124 \;              (n_f=4).
\label{f3} \eea \ees
The numerical results for $f_3(\nu_0;1)$ are approximate, because in order to obtain the exact value, the value of the four-loop anomalous dimension coefficient $\gamma_3$ is needed (which is not known). The value of $f_3(\nu_0;1)$ we give here is obtained by taking $\gamma_3=0$. However, there are indications that this approximation is reasonably good, as it gives in the large-$\beta_0$ approximation a value that is close to the known large-$\beta_0$ value.\footnote{The large-$\beta_0$ value is $f_3^{\rm (LB)}(\nu_0;1)\approx 56.6608 \beta_0^2$, while the term with the highest power of $\beta_0$ in the exact coefficient is $f_3(\nu_0;1)\approx 57.2618 \beta_0^2+\mathcal{O}(\beta_0)$ for $\gamma_3=0$. If we used, instead of $\gamma_3=0$, the values $\gamma_3=\gamma_3^{\rm (LB)}$ ($=-10516., -8348.$, for $n_f=3,4$), we would obtain $f_3(\nu_0;1)=190.373, 147.508$ for $n_f=3,4$, which are close to the case of $\gamma_3=0$.}

The Borel transform $B^{(\nu_0)}[\cF](u;\kappa)$, is defined in our convention as [cf.~Eq.~(\ref{BDdef}) for Borel of $\cD(Q^2)$]
\be
B^{(\nu_0)}[\cF](u;\kappa)   \equiv  \sum_{n=0}^{\infty} \frac{f_n(\nu_0;\kappa)}{n! \beta_0^n} u^n.
\label{BFdef}
\ee
The corresponding inverse Borel is then [cf.~expansion (\ref{Fpt})]
\be
\cF(\sigma) = a(\kappa \sigma)^{\nu_0-1} \frac{1}{\beta_0} \int_0^{\infty} du \exp \left( -\frac{u}{\beta_0 a(\kappa \sigma)} \right) B^{(\nu_0)}[\cF](u;\kappa).
\label{invB} \ee
The Borel transform has then, according to \cite{GandN},\footnote{
\label{SvsBF}The Borel $S(u)$ given in Eq.~(32) of \cite{GandN} corresponds, in our conventions, to our Borel via the relation: $e^{5 u/3} S(u)=4 \beta_0 (d/du)B^{(\nu_0)}[\cF](u;\kappa)$ with $\kappa=1$.}
the following leading $u=1/2$ IR renormalon terms:
\be
B^{(\nu_0)}[\cF](u;1) = \left\{
\frac{S_{+}}{\left( \frac{1}{2} - u \right)^{+\nu_0 + \beta_1/(2 \beta_0^2)}}
+ \frac{S_{0}}{\left( \frac{1}{2} - u \right)^{+\beta_1/(2 \beta_0^2)}}
+ \frac{S_{-}}{\left( \frac{1}{2} - u \right)^{-\nu_0 + \beta_1/(2 \beta_0^2)}}
\right\}
\left[1 + \cO \left( \frac{1}{2} - u \right) \right].
  \label{BFans} \ee
\textcolor{black}{We now follow the steps summarised in Eq.~(\ref{Bsteps}).}

Theorem 2 [Eqs.~(\ref{BDren})-(\ref{tBDren})] \textcolor{black}{for the timelike quantities, i.e., Eqs.~(\ref{BFtBF}),} then implies that the modified Borel of $\cF$ is
\be
\tB^{(\nu_0)}[\cF](u;1) = \left\{
\frac{\tS_{+}}{\left( \frac{1}{2} - u \right)^{\nu_0}}
+ \tS_{0} \ln \left( \frac{1}{2} - u \right)
+ \frac{\tS_{-}}{\left( \frac{1}{2} - u \right)^{-\nu_0}}
  \right\}
  \left[1 + \cO \left( \frac{1}{2} - u \right) \right].
\label{tBFans} \ee
 We note that the logarithmic term appears because it corresponds effectively to the zero power renormalon term
\be
  \ln \left( \frac{1}{2} - u \right) = \lim\limits_{\varepsilon \to +0} (- \frac{1}{\varepsilon}) \left[
    \frac{1}{\left( \frac{1}{2} - u \right)^{\varepsilon}} -1 \right].
  \label{lnlim} \ee
\textcolor{black}{Then, by the relations Eqs.~(\ref{tBFtBF1}), we obtain from the expression (\ref{tBFans})}  
\be
\tB[\cFo](u;\kappa) = \left\{
\frac{\tS_{+}^{(1)}}{\left( \frac{1}{2} - u \right)^{1}}
+ \frac{\tS_{0}^{(1)}}{\left( \frac{1}{2} - u \right)^{-\nu_0+1}}
+ \frac{\tS_{-}^{(1)}}{\left( \frac{1}{2} - u \right)^{-2 \nu_0+1}}
\right\}
\left[1 + \cO \left( \frac{1}{2} - u \right) \right].
\label{tBF1ans} \ee
When we now apply Theorem 5, Eq.~(\ref{tBD1vstBF1}), where we take into account the expansion (\ref{sinexp}) of the proportionality factor, we obtain
\be
\tB[\cDo](u;\kappa) = \left\{
\frac{\tK_{+}^{(1)}(\kappa)}{\left( \frac{1}{2} - u \right)^{1}}
+ \frac{\tK_{+0}^{(1)}(\kappa)}{\left( \frac{1}{2} - u \right)^{-\nu_0+1}}
+ \frac{\tK_{-}^{(1)}(\kappa)}{\left( \frac{1}{2} - u \right)^{-2 \nu_0+1}}
\right\}
\left[1 + \cO \left( \frac{1}{2} - u \right) \right],
\label{tBD1ans} \ee
where $\tK^{(1)}_q = \tS^{(1)}_q (\pi p)/(\sin(\pi p))$.
When we take into account the dependence Eq.~(\ref{tBD1kap}) under the variation of the renormalisation scale parameter $\kappa \equiv \mu^2/Q^2$, this implies the following $\kappa$-dependence of these parameters:
\be
\tK_q^{(1)}(\kappa) = \exp\left(\ln(\kappa) u \right) \tK_q^{(1)} \quad (q=+,0,-).
\label{tKqkap} \ee
This then means that at $\kappa=1$ the modified Borel $\tB[\cDo](u)$ in Eq.~(\ref{tBD1ans}) has three parameters, $\tK_q^{(1)} \equiv \tK_q^{(1)}(1)$ ($q=+,0,-$). However, even at $\kappa=1$, we may want to allow for the dependence of this modified Borel on the redefinition (rescaling) of the momentum, which would give us the following form of $\tB[\cDo]$ (at $\kappa=1$):
\be
\tB[\cDo](u) = \exp(\tK_e^{(1)} u) \left\{
\frac{\tK_{+}^{(1)}}{\left( \frac{1}{2} - u \right)^{1}}
+ \frac{\tK_{0}^{(1)}}{\left( \frac{1}{2} - u \right)^{-\nu_0+1}}
+ \frac{\tK_{-}^{(1)}}{\left( \frac{1}{2} - u \right)^{-2 \nu_0+1}}
\right\}.
\label{tBD1ans4} \ee
As indicated, the possible relative subleading corrections ${\cal O}(1/2-u)$ to this ansatz Eq.~(\ref{tBD1ans4}) will be neglected.
This expression has four parameters: $\tK_q^{(1)}$ ($q=e,+,0,-$).\footnote{
The exponential factor $\exp(\tK_e^{(1)} u)$ in Eq.~(\ref{tBD1ans4}) can be interpreted as $\exp(\tK_e^{(1)} u) = \exp(\tK_e^{(1)}/2)(1 - \tK_e^{(1)} (1/2 - u))$ $+\cO( (1/2-u)^2)$. This means that this factor generates next-to-leading relative corrections $\cO(1/2-u)$ to each of the three leading renormalon terms of Eq.~(\ref{tBD1ans}) in the way that rescaling of the momentum would generate.}
These four parameters can be determined by the knowledge of the original first four coefficients $f_n(\nu_0;\kappa)$ (at $\kappa=1$), Eqs.~(\ref{f0f1})-(\ref{f2f3}).

The relevant characteristic function for the resummations of $\cDo$, $\cD$ and $\cF$ is then, according to Theorem 4 [Eqs.~(\ref{tBD1rencan})-(\ref{FD1pts}]) and Eqs.~(\ref{restK})
\be
F_{\cDo}(t) = \Theta(1-t) t^{1/2} \left\{
\tK_{+}^{(1)} + \frac{\tK_{0}^{(1)}}{\Gamma(-\nu_0+1) (- \ln t)^{\nu_0}} +  \frac{\tK_{-}^{(1)}}{\Gamma(-2 \nu_0+1) (- \ln t)^{2 \nu_0}}
\right\} .
\label{FD1fin}
\ee
The resummations are performed according to the formulas (\ref{restKr}) in pQCD and (\ref{restKA}) in (holomorphic) AQCD, where $F_{\cDo}(t)_{(p,\ts)}$ is replaced by $F_{\cDo}(t)$ of Eq.~(\ref{FD1fin}),
\begingroup \color{black}
and $\tK_e \mapsto \tK_e^{(1)}$
\bes
\label{DD1res1}
\bea
\cDo(Q^2)_{\rm res.} &=& \int_0^{\infty} \frac{dt}{t}  F_{\cDo}(t) \A(t e^{-\tK_e^{(1)}} Q^2),\label{D1res1}
\\
\cD(Q^2)_{\rm res.} &=& \int_0^{\infty} \frac{dt}{t}  F_{\cDo}(t) \tA_{\nu_0}(t e^{-\tK_e^{(1)}} Q^2),\label{Dres1}
\eea \ees
\endgroup
and specifically, the sought resummation of the considered timelike quantity $\cF(\sigma)$ in AQCD is then
\be
\cF(\sigma)_{\rm res.} = \int_0^{\infty}  \frac{dt}{t}  F_{\cDo}(t) \tH_{\nu_0}(t e^{-\tK_e^{(1)}} \sigma),
\label{FrestKA2}   \ee
where the timelike analog $\tH_{\nu_0}(\sigma)$ of the generalised logarithmic derivative coupling $\tA_{\nu_0}(Q^2)$ in AQCD is given in Eqs.~(\ref{tH}).

\begingroup \color{black}
The exchange of the order of integration, over $dt$ and $d \phi$, leading to the result (\ref{FrestKA2}) [cf.~Eqs.~(\ref{Fdef}), (\ref{Dres1}) and (\ref{tHa})] is justified according to Fubini's theorem. This is so because the integrand in the double integration\footnote{
\textcolor{black}{This integrand is: $(F_{\cDo}(t)/t) \tA_{\nu_0}(t e^{- \tK_e^{(1)}} \sigma e^{i \phi}))/(2 \pi)$.}}
is continuous in any finite rectangle $-\pi \leq \phi < +\pi$ and $0 \leq t \leq t_{\rm max}$, and in general we can even make the limit $t_{\rm max} \to + \infty$. In our specific case, we have $t_{\rm max}=1$, cf.~Eq.~(\ref{FD1fin}). We also checked numerically the legitimacy of this exchange.
\endgroup

Since the expansion of $\tB[\cDo](u)$ in powers of $u$ generates the coefficients $\td_n(1;\kappa)$ (with $\kappa=1$), cf.~Eq.~(\ref{tBD1def}), the knowledge of the four parameters $\tK_q^{(1)}$ ($q=e,+,0,-$) needed to obtain the characteristic function $F_{\cDo}(t)$ is equivalent to the knowledge of the first four coefficients $\td_n(1;\kappa)$ ($n=0,1,2,3$; with $\kappa=1$).
In order to obtain these four coefficients, the question is how they are related to the aforementioned (and known) coefficients $f_n(\nu_0;\kappa)$ ($n=0,1,2,3$; with $\kappa=1$). These relations are obtained as follows.

The knowledge of the first four coefficients $f_n(\nu_0;\kappa)$ ($\kappa=1$),
Eqs.~(\ref{f0f1})-(\ref{f2f3}), gives us $\tf_n(\nu_0;\kappa)$ ($\kappa=1$) via the relations of the form Eq.~(\ref{tdd})\footnote{These relations are valid for the expansion coefficients of spacelike and timelike observables.}
\be
\tf_n(\nu_0,\kappa) = \sum_{s=0}^{n} \tk_{n-s}(\nu_0+s) f_s(\nu_0;\kappa),
\label{tff}  \ee
where we recall that the coefficients $\tk_{n-s}(\nu_0+s)$ are given in Ref.~\cite{GCAK}. The coefficients $\td_n(\nu_0;\kappa)$ (with $\kappa=1$; $0 \leq n \leq 3$) are then obtained by inverting the first four equations (\ref{tfnvstdn}). And finally, the coefficients $\td_n(1;\kappa)$ (with $\kappa=1$;  $0 \leq n \leq 3$) are then obtained by the rescaling relations (\ref{tdn1}). This then allows us to obtain the four coefficients $\tK_q^{(1)}$ appearing in the modified Borel Eq.~(\ref{tBD1ans4}) by expanding this expression up to $u^3$, where we recall that $\tB[\cDo](u)=$ $\sum_{n \geq 0} \td_n(1;1) u^n /(n! \beta_0^n)$, and by equating the coefficients at $u^n$ ($n=0,1,2,3$).

If the considered observable (to be resummed) is spacelike, $\cD(Q^2)$, then the procedure is somewhat shorter. Namely, the knowledge of the first four coefficients $d_n(\nu_0;\kappa)$ (with $\kappa=1$) gives us the first four coefficients $\td_n(\nu_0;\kappa)$ (with $\kappa=1$) by using the relations (\ref{tdd}), and the coefficients $\td_n(1;\kappa)$ (with $\kappa=1$; $0 \leq n \leq 3$) are obtained by the rescaling relations (\ref{tdn1}). This allows us to obtain the four parameters $\tK_q^{(1)}$ appearing in the modified Borel (\ref{tBD1ans4}), and thus the characteristic function $F_{\cDo}(t)$ of Eq.~(\ref{FD1fin}) that replaces $F_{\cDo}(t)_{(p,\ts)}$ in the resummations Eqs.~(\ref{DrestKr}) and Eqs.~(\ref{DrestKA}) for $\cD(Q^2)$.

For the holomorphic couplings we will apply in the following the 3$\delta$AQCD, cf.~Eqs.~(\ref{rhoA})-(\ref{tH}) and the discussion of the previous Sec.~\ref{sec:AQCD}, following the approach described here above.\footnote{For some applications of various holomorphic QCD (AQCD) models to QCD phenomenology, we refer to \cite{ShirkEPJC,Nest1,Nest2,NestBook,GCAK,ACKS,KotBSR,CAGCUps,CASM,Mirj,Nestamu,amu3dAQCD,GGKT}.}

In Table \ref{tabdn} we present the values of the first four (exactly known) expansion coefficents $f_n(\nu_0;\kappa)$, $d_n(\nu_0;\kappa)$ and $d_n(1;\kappa)$ (all for $\kappa=1$), for $n_f=3,4$.
\begin{table}
     \caption{The various coefficients appearing in our considered quantities, for $\kappa=1$ and for $n_f=3$ (and in parentheses for $n_f=4$). The coefficients ${\bar f}_n$ are in the $\MSbar$ scheme [which has: ${\bar c}_2=4.47106 (3.04764)$; ${\bar c}_3=20.9902 (15.066)$; ${\bar c}_4=56.5876 (27.3331)$]. All other coefficients are in the LMM scheme [which has: $c_2=9.29703 (6.36801)$; $c_3=71.4538 (50.8025)$; $c_4=201.843 (74.2128)$]. All the values for $n \leq 3$ are exact. The values for $n \geq 4$ come from from the modified Borel Eq.~(\ref{tBD1ans4}). The values $f_n$ and ${\bar f}_n$ were not estimated beyond $n=4$ because we do not have the expressions for the coefficients $k_{n-s}(\nu_0+s)$ [appearing in the relation (\ref{dtd}), valid also for $f_n$'s] for $n-s > 4$.} 
\label{tabdn}
\begin{center}
\scalebox{0.9}{
\noindent
  \begin{ruledtabular}
\begin{tabular}{@{} c|c|c|c|c|c}
 $n$ & ${\bar f}_n(\nu_0)$ & $f_n(\nu_0)$ & $\tf_n(\nu_0)$  & $\td_n(\nu_0)$ &  $\td_n(1)$
\\
\hline
0       &  1 (1)      & 1 (1)        &  1 (1) &   1 (1)  & 1 (1) \\
1       &  2.15741 (2.11227) &   2.15741 (2.11227) &  2.48619 (2.40388) & 2.48619 (2.40388) &   7.45857 (6.67744) \\
2       &  16.8999 (14.6762) & 15.2913 (13.4809) & 14.8747 (12.9312) &  18.5758 (16.4267) & 83.5909 (67.1027) \\
3       &  193.42 (150.119) & 171.127 (134.148) &  131.911 (104.206) & 196.322 (159.291) & 1135.86 (827.158) \\
4       &  3485.26 (2533.53) & 3216.58 (2367.68) & 2235.09 (1696.96) & 3366.32 (2567.56) & 23371.8 (15872.3) \\
5       &  - & - & 47196.3 (33189.2) & 68486.8 (48121.1) & 548647. (341144.) \\
6       & - & - & $1.15082 \times 10^6$ (746631.) & $1.73121 \times 10^6$ ($1.12989 \times 10^6$) & $1.56023 \times 10^7$ ($8.96651 \times 10^6$) \\
7       & - & - & $3.38875 \times 10^7$ ($ 2.04332 \times 10^7$) & $5.12524 \times 10^7$ ($3.09886 \times 10^7$) & $5.10527 \times 10^8$ ($2.70665 \times 10^8$) \\
8       & - & - & $1.14941 \times 10^9$ ($6.42552 \times 10^8$) & $1.74758 \times 10^9$ ($9.79388 \times 10^8$) & $1.89902 \times 10^{10}$ ($9.29815 \times 10^9$)
\end{tabular}
\end{ruledtabular}}
\end{center}
\end{table}

In Table \ref{tabtK} we present the parameters $\tK_q$ of the modified Borel $\tB[\cDo](u)$ Eq.~(\ref{tBD1ans4}) (at $\kappa=1$), in the mentioned LMM scheme, for $n_f=3,4$. We recall that these parameters then determine the characteristic function $F_{\cDo}(t)$ (\ref{FD1fin}).
\begin{table}
  \caption{The parameters $\tK_q^{(1)}$ appearing in the modified Borel transform $\tB[\cDo](u)$ Eq.~(\ref{tBD1ans4}), for $n_f=3$ and $n_f=4$ (in the LMM scheme, and for $\kappa=1$). The corresponding scheme parameter values $c_j$ ($j=2,3$) and the power index $\nu_0$ are included.}
\label{tabtK}
\begin{ruledtabular}
\begin{tabular}{r|rrrr|rr|r}
 $n_f$ & $\tK_e^{(1)}$ & $\tK_{+}^{(1)}$ & $\tK_0^{(1)}$ & $\tK_{-}^{(1)}$ & $c_2$ & $c_3$ & $\nu_0(n_f)$ \\
\hline
$n_f=3$ & -2.23644 & 8.45107 & -16.6796 & 8.39384 & 9.29703 & 71.4538 & $1/3$ \\
$n_f=4$ & -2.36840 & 7.50692 & -14.8007 & 7.45394 & 6.36801 & 50.8025 & $9/25$ \\
\end{tabular}
\end{ruledtabular}
\end{table}

We then evaluate the timelike quantity of interest $\cF(\sigma)$ with the resummation (\ref{FrestKA2}), for $n_f=3,4$. In particular, we evaluate it: (a) at $\sigma=m_c^2$ (with $n_f=3$) where $m_c=(1.67 \pm 0.07)$ GeV is the pole mass of charm quark \cite{PDG2024}; (b) at $\sigma=m_b^2$ (with $n_f=4$) where $m_b=(4.78 \pm 0.06)$ GeV is the pole mass of the bottom quark \cite{PDG2024}.\footnote{\textcolor{black}{In this context, we note that another type of resummation for spacelike and timelike observables was constructed in Refs.~\cite{SM1,SMnu,SMnuaux}. In \cite{SM1} it was constructed for observables with $\nu_0=1$: In the one-loop approximation, it turns out to give the same form as our approach in the one-loop approximation (and that of Neubert \cite{Neubert}), although it is constructed in a different way. But in the case beyond the one-loop, the resummation in \cite{SM1} has a different form which includes in the integrand, besides the mentioned one-loop-type characteristic function (i.e., the one that fulfills the sum rules (\ref{tdn1SR}) with $d_n$ instead of $\td_n$ on the left-hand side), the running coupling $a$ at scales which depend on that running coupling itself. The application of such resummations in general AQCD variants appears to be more complex. In Ref.~\cite{SMnu} (cf.~also \cite{SMnuaux}) the approach of \cite{SM1} was extended to the general observables with noninteger $\nu_0$, in the one-loop approximation.}}  

Furthermore, we also evaluate these quantities with the naive perturbation approach (truncated perturbation series (TPS)) in powers of the $\MSbar$ pQCD coupling:
\be
\cF(\sigma)^{{\rm TPS[N]};\MSbar} = {\bar a}(\sigma)^{\nu_0} + \sum_{j=1}^{N-1} {\bar f}_j {\bar a}(\sigma)^{\nu_0+j},
\label{TPSMSb} \ee
where the bars indicate that these are the quantities in the $\MSbar$ scheme, either with $n_f=3$ or $n_f=4$.

\begin{figure}[htb] 
\centering\includegraphics[width=80mm]{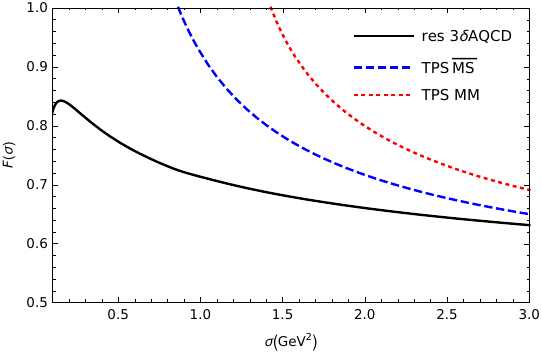}
 \vspace{4pt}
 \caption{\footnotesize The renormalon-resummed $\cF(\sigma)$, as a function of the squared timelike momentum (squared mass) $\sigma$, in 3$\delta$AQCD, for $n_f=3$, $M_1=0.150$ GeV and $\alpha_s^{\MSbar}(M_Z^2)=0.1180$. For comparison, we include also the corresponding pQCD TPS Eq.~(\ref{TPSMSb}) in the $\MSbar$ and (L)MM schemes, with three terms included ($N=3$).}
 \label{FigFsigCentr}
\end{figure}

In Fig.~\ref{FigFsigCentr} we present the results for the resummed $\cF(\sigma)$, Eq.~(\ref{FrestKA}), for $n_f=3$, as a function of $\sigma$ in the interval $0.1 \ {\rm GeV}^2 < \sigma < 3 \ {\rm GeV}^2$.
This is for the (central) choice of the input parameters: $\alpha_s^{\MSbar}(M_Z^2)=0.1180$ and $M_1=0.150$ GeV. For comparison, the simple ($n_f=3$) pQCD TPS in the $\MSbar$ and in the LMM scheme are presented [for the same value of $\alpha_s^{\MSbar}(M_Z^2)=0.1180$] where the truncation is made at $N=3$ (three terms), because these TPS include as the last term the smallest term (further terms in the series start increasing). We can clearly see that the pQCD (TPS) approaches start failing fast when $\sigma$ decreases, while for the (3$\delta$)AQCD resummed approach this is not the case.

\begin{figure}[htb] 
\begin{minipage}[b]{.49\linewidth}
\includegraphics[width=80mm,height=50mm]{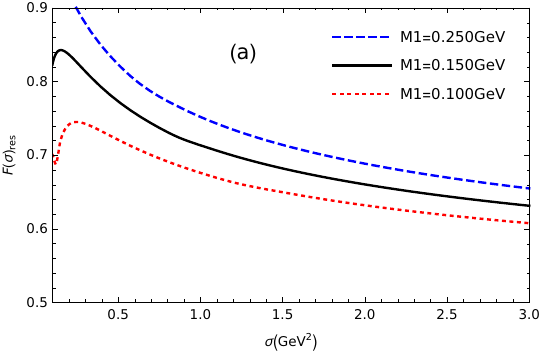}
\end{minipage}
\begin{minipage}[b]{.49\linewidth}
  \includegraphics[width=80mm,height=50mm]{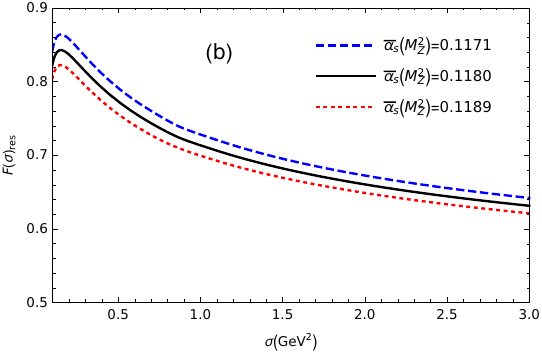}
\end{minipage} \vspace{4pt}
\caption{\footnotesize The resummed values of $\cF(\sigma)$ as in Fig.~\ref{FigFsigCentr}, when (a) the IR-threshold scale $M_1$ is varied and $\alpha_s^{\MSbar}(M_Z^2)=0.1180$; (b) when $\alpha_s^{\MSbar}(M_z^2)$ value is varied and $M_1=0.150$ GeV.}
\label{FigFsigM1Al}
\end{figure}

In Figs.~\ref{FigFsigM1Al}(a), (b), we present the results of the resummed $\cF(\sigma)$, in 3$\delta$AQCD with $n_f=3$, for various values of the threshold parameter $M_1=(0.150^{+0.100}_{-0.050})$ GeV, and for various values of the coupling strengh $\alpha_s^{\MSbar}(M_Z^2)=0.1180 \pm 0.0009$.

The results depend significantly on the variation of the (representative) nonperturbative physics parameter $M_1$ [the IR-threshold of the spectral function of the coupling $\A$, cf.~Eq.~(\ref{rhoA})], especially at low $\sigma$ values. Nonetheless, as we will see, the pQCD results (at $\sigma=m_c^2$) possess a significantly larger uncertainty originating from the IR-renormalon ($u=1/2$) ambiguity.

The numerical results for the (3$\delta$AQCD)-resummed quantities $\cF(m_c^2)_{\rm res.}$, i.e., at $\sigma=m_c^2$ and $n_f=3$, are
\bes
\label{Fmc}
\bea
\cF(m_c^2)_{\rm res.} & = & 0.6365^{-0.0058}_{+0.0060} (m_c)^{+0.0108}_{-0.0105} (\alpha_s)^{+0.0243}_{-0.0245} (M_1)
\label{Fmca} \\
& = & 0.6365 \pm 0.0273.
\label{Fmcb} \eea \ees
In Eq.~(\ref{Fmca}) we denoted separately the uncertainties stemming from the variations of the pole mass $m_c=(1.67 \pm 0.07)$ GeV, the variations of $\alpha_s^{\MSbar}(M_Z^2)=0.1180 \pm 0.009$, and the variations of the (IR-)threshold scale $M_1=(0.150^{+0.100}_{-0.050})$ GeV. In Eq.~(\ref{Fmcb}), the variations were added in quadrature.

Analogously, the numerical results for the (3$\delta$AQD)-resummed quantities $\cF(m_b^2)_{\rm res.}$, at $\sigma=m_b^2$ and $n_f=4$, are
\bes
\label{Fmb}
\bea
\cF(m_b^2)_{\rm res.} & = & 0.4792 \mp 0.0010 (m_b)^{+0.0055}_{-0.0053} (\alpha_s)^{+0.0061}_{-0.0065} (M_1)
\label{Fmba} \\
& = & 0.4792^{+0.0084}_{-0.0083}.
\label{Fmbb} \eea \ees
The uncertainties at '($m_b$)' come from the uncertainty of the $b$ quark pole mass $m_b=(4.78 \pm 0.06)$ GeV.
We can see that the IR-uncertainties '$(M_1)$' are relatively large in the case of $\cF(m_c^2)_{\rm res.}$ (with $n_f=3$), but less so for  $\cF(m_b^2)_{\rm res.}$ (with $n_f=4$).

On the other hand, the naive pQCD TPS approach Eq.~(\ref{TPSMSb}), in the $\MSbar$ scheme, gives for $n_f=3$, at $\sigma=m_c^2$
\bes
\label{FmcT}
\bea
\cF(m_c^2)^{\rm TPS[3]} & = & 0.6604^{-0.0116}_{+0.0130} (m_c)^{+0.0120}_{-0.0115} (\alpha_s) \pm 0.0854 ({\rm TPS})
\label{FmcTa} \\
& = & 0.6604^{+0.0872}_{-0.0869},
\label{FmcTb} \eea \ees
and for $n_f=4$, at $\sigma=m_b^2$
\bes
\label{FmbT}
\bea
\cF(m_b^2)^{\rm TPS[4]} & = & 0.4807 \mp 0.0014 (m_b) \pm 0.0049 (\alpha_s) \pm 0.0184 ({\rm TPS})
\label{FmbTa} \\
& = & 0.4807 \pm 0.0191.
\label{FmbTb} \eea \ees
In both of these cases, the series was truncated at the smallest term (and including that term). This means that in the case of $\sigma=m_c^2$ the series was truncated at $N=3$ (the last term is $\sim {\bar a}^{\nu_0+2}$), and in the case of $\sigma=m_b^2$ it was truncated at $N=4$. The IR-uncertainties in these cases, '(TPS)', were taken to be these last (smallest) terms, which also indirectly reflect the IR-renormalon pQCD uncertainties ($u=1/2$) of these quantities. When we compare the '(TPS)' uncertainties in this pQCD approach for $\cF(m_c^2)$, Eq.~(\ref{FmcTa}), with those of '($ M_1$)' IR-uncertainties in Eq.~(\ref{Fmca}), we see that the latter are much smaller, and this despite choosing relatively large variations of the threshold scale $M_1$ around the pion mass value.

In principle, we could have evaluated  $\cF(\sigma)$ in the renormalon-motivated resummation with the (miniMOM) pQCD coupling, Eq.~(\ref{FrestKr}). However, it turns out that this evaluation gives us significantly lower results; the central results are in this case: $\cF(m_c^2)_{\rm res.,pQCD}=0.4645$ and $\cF(m_b^2)_{\rm res.,pQCD}=0.4264$. The reason for this discrepancy with the above results lies probably in the significantly large Landau cut of the (miniMOM) pQCD coupling $a(Q^2)$, namely this cut is $(0 \leq)$$Q^2 < 1.27 \ {\rm GeV}^2$ for $n_f=3$ and $(0 \leq)$$Q^2 < 0.83 \ {\rm GeV}^2$ for $n_f=4$. This then results in the same large Landau cuts for the coupling $\ta_{\nu_0}(Q^2)$, where\footnote{The results for $\cF(m_q^2)_{\rm res.,pQCD}$ are up to four digits independent of whether or not we include the term $k_4(\nu_0) a(Q^2)^{\nu_0+4}$ in the sum Eq.~(\ref{tanuanu}) that gives us $\ta_{\nu_0}(Q^2)$.} we recall that the timelike pQCD coupling $\ths_{\nu_0}(\sigma)$ appearing in the resummation integral (\ref{FrestKr}) is a contour integral of $\ta_{\nu_0}(\sigma e^{i \phi})$ according to Eq.~(\ref{thnudef}), and the value of this contour integral is for low $\sigma$ values certainly affected by the mentioned large Landau cuts.

We point out that the quantities ${\hat C}(m_q)$ in the work of \cite{GandN}, the scheme invariant factor of the Wilson coefficient of the chromomagnetic factor of the heavy-quark effective Lagrangian, are connected with our (``canonical'') function $\cF(\sigma)$ in the following way:
\be
 {\hat C}(m_q) = \pi^{\nu_0} \cF(m_q^2)\,
 \label{hatCF} \ee
 where we choose to take for the number of (effectively) massless quarks $n_f=3$ when $m_q=m_c$, and $n_f=4$ when $m_q=m_b$. We note that $\nu_0$ depends on $n_f$, i.e., $\nu_0=\nu_0(n_f)$ In particular, the ratios of these two functions at $m_q=m_c$ and $m_q=m_b$ is then
 \be
 \frac{{\hat C}(m_b)}{{\hat C}(m_c)} = \pi^{\nu_0(4)-\nu_0(3)} \frac{ \cF(m_b^2) }{\cF(m_c^2)}.
  \label{rat} \ee
Here, $\nu_0(4)-\nu_0(3)= 2/75$.
The above results Eqs.~(\ref{Fmc})-(\ref{FmbT}) imply for the ratios the following values:
\bes
\label{rathatC}
\bea
\left( \frac{{\hat C}(m_b)}{{\hat C}(m_c)} \right)_{\rm res.} & = & 0.7763^{+0.0071}_{-0.0073} (m_c) \mp 0.0017 (m_b)^{-0.0196}_{+0.0204} (M_1 \text{\&} \alpha_s),
\label{rathatCa} \\
\left( \frac{{\hat C}(m_b)}{{\hat C}(m_c)} \right)_{\rm TPS} & = & 0.7504^{+0.0134}_{-0.0145} (m_c)^{-0.0021}_{+0.0022} (m_b)^{-0.0608}_{+0.0787}({\rm TPS} \text{\&} \alpha_s).
\label{rathatCb} \eea \ees
In these ratios, we considered that the uncertainties coming from the IR regime ['$(M_1)$' or (TPS)] in the numerator and the denominator are  completely correlated, as are the '$(\alpha_s)$' uncertainties. These two types of uncertainties were then added in quadrature.

It turns out that this ratio is the leading order approximation for the following ratio of mass splitting between the ground-state pseudoscalar and vector mesons, in the bottom and charm quark systems \cite{GandN} \textcolor{black}{(cf.~also \cite{ABN})}
\be
\frac{M_{B^*}^2-M_B^2}{M_{D^*}^2-M_D^2} = \frac{{\hat C}(m_b)}{{\hat C}(m_c)}
  \left[ 1 + \Lambda_{\rm eff} \left( \frac{1}{m_c}-\frac{1}{m_b} \right) + \ldots
    \right],
\label{ratrel} \ee
and $\Lambda_{\rm eff}$ in the subleading terms is a combination of the hadronic parameters. This ratio of mass splitting is $0.8776$, using the data \cite{PDG2024}. We now use in this relation the results (\ref{rathatC}) and we extract the value of this hadronic parameter
\bes
\label{Leff}
\bea
\Lambda_{\rm eff} & = & \left( 0.335^{-0.006}_{+0.005} (m_c) \mp 0.004 (m_b)^{+0.075}_{-0.074}(M_1 \text{\&} \alpha_s) \right) \ {\rm GeV}
\nonumber\\
&=& \left( 0.335 \pm 0.075 \right) \ {\rm GeV}  \qquad {\rm (res)},
\label{Leffres} \\
\Lambda_{\rm eff} & = & \left( 0.435^{-0.027}_{+0.028} (m_c) \pm 0.006 (m_b)^{+0.265}_{-0.285} ({\rm TPS} \text{\&} \alpha_s) \right) \ {\rm GeV}
\nonumber\\
&=&
\left( 0.435^{+0.266}_{-0.286} \right) \ {\rm GeV} \qquad {\rm (TPS)}.
\label{LeffTPS} \eea \ees

\textcolor{black}{The quantities ${\hat C}(m_q)$ can formally be considered as observables because they are renormalisation scale (RScl) and scheme (RSch) invariant. The scale $\Lambda_{\rm eff}$, appearing in the relation (\ref{ratrel}), is thus also RScl and RSch invariant, thus formally an observable.}

We can see that the central values in Eqs.~(\ref{Leff}) differ somewhat, but the uncertainties are much larger in the pQCD TPS approach -- they are dominated by the large renormalon ambiguities. In the ($3\delta$AQCD-)resummed case, the uncertainties are much smaller, and are dominated by the variation of the IR-parameter [threshold scale $M_1$ of the spectral function, cf.~Eq.~(\ref{rhoA})], $M_1=(0.150^{+0.100}_{-0.050})$ GeV, which represents a relatively large variation around the pion mass value.
\begingroup \color{black}
In particular, if we use the upper value of the IR threshold scale, $M_1=0.250$ GeV, we obtain in our resummed approach the value of $\Lambda_{\rm eff}=0.408$ GeV, which is closer to the central value of the TPS approach Eq.~(\ref{LeffTPS}).

In the work \cite{GandN}, the quantities ${\hat C}(m_b)$ and ${\hat C}(m_c)$ were calculated as well, as TPS in $\MSbar$, with two terms in ${\hat C}(m_c)$ and three terms in ${\hat C}(m_b)$ (due to strong divergence). They used $\alpha_s^{\MSbar}(m_c^2)=0.36$ and $\alpha_s^{\MSbar}(m_b^2)=0.22$, and obtained for the ratio the value ${\hat C}(m_b)/{\hat C}(m_c) \approx 0.80$ and consequently $\Lambda_{\rm eff} \approx 0.220$ GeV. Uncertainties were not estimated there.
\endgroup

To improve the $\MSbar$ TPS result in the future, it is desirable to have a parametric control on the renormalon contributions in the Hyperasymptotic approximation \cite{Ayala:2019uaw,Ayala:2019hkn}.

\section{Conclusions}
\label{sec:concl}

A renormalon-motivated resummation of QCD observables, very convenient in QCD formulations where the running coupling is free from the Landau singularities (AQCD: holomorphic QCD), was first developed \cite{renmod} for the evaluation of (observable) quantities whose perturbation expansion has integer powers of the coupling.

In order to perform the evaluation, we have to know the renormalon structure of the considered quantity, and the first few coefficients of the perturbation expansion.

In this work we extend this formalism to the QCD observables, either spacelike or timelike, whose perturbation expansion has in general noninteger powers of the coupling.

As an example of specific application, we evaluated with this formalism the scheme-invariant factor ${\hat C}(m_q)$ of the Wilson coefficient of the chromomagnetic operator in the heavy-quark effective Lagrangian, and related quantities. We used in our approach a running coupling $\A(Q^2)$ free from the Landau singularities. The IR-behaviour of the spectral function $\rho_{\A}(\sigma)$ of the (holomorphic) coupling $\A(Q^2)$ is parametrised with a sum of three Dirac delta functions, and we impose on $\A(Q^2)$ at low $|Q^2| \lesssim 1 \ {\rm GeV}^2$ the behaviour as suggested by large-volume lattice calculations. The IR-uncertainty was parametrised by varying significantly the (IR-)threshold scale $M_1^2$ of $\rho_{\A}(\sigma)$ around the value of the squared pion mass $m^2_{\pi}$. The obtained values of the factor functions ${\hat C}(m_c)$ and ${\hat C}(m_b)$ are somewhat different from those of the naive evaluation with the truncated perturbation series (TPS); however, the IR-uncertainties of the obtained results turn out to be much smaller in our approach than in the TPS approach, especially for the low-energy quantity ${\hat C}(m_c)$.

We hope that this formalism can be useful in theoretical evaluations of several other low-energy QCD observables. 

\vspace{0.8cm}

\noindent
\centerline {\bf \small Data availability statement:}

\noindent
All the necessary data that support the findings of this study are included within the article. If desired, some specific numerical programs (in Mathematica) can be obtained from one of the authors (G.C.).

\acknowledgments
This work was supported in part by FONDECYT (Chile) Grants No.~1220095 (G.C.) and 1240329 (C.A.).

\appendix

\section{Proof of Theorem 2}
\label{app:T2}

For $d(\nu_0;\kappa)$ we have, using (\ref{BDren}), the expansion (\ref{binexp}) and the asymptotic formula (\ref{Gammaasym})
\bes
\label{dnT2}
\bea
d_n(\nu_0;\kappa) &=& \frac{\cK(\kappa) p ^{-s_0}}{\Gamma(s_0)} \Gamma(s_0+n)
\left( \frac{\beta_0}{p} \right)^n \left(1 + \cO(1/n) \right)
\label{dnT2a} \\
& = & \frac{\cK(\kappa) p ^{-s_0}}{\Gamma(s_0)}  n! n^{s_0-1} 
\left( \frac{\beta_0}{p} \right)^n \left(1 + \cO(1/n) \right).
\label{dnT2b} \eea \ees
This is a situation similar to the one considered in Appendix B of Ref.~\cite{GCAK}. There, $s_0=\nu_0$ (while here $s_0 \not= \nu_0$ in general), $\beta_0/p \mapsto b$ and $d_n(\nu_0;\kappa) \mapsto \cF_n$. Further, there we have no $\cO(1/n)$ corrections; nonetheless, since the $\cO(1/n)$-terms will be kept here undetermined throughout, we can follow the proof in Appendix B of \cite{GCAK}, with the only modifications arising now from the fact that $s_0 \not= \nu_0$.

The idea of the proof of Theorem 2 is to find the ratio between the coefficients $\td_n(\nu_0;\kappa)$ and $d_n(\nu_0,\kappa)$ for large $n$. For this, the relations (\ref{tdd}) at large $n$ will be investigated here.\footnote{
The notational conventions for the $\beta$-function coefficients $\beta_j$ and $c_j$ are given in Eqs.~(\ref{RGE}) where, as always, $a = \alpha_s/\pi$.}

On the right-hand side (RHS) of Eq.~(\ref{tdd}) we have, {\bf for $s=n$},
$\tk_0(\nu_0+s)=1$, and thus
\be
\frac{\tk_0(\nu_0+n) d_n(\nu_0;\kappa)}{d_n(\nu_0;\kappa)} = 1.
\label{ratsn} \ee
From now on, for convenience, we will use the notations
\be
b \equiv \frac{\beta_0}{p}; \quad K \equiv \cK(\kappa) p^{-s_0}.
\label{notappT2} \ee
With this notation, the behaviour of $d_n(\nu_0;\kappa)$ in the considered case, Eq.~(\ref{dnT2a}), can be written as
\be
d_n(\nu_0;\kappa) = K \frac{\Gamma(s_0+n)}{\Gamma(s_0)} b^n \left( 1 + \cO(1/n) \right).
\label{dnT2ref} \ee

Further, we will repeatedly use Eqs.~(A.17)-(A.21) of Ref.~\cite{GCAK} for the coefficients $\tk_m(\nu)$, and Eq.~(A.13) for the function $\tZ_m(\nu)$ appearing in those expressions
\be
\tZ_m(\nu) = \Gamma(\nu+1) \left( \frac{d}{dx} \right)^m
\left( \frac{\Gamma(1-x)}{\Gamma(\nu+1-x)}  \right){\Bigg |}_{x=0}.
\label{tZm} \ee
Then, on the RHS of Eq.~(\ref{tdd}) we have {\bf for $s=n-1$} 
\bes
\label{snm1}
\bea
\tk_1(\nu_0+n-1) d_{n-1}(\nu_0;\kappa) & = &
- c_1 \nu \left[ \tZ_1(\nu)-1 \right]
\frac{\Gamma(s_0+n-1)}{\Gamma(s_0)} b^{n-1}
\left( 1 + \cO(1/n) \right) K {\Big |}_{\nu=\nu_0+n-1} \; \Rightarrow
\label{dnsnm1} \\
\frac{ \tk_1(\nu_0+n-1) d_{n-1}(\nu_0;\kappa) }{d_n(\nu_0;\kappa)} & = &
- c_1 \frac{1}{b} \left( \frac{d}{dx} -1 \right)
\left( \frac{\Gamma(1-x) \Gamma(\nu_0+n)}{\Gamma(\nu_0+n-x)}  \right){\Bigg |}_{x=0}
\times \left(1 + \cO(1/n) \right).
\label{ratsnm1} \eea \ees
The second identity (\ref{ratsnm1}) is obtained by the use of the identity (\ref{tZm}) for $\tZ_1$ and the known property of the Gamma function, $\Gamma(z+1)=z \Gamma(z)$.

Similarly, on the RHS of Eq.~(\ref{tdd}) we have {\bf for $s=n-2$} 
\bes
\label{ratsnm2}
\bea
\frac{\tk_2(\nu_0+n-2) d_{n-2}(\nu_0;\kappa)}{d_n(\nu_0;\kappa)} & = &
\nu (\nu+1) \left[-c_2 \frac{(\nu-1)}{2 (\nu+1)} + \frac{1}{2} c_1^2
  \left( \tZ_2(\nu) - 2 \tZ_1(\nu+1) +1 \right) \right] 
\nonumber\\
&& \times \frac{\Gamma(s_0+n-2)}{\Gamma(s_0 + n)} \frac{1}{b^2} \left(1 + \cO(1/n) \right)
\label{ratsnm2a} \\
& = &
\frac{1}{2 b^2} \left[ c_1^2 \left( \frac{d}{dx} - 1 \right)^2 - c_2 \right]
\left( \frac{\Gamma(1-x) \Gamma(\nu_0+n)}{\Gamma(\nu_0+n-x)}  \right){\Bigg |}_{x=0}
\times \left(1 + \cO(1/n) \right).
\label{ratsnm2b} \eea \ees
In arriving to Eq.~(\ref{ratsnm2b}), we in addition used the relation
\be
\tZ_2(\nu_0+n-2) = \tZ_2(\nu_0+n-1) \left( 1 + \cO(1/n) \right).
\label{tZ2asym} \ee

Analogously, on the RHS of Eq.~(\ref{tdd}) we obtain {\bf for $s=n-3$} 
\bea
\frac{\tk_3(\nu_0+n-3) d_{n-3}(\nu_0;\kappa)}{d_n(\nu_0;\kappa)} & = &
    \frac{1}{6 b^3} \left[ -c_1^3 \left( \frac{d}{dx} - 1 \right)^3 + 3 c_1 c_2 \left( \frac{d}{dx} - \frac{1}{6} \right) - \frac{1}{2} c_3 \right]
    \left( \frac{\Gamma(1-x) \Gamma(\nu_0+n)}{\Gamma(\nu_0+n-x)}  \right) {\Bigg |}_{x=0}
    \nonumber\\
&& \times \left(1 + \cO(1/n) \right).
\label{ratsnm3} \eea 
and  {\bf for $s=n-4$}
\bea
\frac{\tk_4(\nu_0+n-4) d_{n-4}(\nu_0;\kappa)}{d_n(\nu_0;\kappa)} & = &
\frac{1}{24 b^4} {\Bigg \{}  c_1^4 \left( \frac{d}{dx} - 1 \right)^4 - 6 c_1^2 c_2 \left[ \left( \frac{d}{dx} - \frac{1}{6} \right)^2 + \frac{31}{36} \right] +
2 c_1 c_3 \left( \frac{d}{dx} + \frac{1}{6} \right)
\nonumber\\
&& + \frac{(13 c_2^2-c_4)}{3} {\Bigg \}}
\left( \frac{\Gamma(1-x) \Gamma(\nu_0+n)}{\Gamma(\nu_0+n-x)} \right) {\Bigg |}_{x=0}
\times \left(1 + \cO(1/n) \right).
\label{ratsnm4} \eea

\vspace{0.5cm}

\noindent
{\bf Appendix A.1}

\vspace{0.3cm}

\noindent
The rest of this proof consists of two parts.
The first part of the rest of this proof considers the effects of the terms $c_1$ ($=\beta_1/\beta_0$), regarding $c_n=0$ ($n=2,3,\ldots$).

In this part we now follow entirely Appendix B.1 of Ref.~\cite{GCAK}. This is so because the asymptotic results of the previous part of the present Appendix, Eqs.~(\ref{ratsnm1}), \ldots, (\ref{ratsnm4}), are the same as in App.~B of \cite{GCAK} (although we have here $s_0 \not= \nu_0$ in general).

When taking the terms in powers of $c_1$ appearing in Eqs.~(\ref{ratsnm1}), \ldots, (\ref{ratsnm4}), we obtain
\bea
\td_n(\nu_0;\kappa)^{(c_1)} & = & d_n(\nu_0;\kappa) \sum_{m=0}^n \frac{(-1)^m c_1^m}{b^m m!} \left( \frac{d}{dx} - v \right)^m
\left( \frac{\Gamma(1-x) \Gamma(\nu_0+n)}{\Gamma(\nu_0+n-x)}  \right) {\Bigg |}_{x=0}
\times \left(1 + \cO(1/n) \right),
\label{tdnc1sum} \eea
where $v=1$
and the superscript '$(c_1)$' denotes that only those terms are taken into account that are nonzero when $c_j=0$ for $j \geq 2$.
The last term in Eq.~(\ref{tdnc1sum}) we represent by using the integral form of the mathematical beta-function $B(u,w)$
\bea
\frac{\Gamma(1-x) \Gamma(\nu_0+n)}{\Gamma(\nu_0+n-x)}  & = & (\nu_0+n-1) B(1-x,\nu_0+n-1) =  (\nu_0+n-1) \int_{0}^1 dy y^{-x} (1-y)^{\nu_0+n-2}.
\label{beta} \eea
Application of $( d/dx - v)^m$ to the identity (\ref{beta}) means that this operator is applied to the factor $y^{-x} = \exp(-x \ln y)$, and this leads, after some algebra, to the identity
\bea
\left( \frac{d}{dx} - v \right)^m
\left( \frac{\Gamma(1-x) \Gamma(\nu_0+n)}{\Gamma(\nu_0+n-x)}  \right){\Bigg |}_{x=0}
& = & (\nu_0+n-1) \int_{0}^1 dy (- \ln y - v)^m (1-y)^{\nu_0+n-2}.
\label{derbeta} \eea
When we use this identity in Eq.~(\ref{tdnc1sum}), we obtain
\bes
\label{tdnc1as}
\bea
\td_n(\nu_0;\kappa)^{(c_1)} & = & d_n(\nu_0;\kappa) (\nu_0+n-1) \int_{0}^1 dy (1-y)^{\nu_0+n-2}
\sum_{m=0}^n \frac{(\ln y + v)^m c_1^m}{b^m m!} {\Big |}_{v=1}
\left(1 + \cO(1/n) \right)
\label{tdnc1asa} \\
& \approx & d_n(\nu_0;\kappa) (\nu_0+n-1) \exp(v c_1/b)|_{v=1} 
\int_{0}^1 dy y^{c_1/b} (1-y)^{\nu_0+n-2} \left(1 + \cO(1/n) \right)
\label{tdnc1asb} \\
&=&  d_n(\nu_0;\kappa) \exp(c_1/b)  \frac{\Gamma \left(1 + \frac{c_1}{b} \right) \Gamma( \nu_0+n) }{\Gamma \left( \nu_0 + n + \frac{c_1}{b} \right)}
\left(1 + \cO(1/n) \right).
\label{tdnc1asc} \eea \ees
In the step from Eq.~(\ref{tdnc1asa}) to (\ref{tdnc1asb}) we replaced the sum $\sum_{m=0}^n$ by $\sum_{m=0}^{\infty}$, and in Eq.~(\ref{tdnc1asc}) we explicitly used $v=1$. It can be checked numerically that this approximation results in relative errors which diminish with increasing $n$ significantly faster than $\cO(1/n)$.

If we use on the RHS of Eq.~(\ref{tdnc1asc}) the asymptotic formula (\ref{Gammaasym}) for the Gamma functions, we immediately obtain
\bea
\td_n(\nu_0;\kappa)^{(c_1)} & = & d_n(\nu_0;\kappa) e^{c_1/b} \Gamma\left(1 + \frac{c_1}{b} \right) n^{-c_1/b} \left(1 + \cO(1/n) \right)
\nonumber \\
& = &  d_n(\nu_0;\kappa) e^{p \beta_1/\beta_0^2} \Gamma\left(1 + \frac{p \beta_1}{\beta_0^2} \right) n^{-p \beta_1/\beta_0^2} \left(1 + \cO(1/n) \right),
\label{tdnc1asd} \eea
where we used in the last identity the explicit expression for $b$ Eq.~(\ref{notappT2}) and $c_1=\beta_1/\beta_0$. The obtained asymptotic relation (\ref{tdnc1asd}), when combined with the asymptotic expression (\ref{dnT2b}), then gives
\bea
\td_n(\nu_0;\kappa)^{(c_1)} & = & 
\frac{\cK(\kappa) p ^{-s_0} e^{p \beta_1/\beta_0^2} \Gamma(1 + p \beta_1/\beta_0^2)}{\Gamma(s_0)}  n! n^{(s_0-p \beta_1/\beta_0^2) -1} 
\left( \frac{\beta_0}{p} \right)^n \left(1 + \cO(1/n) \right).
\label{tdnc1asfin} \eea
As shown above around Eq.~(\ref{dnT2b}), the asymptotic expression (\ref{dnT2b}) for $d_n(\nu_0;\kappa)$ is equivalent to the form Eq.~(\ref{BDren}) of the Borel $B[\cD](u;\kappa)$. Comparing the obtained asymptotic expression (\ref{tdnc1asfin}) for $\td_n(\nu_0;\kappa)^{(c_1)}$ with that of $d_n(\nu_0;\kappa)$ Eq.~(\ref{dnT2b}), we see that the modified Borel $\tB^{(\nu_0)}[\cD](u;\kappa)$ has the same structure, with the difference that the index $s_0$ is replaced by $\ts_0=s_0-p \beta_1/\beta_0^2$.

  This proves Theorem 2 for the case when the effects of the higher beta-coefficients are neglected (i.e., $c_2=c_3=\cdots=0$).\footnote{The radiative corrections $\cO(1/n)$ in the result (\ref{tdnc1asfin}) clearly correspond to $s_0 \mapsto (s_0-1)$, i.e., in the modified Borel $\tB^{(\nu_0)}[\cD](u;\kappa)$ they correspond to the relative corrections $\cO(p-u)$, cf.~Eq.~(\ref{tBDren}). These corrections are in general nonzero even if the relative corrections $\cO(p-u)$ in the Borel $B^{(\nu_0)}[\cD](u;\kappa)$ are taken to be zero.}

\vspace{0.5cm}

\noindent
{\bf Appendix A.2}

\vspace{0.3cm}

\noindent

Now we extend the proof of Theorem 2 to the case when $c_2, c_3, \ldots \not=0$. For this, we follow the steps formulated in Appendix B.2 of Ref.~\cite{GCAK}. Some of the steps involve ``educated guess'' extrapolations (especially for the cases $c_3 \not= 0$ and $c_4 \not=0$).

First the terms that are nonzero when $c_2 \not=0$, in Eqs.~(\ref{ratsn})-(\ref{ratsnm4})
\bea
\frac{\td_n(\nu_0;\kappa)^{(c_2)}}{d_n(\nu_0;\kappa)} & = &
\left\{ - \frac{c_2}{2 b^2} + \frac{c_1 c_2}{2 b^3} \left( \frac{d}{dx} - \frac{1}{6} \right) - \frac{c_1^2 c_2}{4 b^4} \left( \frac{d}{dx} - \frac{1}{6} \right)^2 + \ldots \right\}
\left( \frac{\Gamma(1-x) \Gamma(\nu_0+n)}{\Gamma(\nu_0+n-x)}  \right){\Bigg |}_{x=0}.
\nonumber\\ &&
\times \left(1 + \cO(1/n) \right).
\label{ratc2a} \eea 
Here we excluded the curly brackets the term $-(31/36) (c_1^2 c_2)/(4 b^4)$, we will combine it with the term $\sim c_4/b^4$ later. The pattern that we see in Eq.~(\ref{ratc2a}) gives
\bea
\frac{\td_n(\nu_0;\kappa)^{(c_2)}}{d_n(\nu_0;\kappa)} & = &
- \frac{c_2}{2 b^2} \sum_{m=0}^{n-2} \frac{(-c_1)^m}{m! b^m} \left( \frac{d}{dx} - \frac{1}{6} \right)^m \left( \frac{\Gamma(1-x) \Gamma(\nu_0+n)}{\Gamma(\nu_0+n-x)}  \right){\Bigg |}_{x=0} \left(1 + \cO(1/n) \right).
\label{ratc2b} \eea 
This is as Eq.~(\ref{tdnc1sum}), except that now $v=1$$\mapsto v=1/6$. We repeat the steps Eqs.~(\ref{beta})-(\ref{tdnc1asd}), and obtain
\bea
\frac{\td_n(\nu_0;\kappa)^{(c_2)}}{d_n(\nu_0;\kappa)} & = &
-\frac{c_2}{2 b^2} e^{(1/6) c_1/b}
\frac{\Gamma(1+c_1/b) \Gamma(\nu_0+n)}{\Gamma(\nu_0+n+c_1/b)}
\left( 1 + \cO(1/n) \right)
\nonumber\\
& = &
-\frac{c_2}{2 b^2} e^{(1/6) c_1/b} \Gamma(1 + c_1/b)  n^{- c_1/b}
\left( 1 + \cO(1/n) \right) \; \Rightarrow
\nonumber\\
\td_n(\nu_0;\kappa)^{(c_2)} & = & 
-\frac{c_2}{2 b^2} \td_n(\nu_0;\kappa)^{(c_1)} e^{-(5/6) (c_1/b)} \left( 1 + \cO(1/n) \right) ,
\label{tdnc2} \eea 
where in the last identity we used the result (\ref{tdnc1asd}). This identity is important, because it means that the effects of $c_2 \not= 0$ only rescale the full $\td_n(\nu_0;\kappa)$ [in comparison to $\td_n(\nu_0;\kappa)^{(c_1)}$] by a factor that is independent of $n$ (in the leading order of large $n$). Below we will see that the effects of $c_3 \not=0$ and $c_4 \not= 0$ result in a similar kind of rescaling of  $\td_n(\nu_0;\kappa)$.

The terms in Eqs.~(\ref{ratsn})-(\ref{ratsnm4}) that are nonzero when $c_3 \not=0$, are
\bes
\label{ratc3}
\bea
\frac{\td_n(\nu_0;\kappa)^{(c_3)}}{d_n(\nu_0;\kappa)} & = &
\left\{ - \frac{c_3}{12 b^3} + \frac{c_1 c_3}{12 b^4} \left( \frac{d}{dx} + \frac{1}{6} \right) + \ldots \right\}
\left( \frac{\Gamma(1-x) \Gamma(\nu_0+n)}{\Gamma(\nu_0+n-x)}  \right){\Bigg |}_{x=0} \left(1 + \cO(1/n) \right)
\label{ratc3a} \\
& = & 
- \frac{c_3}{12 b^3} \sum_{m=0}^{n-3} \frac{(-c_1)^m}{m! b^m} \left( \frac{d}{dx} + \frac{1}{6} \right)^m \left( \frac{\Gamma(1-x) \Gamma(\nu_0+n)}{\Gamma(\nu_0+n-x)}  \right){\Bigg |}_{x=0} \left(1 + \cO(1/n) \right).
\label{ratc3b} \eea \ees
Following the same steps as before, we obtain
\bea
\td_n(\nu_0;\kappa)^{(c_3)} & = & 
-\frac{c_3}{12 b^3} \td_n(\nu_0;\kappa)^{(c_1)} e^{-(7/6) (c_1/b)} \left( 1 + \cO(1/n) \right) .
\label{tdnc3} \eea
When we consider the remaining terms, $\sim 1/b^4$, in Eqs.~(\ref{ratsn})-(\ref{ratsnm4}), we obtain
\bea
\frac{\td_n(\nu_0;\kappa)^{(c_4)}}{d_n(\nu_0;\kappa)} & = &
\frac{(-1)}{144 b^4} ( 2 c_4 - 26 c_2^2+31 c_1^2 c_2 ) \times 1 \times
\left(1 + \cO(1/n) \right).
\label{ratc4lead} \eea
When we add additional terms $\sim 1/b^5$, $\sim 1/b^6$, etc., we expect to obtain analogously an expression of the form
\bes
\label{ratc4}
\bea
\frac{\td_n(\nu_0;\kappa)^{(c_4)}}{d_n(\nu_0;\kappa)} & = &
\frac{(-1)}{144 b^4} (2 c_4 - 26 c_2^2+31 c_1^2 c_2) e^{-\beta^{(4)} c_1/b}
\frac{\Gamma(1 + c_1/b) \Gamma(\nu_0+n)}{\Gamma(\nu_0+n + c_1/b)} 
\left(1 + \cO(1/n) \right)
\label{ratc4a} \\
& = &\frac{(-1)}{144 b^4} (2 c_4 - 26 c_2^2+31 c_1^2 c_2) e^{-\beta^{(4)} c_1/b}
\Gamma(1 + c_1/b) n^{-c_1/b} \left(1 + \cO(1/n) \right),
\label{ratc4b} \eea \ees
where $\beta^{(4)} \sim 1$ (and probably $\beta^{(4)} >0$).
Using in Eq.~(\ref{ratc4b}) the results (\ref{tdnc1asd}), this gives us
\bea
\td_n(\nu_0;\kappa)^{(c_4)} & = & 
\td_n(\nu_0;\kappa)^{(c_1)}
\frac{(-1)}{144 b^4} (2 c_4 - 26 c_2^2+31 c_1^2 c_2) e^{-(\beta^{(4)}+1) c_1/b}
\left(1 + \cO(1/n) \right).
\label{tdnc4} \eea

Combining the relations (\ref{tdnc2}), (\ref{tdnc3}) and (\ref{tdnc4}), we obtain
\bea
\lefteqn{
  \td_n(\nu_0;\kappa)  =  \sum_{q=1}^{\infty}  \td_n(\nu_0;\kappa)^{(c_q)}
  }
  \nonumber\\
  & = & \td_n(\nu_0;\kappa)^{(c_1)}
  \left[ 1 - \frac{c_2}{2 b^2} e^{-(5/6)(c_1/b)}
    - \frac{c_3}{12 b^3} e^{-(7/6)(c_1/b)}
    - \frac{(2 c_4 - 26 c_2^2 + 31 c_1^2 c_2)}{144 b^4} e^{-(\beta^{(4)} +1)(c_1/b)}
    - \ldots \right],
    \label{tdnfin} \eea
where we recall that $b=\beta_0/p$ and $c_1=\beta_1/\beta_0$. To obtain the excact value of the number $\beta^{(4)}$ in the exponent in Eq.~(\ref{tdnfin}), we would need to extend the explicit analysis of Eqs.~(\ref{ratsn})-(\ref{ratsnm4}) to the case of $s=n-5$, i.e., the inclusion of the explicit expression for $\tk_5(\nu_0+n-5)$.

In the specific case considered in the main text of this work, we have $n_f=3$ and $p=1/2$, therefore $1/b =1/4.5 =0.2222$ is small and the sum in Eq.~(\ref{tdnfin}) is expected to converge well.\footnote{
For $p=1/2$ and $n_f=3$, the sum in the brackets on the RHS of Eq.~(\ref{tdnfin}) is: $1 - 0.0794$ $-0.0121$ $-0.0005 \times 0.7^{1+\beta^{(4)}}$.}

What the result (\ref{tdnfin}) shows is that the asymptotic behaviour of the coefficients $\td_n(\nu_0;\kappa)$ is determined by the $c_1$-terms only, i.e., by the behaviour of $\td_n(\nu_0;\kappa)^{(c_1)}$ Eq.~(\ref{tdnc1asfin})
\bea
\td_n(\nu_0;\kappa) & = &
{\cal C}(\kappa)  n! n^{(s_0-p \beta_1/\beta_0^2) -1}          %
\left( \frac{\beta_0}{p} \right)^n \left(1 + \cO(1/n) \right), %
\label{tdnasfin} \eea    
where ${\cal C}(\kappa)$ is a constant independent of $n$. We recall that the asymptotic behaviour of $d_n(\nu_0;\kappa)$, Eq.~(\ref{dnT2b}), reflects the (i.e., is equivalent to the) form Eq.~(\ref{BDren}) of $B^{(\nu_0)}[\cD](u;\kappa)$ of Theorem 2. Our result Eq.~(\ref{tdnasfin}) shows that $\td_n(\nu_0;\kappa)$ has the same asymptotic behaviour, with the only difference that $s_0 \mapsto s_0 - p \beta_1/\beta_0^2$, which means that the corresponding modified Borel $\tB^{(\nu_0)}[\cD](u;\kappa)$ of Eq.~(\ref{tBDren}) has the power index $\ts_0$ as claimed in Eq.~(\ref{s0vsts0}) of the Theorem.

This concludes the proof of Theorem 2.

\section{Proof of Theorem 4}
\label{app:T4}

When the form of $\tB[\cDo](u)$ is the renormalon term as given in Eq.~(\ref{tBD1rencan}), then the inverse Mellin transformation thereof, Eq.~(\ref{FD1}), gives us the corresponding characteristic function
\bea
F_{\cDo}(t) & = & \frac{1}{2 i} \int_{u=-i \infty}^{+i \infty}
\frac{du e^{u \ln t}}{(p-u)^{\ts}},
\label{FD1a} \eea
where we took $u_0=0$. We recall that $0<\ts < 1$. We introduce the change of the integration variable $z=i u$, and this gives
\bea
F_{\cDo}(t) & = & \frac{1}{2} \int_{z=-\infty}^{+\infty}
\frac{dz e^{-i z \ln t}}{(i z + p)^{\ts}},
\label{FD1b} \eea
The cut of the integrand, in the complex $z$-plane, is for $(i z + p) \leq 0$,
i.e., for $z=i |z|$ with $|z| \geq p$ (i.e., on the semiaxis along the positive imaginary axis).

When $t > 1$, we have $\ln t = |\ln t|$, we can close the contour of integration of the integral (\ref{FD1b}) with the (large) semicircle in the lower half plane because of the exponential suppression there; Jordan Lemma ensures that the contribution along the semicircle of radius $R$, when $R \to \infty$, is zero. No singularities are enclosed, and this then gives us by the Cauchy theorem
\be
F_{\cDo}(t) = 0 \qquad (t>1).
\label{FD1tg1} \ee

\begin{figure}[htb] 
\centering\includegraphics[width=80mm]{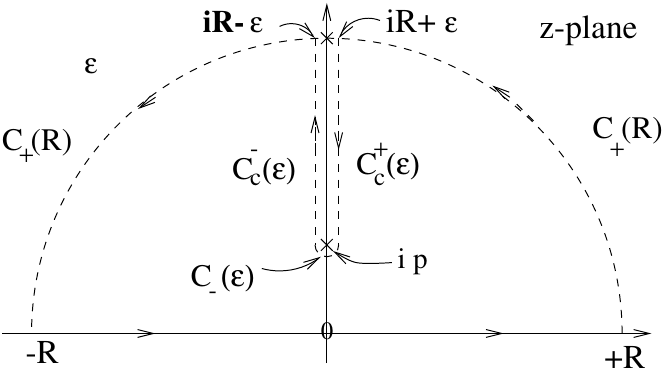}
 \vspace{4pt}
 \caption{\footnotesize The closed contour in the complex $z$-plane for the integral (\ref{FD1b}) for the case $0 < t < 1$. The limits $R \to + \infty$ and $\varepsilon \to +0$ are taken.}
\label{figcontFD1}
\end{figure}

When $0 \leq t < 1$, we have $\ln t = - |\ln t|$, hence the integrand is exponentially suppressed if we close the contour with the (large) semicircle in the upper half plane. Jordan Lemma ensures that the contribution along the semicircle of radius $R$, when $R \to \infty$, is zero.
Nonetheless, due to the aforementioned cut of the integrand along the positive imaginary axis ($z=i |z|$, $|z| \geq p$), we have to avoid this cut in order to apply the Cauchy theorem.
The necessary contour is presented in Fig.~\ref{figcontFD1}.
Since the closed contour avoids enclosing singularities, by Cauchy theorem the integral along the contour is zero
\be
\left\{ \int_{-R}^R + \int_{\cC_{+}(R)} +  \int_{\cC_c^{+}(\varepsilon)} + \int_{\cC_c^{-}(\varepsilon)} + \int_{\cC_{-}(\varepsilon)} \right\}
\frac{dz e^{+i z |\ln t|}}{2 (i z + p)^{\ts}} = 0.
  \label{intcontsum} \ee
As mentioned, the contribution along (both parts of) the upper semicircle $\cC_{+}(R)$ gives zero (when $R \to +\infty$) due to Jordan Lemma.
We now show that the contribution around the small semicicle $\cC_{-}(\varepsilon)$ or radius $\varepsilon$ (around the point $z=i p$) also gives zero (when $\varepsilon \to +0$). On that semicricle, we have $z=i p + \varepsilon \exp(i \phi)$ ($-\pi < \phi <0$) and thus the contribution there is
\bea
\frac{1}{2} \int_{\cC_{-}(\varepsilon)} \frac{dz e^{i z |\ln t|}}{(i z + p)^{\ts}}
& = & \frac{1}{2} i \varepsilon \int_{\phi=0}^{-\pi}
\frac{d \phi e^{i \phi} \exp \left[ -p |\ln t| + \cO(\varepsilon) \right]}{(i \varepsilon e^{i \phi})^{\ts}}
  \nonumber \\
  & = & \varepsilon^{1 - \ts}  \frac{(-1)}{2 (1 - \ts)} e^{-i \ts \pi/2} e^{-p |\ln t|}
 \frac{1}{\ts} \left( 1 + e^{i \ts \pi} \right) \left( 1 + \cO(\varepsilon) \right) \sim  \varepsilon^{1 - \ts}.
\label{intCeps} \eea
This goes clearly to zero when $\varepsilon \to +0$, because $0 < \ts < 1$.

\begin{figure}[htb] 
\centering\includegraphics[width=80mm]{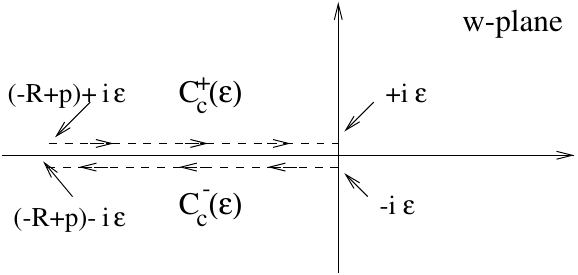}
 \vspace{4pt}
 \caption{\footnotesize The paths $\cC_{c}^{\pm}(\varepsilon)$ in the complex $w$-plane, where $w=p + iz$. The limit $\varepsilon \to =)$ is taken.}
 \label{figcutw}
\end{figure}

The contributions of the integral along both sides of the cut, $\cC_{c}^{\pm}(\varepsilon)$ (cf.~Fig.~\ref{figcontFD1}) can now be evaluated by introducing a new integration variable $w = i z + p$ and its absolute value $|w|$. The cut is now in the complex $w$-plane along the negative semiaxis, cf.~Fig.~\ref{figcutw}.
We have $dz = - i dw$, and since on the two paths $\cC_{c}^{\pm}(\varepsilon)$ we have $w=-|w| \pm i \varepsilon$ along this cut, therefore $dz = +i d|w|$.
Along  $\cC_{c}^{+}(\varepsilon)$ we have: $w=(-R+p)+i \varepsilon \to + i \varepsilon$, i.e., $w = |w| e^{+i (\pi-\varepsilon')}$. Along $\cC_{c}^{-}(\varepsilon)$ we have: $w=-i \varepsilon \to (-R+p)-i \varepsilon$, i.e., $w = |w| e^{-i (\pi-\varepsilon')}$. Here, $\varepsilon \to +0$ and $\varepsilon' \to +0$. The contributions are then
\bes
\label{intc}
\bea
\lefteqn{\left\{ \int_{\cC_c^{+}(\varepsilon)} + \int_{\cC_c^{-}(\varepsilon)} \right\}
\frac{dz e^{i z |\ln t|}}{2 (i z + p)^{\ts}}
}
\nonumber\\
&=& \frac{i}{2} {\Big \{}
\int_{|w|=R-p}^0 d|w| \frac{\exp \left( |w| e^{+i \pi} |\ln t| \right) e^{-p |\ln t|}}{\left( |w| e^{+i \pi} \right)^{\ts}} +
\int_{|w|=0}^{R-p} d|w| \frac{\exp \left( |w| e^{-i \pi} |\ln t| \right) e^{-p |\ln t|}}{\left( |w| e^{-i \pi} \right)^{\ts}} {\Big \}}
\label{intc1} \\
& = & \frac{i}{2} e^{-p |\ln t|} \int_{|w|=0}^{+\infty} d |w|
\frac{e^{-  |w| |\ln t|}}{|w|^{\ts}} \left( e^{i \ts \pi} - e^{-i \ts \pi} \right)
= (-1)  e^{-p |\ln t|} \sin(\ts \pi) \int_{|w|=0}^{+\infty} d|w| |w|^{-\ts} e^{- i |\ln t| |w|} 
\label{intc2} \\
& = & \frac{(-1)  e^{-p |\ln t|}}{|\ln t|^{1- \ts}} \sin(\ts \pi) \Gamma(1-\ts)
  = (-1) \frac{\pi  e^{-p |\ln t|}}{\Gamma(\ts) |\ln t|^{1- \ts}}.
\label{intc3} \eea \ees
In Eq.~(\ref{intc2}) we took into account $R \to + \infty$.
We now take into account this result in the Cauchy identity (\ref{intcontsum}), and the fact that the contributions along the large semicircle $\cC_{+}(R)$ and along the small semicircle $\cC_{-}(\varepsilon)$ are zero [Eq.~(\ref{intCeps})]. This gives us the final result for $0<t<1$
\bea
F_{\cDo}(t)_{(p,\ts)} &=&
\frac{1}{2} \int_{-\infty}^{+\infty} \frac{dz e^{+i z |\ln t|}}{(i z + p)^{\ts}}
  = (-1) \left\{ \int_{\cC_c^{+}(\varepsilon)} + \int_{\cC_c^{-}(\varepsilon)} \right\}
  \frac{dz e^{+i z |\ln t|}}{2 (i z + p)^{\ts}}
\nonumber\\
& = & \frac{\pi  e^{-p |\ln t|}}{\Gamma(\ts) |\ln t|^{1- \ts}} = \frac{\pi t^p}{\Gamma(\ts) (-\ln t)^{1- \ts}}
\quad (0 \leq t \leq 1).
\label{FD1tl1} \eea 
We note that in this case $e^{-p |\ln t|} = e^{p \ln t} = t^p$. The result Eqs.~(\ref{FD1tg1}) and (\ref{FD1tl1}) then give us the final result Eq.~(\ref{FD1pts}) for the characteristic function $F_{\cDo}(t)_{(p,\ts)}$.

This concludes the proof of Theorem 4.

The Theorem was proven for the case when the power index is $0 < \ts < 1$.
In Ref.~\cite{renmod}, the characteristic function was evaluated for the case of the IR renormalons with $\ts=1$ and $\ts=2$ (simple and double poles). It is interesting that the formula of Theorem 4 can be applied even in these cases $\ts=1,2$.

Further, in the limit $\ts \to 0$ we can take into account that
\be
\lim\limits_{\ts \to 0} \pi \frac{(-1)}{\ts} \left( \frac{1}{(p-u)^{\ts}} - 1 \right) - \pi \ln p = \pi \ln \left( 1-\frac{u}{p} \right).
  \label{tsto0} \ee
  This means that in such a case we have to take in the result (\ref{FD1tl1}) the limit $\ts \to 0$. Furthermore, $1/(\ts \Gamma(\ts)) \to 1$ when $\ts \to 0$.  However, since such $\tB[\cDo](u)_{(p,0)}$ does not produce the constant term $\sim u^0$ in its expansion, the corresponding $\cDo(Q^2)$ does not contain the leading term $a(Q^2)^1$, and this term has to be subtracted. It turns out that we then get 
  \bes
  \label{ps0}
\bea
 \tB[\cDo](u)_{(p,0)} &=& \pi \ln \left( 1 - \frac{u}{p} \right) \; \Rightarrow
\label{ps0a} \\ 
\cDo(Q^2)_{(p,0)} &=& \pi \int_0^1 \frac{dt}{t} \frac{t^p}{\ln t} \left[ a(t Q^2) - a(Q^2) \right],
\label{ps0b} \\
\cD(Q^2)_{(p,0)} &=& \pi \int_0^1 \frac{dt}{t} \frac{t^p}{\ln t} \left[ \ta_{\nu_0}(t Q^2) - \ta_{\nu_0}(Q^2) \right].
\label{ps0c} \eea \ees
The corresponding expansion (\ref{Dptlpt}) of $\cD(Q^2)_{(p,0)}$ has $d_0(\nu_0;\kappa) = \td_0(\nu_0;\kappa) = 0$ in such a case (while in the canonical case this coefficient was equal to unity). 

As an aside, we mention that quite an analogous proof can be performed for the version of Theorem 4 for the UV renormalon (at $u=-p < 0$):
\bes
\label{Th4UV}
\bea
\tB[\cDo](u)_{(-p,\ts)} &=& \frac{\pi}{(p+u)^{\ts}}, \; \Rightarrow
\label{Th4UVa} \\
F_{\cDo}(t) &=& \Theta(t - 1) \frac{\pi}{\Gamma(\ts) \; t^p (\ln t)^{1- \ts}}
\label{Th4UVb} \eea \ees
where $p>0$ and $0 < \ts < 1$. In Ref.~\cite{renmod}, the UV cases for $\ts=1$ and $\ts=2$ were investigated, and it turns out that the formula (\ref{Th4UVb}) can be applied even in these cases $\ts=1, 2$.

\section{Renormalisation scheme dependence of the expansion coefficients}
\label{app:RSch}

In this Appendix we present the change of the expansion coefficients $d_n(\nu_0;\kappa)$ that appear in the expansion (\ref{Dpt}), or $f_n(\nu_0;\kappa)$ in the expansion (\ref{Fpt}), for the general case when $\nu_0 \not= 1$ in general. The relations will be equal for the spacelike coefficients $d_n$ and timelike coefficients $f_n$, because they are based on the same principle of the renormalisation scheme independence of the observables.

The renormalisation scheme dependence of the coefficients and the couplings is parametrised by the parameters $c_j \equiv \beta_j/\beta_0$ ($j=2,3,\ldots$) appearing in the RGE (\ref{RGE}).\footnote{We note that in the mass-independent schemes, such as $\MSbar$, the first two coefficients $\beta_0$ and $c_1$ are universal.} First we note that the running coupling $a(\mu^2; c_2, c_3, \ldots) \equiv a$ has the scheme-dependence given by the following relations (cf.~App.~A of \cite{Stevenson}, and App.~A of \cite{GCRK63}):
\bes
\label{aRSch}
\bea
\frac{\partial a}{\partial c_2} & = & a^3 + \frac{c_2}{3} a^5 + {\cal O}(a^6),
\label{ac2} \\
\frac{\partial a}{\partial c_3} & = & \frac{1}{2}a^4 - \frac{c_1}{6} a^5 + {\cal O}(a^6),
\label{ac3} \\
\frac{\partial a}{\partial c_4} & = & \frac{1}{3} a^5 + {\cal O}(a^6).
\label{ac4} \eea \ees
When we apply these relations to the physical condition of the scheme dependence of the observable $\cD(Q^2)$ of Eq.~(\ref{Dpt}), namely $\partial \cD(Q^2)/\partial c_j$ ($j=2,3,4$), we immediately obtain\footnote{Note that $\kappa \equiv \mu^2/Q^2$ is considered fixed here, and we denote everywhere the values of parameters and coefficients in the $\MSbar$ scheme with bar. For simplicity, we also denoted $d_n(\kappa; {\bar c}_2,\ldots,{\bar c}_n)$ simply as ${\bar d}_n(\kappa)$.}
\bes
\label{dnRSch}
\bea
d_1(\kappa) & = & {\bar d}_1(\kappa), \qquad d_2(\kappa; c_2) = {\bar d}_2(\kappa) - \nu_0 (c_2 - {\bar c}_2),
\label{d1d2RSch} \\
d_3(\kappa; c_2, c_3) & = & {\bar d}_3(\kappa) - (\nu_0+1) (c_2 - {\bar c}_2) {\bar d}_1(\kappa) - \frac{1}{2} \nu_0 (c_3 - {\bar c}_3),
\label{d3RSch} \\
d_4(\kappa; c_2, c_3, c_4) & = & {\bar d}_4(\kappa) - (\nu_0+2) (c_2 - {\bar c}_2)  {\bar d}_2(\kappa)- \frac{1}{2} (\nu_0+1) (c_3 - {\bar c}_3) {\bar d}_1(\kappa) 
\nonumber \\ &&
- \frac{1}{6} \nu_0 (c_2^2 - {\bar c}_2^2)
+ \frac{1}{2} \nu_0 (\nu_0+2) (c_2 - {\bar c}_2)^2 + \frac{1}{6} \nu_0 c_1 (c_3 - {\bar c}_3) - \frac{1}{3} \nu_0 (c_4 - {\bar c}_4).
\label{d4RSch}
\eea \ees
As mentioned, the same relations are valid for the coefficients $f_n$.
The corresponding scheme transformation relations for the coefficients $\td_n(\kappa;c_2,\ldots,c_n)$ [or $\tf_n(\kappa;c_2,\ldots,c_n)$] are obtained then by the direct use of the above transformations and the relations (\ref{tdd}) [or: (\ref{tff})], where we note that the transformation coefficients $\tk_{n-s}(\nu_0+s)$ there are independent of the scheme parameters $c_j$ ($j \geq 2$), and even independent of the renormalisation scale parameter $\kappa$, cf.~\cite{GCAK}.

In our implementation of our specific case, we know ${\bar f}_n$ (at $\kappa=1$) for $n=0,1,2,3$ [cf.~Eqs.~(\ref{f0f1})-(\ref{f2f3})], i.e., we implement only the relations (\ref{d1d2RSch})-(\ref{d3RSch}), for $f_n$'s, where $c_2$ and $c_3$ are the scheme parameters of the MiniMOM scheme with $n_f=3$ \cite{MiniMOM}.

\begingroup \color {black}
\section{Mathematical and numerical aspects of the formalism}
\label{app:matnum}

In this Appendix we discuss some mathematical and numerical aspects of the described formalism of resummation.

One of the central results of the resummation procedure is the resummation formula Eq.~(\ref{D1res}) for $\cDo(Q^2)$ and its $\nu_0$-extension Eq.~(\ref{Dres}). The (characteristic) function $F_{\cDo}(t)$ appearing in the resummations Eqs.~(\ref{D1res}) and Eq.~(\ref{Dres}) was obtained in such a way as to fufill the sum rules Eqs.~(\ref{tdn1SR}). Namely, when these sum rules were muliplied by $u^n/(n! \beta_0^n)$ and summed over $n$, and the order of the summation and the integration (over $dt$) was exchanged, the relation Eq.~(\ref{tBD1ukap}) was obtained, i.e., the Mellin transform of $F_{\cDo}(t)$ turned out to be the modified Borel transform $\tB[\cDo](u;\kappa)$ that had been defined by the power expansion Eq.~(\ref{tBD1def}). The inversion of this formula (inverse Mellin) then gave us immediately the (characteristic) function $F_{\cDo}(t)$ Eq.~(\ref{FD1}) in terms of $\tB[\cDo](u;\kappa)$.

  In this above procedure, one may wonder whether the exchange of the order of summation (over $n$) and integration (over $dt$) was justified. Sufficient condition for the legitimacy of this exchange is, according to Fubini's theorem for series, that the series is absolutely convergent in the integral sense, i.e., if the sum of the integrals (over the absolute values of the integrands) converges. It is unclear if this condition is always fufilled in practice. Nonetheless, for the modified Borel $\tB[\cDo](u)$ considered in our application, i.e., Eq.~(\ref{tBD1ans4}) (and Table \ref{tabtK}), we checked numerically that the obtained characteristic function $F_{\cDo}(t)$ Eq.~(\ref{FD1fin}) fulfills all the sum rules (\ref{tdn1SR}).\footnote{\label{SRmod}
  Where, due to the overall exponential factor $\exp(\tK^{(1)}_e u)$ in $\tB[\cDo](u)$, we have to replace $\ln(t/\kappa)$ in the integrands in Eqs.~(\ref{tdn1SR}) by $\ln(t \exp(-\tK^{(1)}_e)/\kappa)$.}
In Fig.~\ref{figFD1}, this characteristic function is presented.  
\begin{figure}[htb] 
\centering\includegraphics[width=80mm]{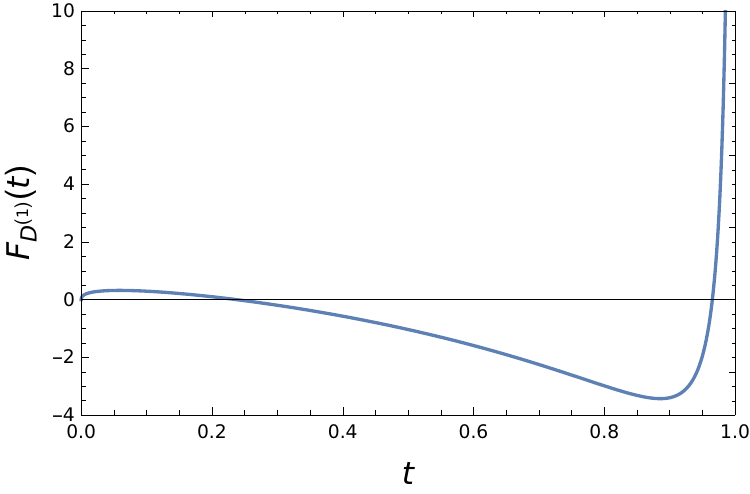}
 \vspace{4pt}
 \caption{\footnotesize The characteristic function $F_{\cDo}(t)$ of Eq.~(\ref{FD1fin}), with the values of $\tK_q^{(1)}$ parameters ($q=\pm,0$) as given in Table \ref{tabtK}, for $n_f=3$.}
 \label{figFD1}
\end{figure}
  
  Another exchange of the order of summation and integration was referred to when argued that the resummation (\ref{D1res}) for $\cDo(Q^2)$ [and analogously (\ref{Dres}) for $\cD(Q^2)$] is correct. Namely, this can be checked by Taylor-expanding the coupling (in the integrand) at $t Q^2$ there around the point $\kappa Q^2$ [cf.~Eqs.~(\ref{atintakap}) and (\ref{tatintakap})], exchanging the order of the (Taylor-)summation and integration over $dt$, and then using the (abovementioned) sum rules (\ref{tdn1SR}); this then led to the original expansion (\ref{D1def}) for $\cDo(Q^2)$ and (\ref{Dlpt}) for $\cD(Q^2)$.

  We point out that these claims are formal, but mathematically problematic, for two reasons: (a) in pQCD, the integrands in the resummation integrals (\ref{D1res}) and (\ref{Dres}) are ill-defined at low positive $t Q^2$ due to Landau singularities of the pQCD couplings $a(t Q^2)$ and $\ta_{\nu_0}(t Q^2)$, and even their Taylor-expansions (around point $\kappa Q^2$) are ill-defined and in general not convergent; (b) the formally obtained expansions (\ref{D1def}) and  (\ref{Dlpt}) are asymptotically divergent for all $Q^2$.

  Here we will discuss how the points (a) and (b) are addressed once we replace the pQCD coupling $a(t Q^2)$ by an AQCD coupling $\A(t Q^2)$  [and $\ta_{\nu_0}(t Q^2)$ by $\tA_{\nu_0}(t Q^2)$], such as the 3$\delta$AQCD coupling that is described in Sec.~\ref{sec:AQCD} and (indirectly) applied in Sec.~\ref{sec:impl}. We will argue in this (AQCD) version how the exchange of order works for $\cDo(Q^2)$ resummation Eq.~(\ref{D1res}), but the argument for $\cD(Q^2)$ resummation Eq.~(\ref{Dres}) is almost the same.

  In the specific case of Sec.~\ref{sec:impl}, the resummation of $\cDo(Q^2)$ has the form Eq. ~(\ref{D1res1}),
where the characteristic function $F_{\cDo}(t)$ is given in Eq.~(\ref{FD1fin}) and the values of the four parameters $\tK_{q}^{(1)}$ ($q=e, \pm, 0$) are given in Table \ref{tabtK}. We now apply Taylor-expansion of $\A(t e^{-\tK_e} Q^2)$ around $\A(\kappa Q^2)$ [i.e., the AQCD version of the Taylor-expansion Eq.~(\ref{atintakap}), where $t \mapsto t e^{-\tK_e^{(1)}}$], and the renormalisation scale parameter is assumed to be $\kappa \sim 1$
\be
\cDo(Q^2)_{\rm res.} = \int_0^{\infty} \frac{dt}{t}  F_{\cDo}(t) \left[
  \A(\kappa Q^2) + \sum_{n=1}^{\infty}  (- \beta_0)^n \ln^n \left( \frac{t e^{-\tK_e^{(1)}}}{\kappa} \right) \tA_{n+1}(\kappa Q^2) \right].
\label{D1res2} \ee
In contrast to the pQCD case, the Taylor-expansion in the integrand is a series with well defined terms, for any complex $Q^2$ outside the timelike axis $Q^2 < 0$. It is convergent for any $(0<t<1)$ that is at least not very close to zero, and there it reproduces the value of $\A(t e^{-\tK_e^{(1)}} Q^2)$; we will not go into details of this claim, but we checked it numerically.


At this point, when we exchange the order of integration and summation in Eq.~(\ref{D1res2}) and apply the sum rules (\ref{tdn1SR})\footnote{
We recall that those sum rules have been numerically checked in our case, for any $n$. See also footnote \ref{SRmod}.}, this then leads to the (original) expansion in logarithmic derivatives Eq.~(\ref{D1def}), in our AQCD framework
\be
\cDo(Q^2) \equiv \sum_{n=0}^{\infty} \td_n(1;\kappa) \tA_{1+n}(\kappa Q^2).
 \label{D1defA} \ee 
 This last step of exchanging the order of (Taylor) summation and integration (over $dt$) is a central step in proving, or motivating, our resummation formula (\ref{D1res}) [or analogously: (\ref{Dres})] in AQCD. As we see, this step is delicate, and we see no clear way to mathematically prove it, because it leads from a convergent well defined integral expression (in AQCD) to an expansion (\ref{D1defA}) that is in general asymptotically divergent.

 We can repeat all these steps in an almost equal manner for the quantity $\cD(Q^2)$ whose general resummation formula is in Eq.~(\ref{Dres}). In our more specific AQCD case of Sec.~\ref{sec:impl}, the resummation of $\cD(Q^2)$ is given in Eq.(\ref{Dres1}).
 
 The question that may now naturally arise is, whether the obtained resummations (\ref{D1res1}) and (\ref{Dres1}), within QCD free of Landau singularities (AQCD), are the only possible (unique) resummations of the asymptotically divergent expansions Eq.~(\ref{D1defA}) for $\cDo(Q^2)$ and the (AQCD-)version of the expansion (\ref{Dlpt}) for $\cD(Q^2)$
 \be
\cD(Q^2) =  \sum_{n=0}^{\infty} \td_n(\nu_0;\kappa) \tA_{\nu_0+n}(\kappa Q^2).
 \label{DlptA} \ee
 We do have numerical indications that these resummations are unique, within any given AQCD. Namely, we can construct, when knowing in the truncated series (\ref{D1defA}) the first $2 N$ (even) number of coefficients ($\td_n$, $n=0,\ldots,2 N -1$), a resummation ${\cal G}_{\cDo}^{[N/N]}(Q^2)$ that can be regarded as an extension (generalisation) of diagonal Pad\'e approximants. These approximants were introduced in \cite{BGApQCD1}\footnote{
 They were extended to the case of truncated series with odd number of terms $2 N -1$ in \cite{BGApQCD2}.}
 and were applied only with mixed success in pQCD, due to the Landau singularities of the coupling. Later, these approximants were applied in AQCD variants \cite{BGA,anOPE,Techn}, successfully, due to the absence of the Landau singularities of the coupling. We refer for the detailed construction and further explanations to these references, as well as to \cite{renmod} (Sec.~IV there). These approximants have a relatively simple form
 \be
 {\cal G}_{\cDo}^{[N/N]}(Q^2) = \sum_{j=1}^N \tal_j \A(\kappa_j Q^2),
 \label{GNN} \ee
 where, operationally, the $2 N$ parameters $\tal_j$ and $\kappa_j$ are determined by the first $2 N$ coefficients $\td_n(1;\kappa)$ ($n=0,\ldots, 2 N - 1$), and are completely independent of the renormalisation scale parameter $\kappa$ used in the truncated series (\ref{D1defA}). These $2 N$ parameters are determined by requiring that this approximant formally fulfill the $2 N$-order approximation requirement, namely that the Taylor expansions of $\A(\kappa_j Q^2)$ terms in Eq.~(\ref{GNN}) around $\A(\kappa Q^2)$ lead to the reproduction of the original truncated series up to (and including) ${\cal O}(\tA_{2 N})$
\be
{\cal G}_{\cDo}^{[N/N]}(Q^2) - \sum_{n=0}^{2 N-1} \td_n(1;\kappa) \tA_{1+n}(\kappa Q^2) = {\cal O}(\tA_{1 + 2 N}).
\label{GNNapp} \ee
We recall that in the perturbative regime (large $|Q^2|$) we have $\tA_{1 + 2 N}(\kappa Q^2) \approx a(\kappa Q^2)^{1 + 2 N}$.
In practice, we Taylor-expand $\A(\kappa_j Q^2)$ around $\A(\kappa Q^2)$ using the AQCD version of Eq.~(\ref{atintakap}) [i.e., replacing there $\ta_{n+1} \mapsto \tA_{n+1}$, $a \mapsto \A$, and $t \mapsto \kappa_j$], and the requirement (\ref{GNNapp}) then leads to $2 N$ relations
\be
\sum_{j=1}^N \tal_j \ln^n \left( \frac{\kappa_j}{\kappa} \right) = \frac{(-1)^n}{\beta_0^n} \td_n(1; \kappa) \quad (n=0, \ldots, 2 N - 1).
\label{rel2N} \ee
These $2 N$ relations then allow us to obtain the $2 N$ parameters $\tal_j$ and $\kappa_j$ ($j=1,\ldots,N$) if we know the first $2 N$ coefficients $\td_n(1; \kappa)$ ($n=0,\ldots, 2 N - 1$). It was proven in \cite{BGApQCD1} that the obtained values of the $2 N$ (dimensionless) parameters $\tal_j$ and $\kappa_j$ are completely independent of the renormalisation scale parameter $\kappa \equiv \mu^2/Q^2$.\footnote{
This can be proven also in the following way: The above relation (\ref{rel2N}) for $n-k$ and $\kappa=1$ is multiplied by $\beta_0^n (\ln^k \kappa) (-1)^k  \binom{n}{k}$ and summed on both sides over $k=0,\ldots, n$. This then gives us the left-hand side of Eq.~(\ref{rel2N}) multiplied by $\beta_0^n$, and on the right-hand side we obtain $(-1)^n \sum_{0 \leq k \leq n} \binom{n}{k} (\beta_0 \ln \kappa)^k \td_{n-k}(1;1)$. This expression was shown to be equal exactly to $(-1)^n \td_n(1; \kappa)$ in \cite{renmod} [Eq.~(16) there]. We thus obtain the identities (\ref{rel2N}) for $\kappa$, with the same $\tal_j$ and $\kappa_j$ that we had for $\kappa=1$. Therefore, $\tal_j$ and $\kappa_j$ are exactly independent of the chosen renormalisation scale parameter $\kappa$.}
Since these parameters are dimensionless, they are then independent of both $Q^2$ and $\mu^2$. In an earlier work \cite{Gardi}, the (usual) diagonal Pad\'e $[N/N](a)$ was applied to truncated power expansions (in pQCD), and it was shown that these approximants were renormalisation scale independent in the one-loop approximation (but not beyond).

It is now straightforward to show that, since ${\cal G}_{\cDo}^{[N/N]}(Q^2)$ of Eq.~(\ref{GNN}) approximates formally correctly $\cDo(Q^2)$, i.e., Eq.~(\ref{GNNapp}) is fulfilled, then the corresponding approximant
 \be
 {\cal G}_{\cD}^{[N/N]}(Q^2) = \sum_{j=1}^N \tal_j \tA_{\nu_0}(\kappa_j Q^2),
 \label{GNND} \ee
 with the same values of $\tal_j$ and $\kappa_j$, formally correctly reproduces the truncated expansion of $\cD(Q^2)$ Eq.~(\ref{DlptA}) up to (and including) ${\cal O}(\tA_{2 N -1 +\nu_0})$
\be
{\cal G}_{\cD}^{[N/N]}(Q^2) - \sum_{n=0}^{2 N-1} \td_n(\nu_0;\kappa) \tA_{\nu_0+n}(\kappa Q^2) = {\cal O}(\tA_{\nu_0 + 2 N}).
\label{GNNDapp} \ee
We recall that formally, at high $|Q^2|$, we have $\tA_{\nu_0 + 2 N}(\kappa Q^2) \approx a(\kappa Q^2)^{\nu_0+ 2 N}$. This relation (\ref{GNNDapp}) can be proven in the following way. We Taylor-expand the terms $\tA_{\nu_0}(\kappa_j Q^2)$ in ${\cal G}_{\cD}^{[N/N]}(Q^2)$ around $\tA_{\nu_0}(\kappa Q^2)$, using the AQCD version of Eq.~(\ref{tatintakap}) [i.e., where $\ta_{\nu_0+n} \mapsto \tA_{\nu_0+n}$, $\ta_{\nu_0} \mapsto \tA_{\nu_0}$, and $t \mapsto \kappa_j$]. The requirement (\ref{GNNDapp}) then leads to the very same set of $2 N$ relations written in (\ref{rel2N}).\footnote{
We recall that we have to take into arrount the simple scaling relation (\ref{tdn1}) between $\td_n(\nu_0;\kappa)$ and $\td_n(1;\kappa)$.}
But the relations (\ref{rel2N}) are fulfilled, as the parameters $\tal_j$ and $\kappa_j$ also appear in ${\cal G}_{\cDo}^{[N/N]}(Q^2)$, Eq.~(\ref{GNN}).

Now the approximants ${\cal G}_{\cF}^{[N/N]}(\sigma)$ for the corresponding timelike quantity $\cF(\sigma)$, Eq.~(\ref{Fdef}), are obtained by contour integration of the terms in Eq.~(\ref{GNND}) term-by-term, giving
 \be
 {\cal G}_{\cF}^{[N/N]}(\sigma) = \sum_{j=1}^N \tal_j \tH_{\nu_0}(\kappa_j \sigma),
 \label{GNNF} \ee
 where the timelike couplings $\tH_{\nu_0}(\kappa_j \sigma)$ are defined in Eqs.~(\ref{tH}).\footnote{\label{restrGNNF}
Some of the approximants ${\cal G}_{\cDo}^{[N/N]}(Q^2)$ have nonreal parameters $\kappa_j$ (in complex conjugate pairs), and that poses no problem for the evaluation of ${\cal G}_{\cDo}^{[N/N]}(Q^2)$ and ${\cal G}_{\cD}^{[N/N]}(Q^2)$, but poses a problem for the evaluation of ${\cal G}_{\cF}^{[N/N]}(\sigma)$. We note that in Eq.~(\ref{F1def}), it is assumed that $\sigma$ is positive ($\sigma > 0$) in the contour integral. In our cases, $\sigma \mapsto \kappa_j \sigma$. In our cases, it turns out that for $6 \leq N \leq 11$ there is a pair of complex nonreal $\kappa_j$'s, and therefore we do not evaluate  ${\cal G}_{\cF}^{[N/N]}(\sigma)$ for such $N$ values. All $\kappa_j$, when having real values, are positive (though some are very close to zero).}
 Again, analogous formal approximation relations can be shown to hold
\be
{\cal G}_{\cF}^{[N/N]}(Q^2) - \sum_{n=0}^{2 N-1} \td_n(\nu_0;\kappa) \tH_{\nu_0+n}(\kappa \sigma) = {\cal O}(\tH_{\nu_0 + 2 N}),
\label{GNNFapp} \ee
where we recall that for high (perturbative) $\sigma$ we have $\tH_{\nu_0 + 2 N}(\sigma) \approx \ths_{\nu_0 + 2 N}(\sigma) \approx \ta_{\nu_0 + 2 N}(\sigma) \approx a(\sigma)^{\nu_0+ 2 N}$.

We now apply all these approximants to our case of Sec.~\ref{sec:impl} in 3$\delta$AQCD, for various values of $Q^2$ (and $\sigma$) and various increasing $N$ indices, and compare them to the corresponding resummed values obtained via the integrals (\ref{D1res1}), (\ref{Dres1}) and (\ref{FrestKA2}), i.e., for the case of the modified Borel transform $\tB[\cDo](u)$ of Eq.~(\ref{tBD1ans4}) with the values of parameters $\tK_q^{(1)}$ given in Table \ref{tabtK} (for $n_f=3$).
We recall that all the evaluations are performed in the mentioned Lambert MiniMOM (LMM) scheme of the 3$\delta$AQCD framework, and for the 3$\delta$AQCD coupling we choose the central case parameters: $n_f=3$, $\alpha_s^{\MSbar}(M_Z^2)=0.1180$; $M_1=0.150$ GeV, cf.~Table \ref{tab3dnf3}.
In addition, we also include, for comparison, the results of the simple pQCD truncated series (in the LMM scheme) for $\cDo$ and $\cD$
\bes
\label{TPSp}
\bea
\cDo(Q^2)^{[N_{\rm max}]} &=& \sum_{n=0}^{N_{\rm max}-1} d_n(1; \kappa) a(\kappa Q^2)^{1 + n},
\label{D1p}
\\
\cD(Q^2)^{[N_{\rm max}]} &=& \sum_{n=0}^{N_{\rm max}-1} d_n(\nu_0; \kappa) a(\kappa Q^2)^{\nu_0+ n},
\label{Dp} \eea \ees

All these results are presented in Tables \ref{tabTPSD1}-\ref{tabGNNF}. The pQCD results were obtained using the corresponding underlying pQCD coupling $a(Q^2)$ (in the same LMM scheme).
\begin{table}
  \caption{The values of the truncated expansions for $\cDo(Q^2)$, for $Q^2= 2 \ {\rm GeV}^2$ and $Q^2=1. \times \exp(i 0.5) \ {\rm GeV}^2$, for pQCD [cf.~Eqs.(\ref{D1p})] and 3$\delta$AQCD [cf.~Eq.~(\ref{D1defA})], for various numbers of terms involved ($N_{\rm max}$)  }
  \label{tabTPSD1}
\begin{ruledtabular}
\begin{tabular}{l|l|lllll}
 $Q^2$ (${\rm GeV}^2$) & type & $N_{\rm max}=1$ & $N_{\rm max}=2$ & $N_{\rm max}=3$ & $N_{\rm max}=4$ & $N_{\rm max}=5$  \\
\hline
$2.$ & pQCD & 0.1339 & 0.2676 & 0.4999 & 1.0064 & 2.5279 \\
$2.$ & AQCD & 0.1246 & 0.2610 & 0.4130 & 0.3658 & -0.6394 \\
\hline
$1.\times \exp(i 0.5)$ & pQCD & $(0.1453 - 0.0727 i)$ & $(0.2633 - 0.2303 i)$ & $(0.3370 - 0.6389 i)$ & $(0.0278 - 1.6928 i)$ & $(-2.7000 - 4.6237 i)$ \\
$1.\times \exp(i 0.5)$ &  AQCD & $(0.1559 - 0.0263 i)$ & $(0.3376 - 0.0433 i)$ &  $(0.4478 + 0.0673 i)$ & $(-0.0228 + 0.3928 i)$ & $(-1.8882 + 0.0495 i)$ \\
\end{tabular}
\end{ruledtabular}
\end{table}
\begin{table}
  \caption{The same as in Table \ref{tabTPSD1}, but now for $\cD(Q^2)$ [cf.~Eqs.~(\ref{Dp}) and (\ref{DlptA})].}
\label{tabTPSD}
\begin{ruledtabular}
\begin{tabular}{l|l|lllll}
 $Q^2$ (${\rm GeV}^2$) & type & $N_{\rm max}=1$ & $N_{\rm max}=2$ & $N_{\rm max}=3$ & $N_{\rm max}=4$ & $N_{\rm max}=5$  \\
\hline
$2.$ & pQCD & 0.5116 & 0.6593 & 0.8334 & 1.1343 & 1.9160 \\
$2.$ & AQCD & 0.4752 & 0.6397 & 0.8097 & 0.9355 & 0.6699 \\
\hline
$1.\times \exp(i 0.5)$ & pQCD & (0.5391 - 0.0840 i) & (0.6949 - 0.1950 i) & (0.8232 - 0.4366 i) & (0.8369 - 1.0099 i) & (0.0665 - 2.6456 i) \\
$1.\times \exp(i 0.5)$ &  AQCD & (0.5120 - 0.0326 i) &  (0.7287 - 0.0721 i) & (0.9410 - 0.0648 i) & (0.9360 + 0.1255 i) & (-0.0413 + 0.4880 i) \\
\end{tabular}
\end{ruledtabular}
\end{table}

\begin{table}
  \caption{The same as in Tables \ref{tabTPSD1} and \ref{tabTPSD}, but now for the Pad\'e-related approach (\ref{GNN}) and (\ref{GNND}): ${\cal G}^{[N/N]}(Q^2)$, for $\cDo(Q^2)$ and $\cD(Q^2)$, respectively, for increasing index $N$. In the last column, for comparison, the resummed values are given, according to Eqs.~(\ref{D1res1}) and (\ref{Dres1}).}
\label{tabGNND1D}
\begin{ruledtabular}
\begin{tabular}{l|l|lllll|l}
 $Q^2$ [${\rm GeV}^2$] & quantity & $N=2$ & $N=8$ & $N=12$ & $N=15$ & resummed \\
\hline
$2.$ & ${\cal G}_{\cDo}^{[N/N]}(Q^2)$ & 0.1454 & 0.1864 & 0.1826 & 0.1822  & 0.1823 ($\cDo(Q^2)$) \\
 $2.$ &  ${\cal G}_{\cD}^{[N/N]}(Q^2)$  & 0.6300 & 0.6576 & 0.6597 & 0.6591 & 0.6592 ($\cD(Q^2)$) \\
\hline
$1.\times \exp(i 0.5)$ &  ${\cal G}_{\cDo}^{[N/N]}(Q^2)$ & $(0.1582 - 0.0038 i)$ & $(0.1933 + 0.0053 i)$ & $(0.1849 + 0.0057 i)$ & $(0.1862 + 0.0047 i)$ & $(0.1855 + 0.0050 i)$ \\
$1.\times \exp(i 0.5)$ & ${\cal G}_{\cD^{[N/N]}}(Q^2)$ &  $(0.6609 - 0.0222 i)$ & $(0.6995 - 0.0265 i)$ & $(0.6978 - 0.0241 i)$ & $(0.6976 - 0.0246 i)$ & $(0.6976 - 0.0244 i)$ \\
\end{tabular}
\end{ruledtabular}
\end{table}

\begin{table}
  \caption{The same as in Table \ref{tabGNND1D}, but now the approximants ${\cal G}^{[N/N]}_{\cF}(\sigma)$ for the corresponding timelike quantity $\cF(\sigma)$, for $\sigma = m_c^2$ ($\approx 2.789 \ {\rm GeV}^2)$ and $\sigma = 1 \ {\rm GeV}^2$, using the Pad\'e-related approach (\ref{GNNF}), for increasing index $N$. No values for $6 \leq N \leq 11$ are given, due to the problem discussed in footnote \ref{restrGNNF}. In the last column, for comparison, the resummed values are given, according to Eq.~(\ref{FrestKA2}).}
\label{tabGNNF}
\begin{ruledtabular}
\begin{tabular}{l|lllllll|l}
 $\sigma$ [${\rm GeV}^2$] & $N=2$ & $N=3$ & $N=5$ & $N=12$ & $N=13$ & $N=14$ & $N=15$ & ${\cal F}(\sigma)_{\rm res.}$  \\
\hline
$m_c^2$ & 0.556323 &  0.71964 & 0.691541 &  0.6309 & 0.6319 & 0.6330 & 0.6351 & 0.6365 \\
 $1. \ {\rm GeV}^2$ &  0.5098 & 0.5179 & 0.5552 & 0.7437 & 0.7334 & 0.7190 & 0.7071 & 0.7134 \\
\end{tabular}
\end{ruledtabular}
\end{table}

We can see in Tables \ref{tabTPSD1}-\ref{tabTPSD} that in general the truncated series in the logarithmic derivatives, Eqs.~(\ref{D1defA}) and (\ref{DlptA}), are asymptotically divergent and have to be truncated at its smallest term - the fourth term, sometimes the third term. The pQCD truncated series, Eqs.~(\ref{TPSp}), behave often even worse and should be truncated at the second or the third term.
The truncated series have large (truncation) uncertainties and differ significantly from the resummed values $\cDo(Q^2)_{\rm res.}$ and $\cD(Q^2)_{\rm res.}$ (given in the last column in Table \ref{tabGNND1D}).

However, and crucially, we can see in Table \ref{tabGNND1D} that the approximants (\ref{GNN}) and (\ref{GNND}), i.e., the extended (generalised) versions of the diagonal Pad\'e's, show numerical convergence to the resummed values as their index $N$ increases. For very large $N$ ($N > 12$) we do see that a certain level of weak oscillation sets in, around the resummed value.
These oscillations are more pronounced for the timelike quantity $\cF(\sigma)$, as seen in Table \ref{tabGNNF}.

The mentioned (small) oscillations in ${\cal G}_{\cDo}^{[N/N]}(Q^2)$ and ${\cal G}_{\cD}^{[N/N]}(Q^2)$, and somewhat larger oscillations in ${\cal G}_{\cF}^{[N/N]}(\sigma)$, when $N$ is large and increases, arise apparently due to numerical instabilities, because several values of $\kappa_j$ are very small (close to zero), and there are cancellations of several terms involved. The considered series are definitely difficult to deal with for these Pad\'e-related approximants, because the series is very divergent, and this is reflected in the fact that the renormalons are at $u=1/2$, i.e., very close to zero.\footnote{
In general, the approximants converge faster for $\cD(Q^2)$ than for $\cDo(Q^2)$. This is so because the series of $\cDo(Q^2)$ is even more divergent than that for $\cD(Q^2)$. This can be seen also in Theorem 3, where the power of the singularity at $u=1/2$ for $\tB[\cDo](u)$ ($\ts_0 + 1 - \nu_0$) is larger than the corresponding power ($\ts_0$) for $\tB[\cD](u)$. We note that $\nu_0=1/3$ when $n_f=3$. This can also been seen by comparing the last two columns of Table \ref{tabdn}.}
 In \cite{renmod} (Sec.~IV there), these approximants were applied to the Adler function, whose first IR renormalon is at $u=2$ (and, of course, $\nu_0=1$ there), and it turned out that these approximants converge numerically perfectly to the resummed values of the Adler function.   

We thus conclude that we have indications (via comparison with Pad\'e-related approximants) that our formalism for the resummations, Eqs.~(\ref{D1res1}), (\ref{Dres1}) and (\ref{FrestKA2}), leads to resummed values that are unique in AQCD frameworks, i.e., in the QCD frameworks whose couplings have no Landau singularities.

\endgroup

\end{document}